\pdfoutput=1
\RequirePackage{ifpdf}
\ifpdf
\documentclass[pdftex]{sigma}
\else
\documentclass{sigma}
\fi

\usepackage{bm,cite}

\numberwithin{equation}{section}

\numberwithin{theorem}{section}
\numberwithin{proposition}{section}
\numberwithin{lemma}{section}
\numberwithin{corollary}{section}
\numberwithin{definition}{section}
\numberwithin{example}{section}
\numberwithin{remark}{section}
\numberwithin{note}{section}

\begin{document}
\allowdisplaybreaks

\renewcommand{\thefootnote}{$\star$}

\renewcommand{\PaperNumber}{103}

\FirstPageHeading

\ShortArticleName{Whitham's Method and Dubrovin--Novikov Bracket}

\ArticleName{Whitham's Method and Dubrovin--Novikov Bracket\\ in
Single-Phase and Multiphase Cases\footnote{This
paper is a~contribution to the Special Issue ``Geometrical Methods in Mathematical Physics''.
The full collection is available
at
\href{http://www.emis.de/journals/SIGMA/GMMP2012.html}{http://www.emis.de/journals/SIGMA/GMMP2012.html}}}

\Author{Andrei Ya.~MALTSEV}

\AuthorNameForHeading{A.Ya.~Maltsev}

\Address{L.D.~Landau Institute for Theoretical Physics,
1A Ak.~Semenova Ave.,\\ Chernogolovka, Moscow reg., 142432, Russia}
\Email{\href{mailto:maltsev@itp.ac.ru}{maltsev@itp.ac.ru}}

\ArticleDates{Received April 23, 2012, in f\/inal form December 11, 2012; Published online December 24, 2012}

\Abstract{In this paper we examine in detail the procedure of averaging
of the local f\/ield-theoretic Poisson brackets proposed by B.A.~Dubrovin
and S.P.~Novikov for the method of Whitham.
The main attention is paid to
the questions of justif\/ication and the conditions of applicability of the
Dubrovin--Novikov procedure.
Separate consideration is given to special
features of single-phase and multiphase cases.
In particular, one of the
main results is the insensitivity of the procedure of bracket averaging
to the appearance of ``resonances'' which can arise in the multi-phase
situation.}


\Classification{37K05; 35B10; 35B15; 35B34; 35L65}

\renewcommand{\thefootnote}{\arabic{footnote}}
\setcounter{footnote}{0}

\section{Introduction}

As is well-known, the Whitham method \cite{whith1,whith2,whith3}
is associated with slow modulations of periodic or quasiperiodic
\mbox{$m$-phase} solutions of nonlinear systems
\begin{gather}\label{insyst}
F^{i}(\bm{\varphi},\bm{\varphi}_{t},\bm{\varphi}_{x},\dots)
=0
,\qquad i=1,\dots,n,\qquad
\bm{\varphi}=\big(\varphi^{1},\dots,\varphi^{n}\big),
\end{gather}
which are usually represented in the form
\begin{gather}\label{phasesol}
\varphi^{i}(x,t)=\Phi^{i}\big({\bf k}({\bf U})x+\bm{\omega}({\bf U})t+\bm{\theta}_{0},{\bf U}\big).
\end{gather}

Let us note that we will consider here systems with one spatial
variable $x$ and one time variable $t$.
In these notations the
functions ${\bf k}({\bf U})$ and $\bm{\omega}({\bf U})$ play the
role of the ``wave numbers'' and
``frequencies'' of \mbox{$m$-phase} solutions, while the parameters
$\bm{\theta}_{0}$ represent the ``initial phase shifts''.
The parameters ${\bf U}=(U^{1},\dots,U^{N})$
can be chosen in an arbitrary way, we just assume that they do not
change under shifts of the initial phases of solutions
$\bm{\theta}_{0}$.

The functions $\Phi^{i}(\bm{\theta})$ satisfy the system
\begin{gather}\label{PhaseSyst0}
F^{i}\big({\bf\Phi},\omega^{\alpha}
{\bf\Phi}_{\theta^{\alpha}},
k^{\beta}{\bf\Phi}_{\theta^{\beta}},
\dots\big)\equiv0,\qquad i=1,\dots,n,
\end{gather}
and we have to choose for each value of ${\bf U}$ some
function ${\bf\Phi}(\bm{\theta},{\bf U})$ as having ``zero
initial phase shift''.
The corresponding set of \mbox{$m$-phase} solutions of
\eqref{insyst} can be then represented in the form~\eqref{phasesol}.
For \mbox{$m$-phase} solutions of~\eqref{insyst} we have in this case
${\bf k}({\bf U})=(k^{1}({\bf U}),\dots,k^{m}({\bf U}))$,
$\bm{\omega}({\bf U})=(\omega^{1}({\bf U}),\dots,
\omega^{m}({\bf U}))$,
$\bm{\theta}_{0}=(\theta^{1}_{0},\dots,\theta^{m}_{0})$,
where ${\bf U}=(U^{1},\dots,U^{N})$ are the parameters of a
solution.
We will require also that all the functions
$\Phi^{i}(\bm{\theta},{\bf U})$ are $2\pi$-periodic with respect
to each $\theta^{\alpha}$, $\alpha=1,\dots,m$.

Consider a~set $\Lambda$ of functions
$\bm{\Phi}(\bm{\theta}+\bm{\theta}_{0},{\bf U})$,
depending smoothly on the parameters ${\bf U}$ and satisfying
system~\eqref{PhaseSyst0} for all ${\bf U}$.

In the Whitham approach the parameters ${\bf U}$ and $\bm{\theta}_{0}$
become slowly varying functions of $x$ and $t$,
${\bf U}={\bf U}(X,T)$,
$\bm{\theta}_{0}=\bm{\theta}_{0}(X,T)$,
where $X=\epsilon x$, $T=\epsilon t$
($\epsilon\rightarrow0$).

For the construction of the corresponding asymptotic solution
the functions ${\bf U}(X,T)$ must sa\-tisfy some system of
dif\/ferential equations (the Whitham system).
In the simplest case
(see~\cite{luke}), we try to f\/ind asymptotic solutions
\begin{gather}\label{whithsol}
\varphi^{i}(\bm{\theta},X,T)=\sum_{k\geq0}
\Psi^{i}_{(k)}\left({{\bf S}(X,T)\over\epsilon}+\bm{\theta},X,T\right)\epsilon^{k}
\end{gather}
with $2\pi$-periodic in $\bm{\theta}$ functions
$\bm{\Psi}_{(k)}$ satisfying system~\eqref{insyst}, i.e.
\begin{gather*}
F^{i}\left(\bm{\varphi},\epsilon\bm{\varphi}_{T},
\epsilon\bm{\varphi}_{X},\dots\right)=0
,\qquad i=1,\dots,n.
\end{gather*}

The function ${\bf S}(X,T)=(S^{1}(X,T),\dots,S^{m}(X,T))$
is called the ``modulated phase'' of solution~\eqref{whithsol}.

Assume now that the function
$\bm{\Psi}_{(0)}(\bm{\theta},X,T)$ belongs to the family $\Lambda$
of \mbox{$m$-phase} solutions of~\eqref{insyst}
for all $X$ and $T$.
We have then
\begin{gather}\label{psi0}
\bm{\Psi}_{(0)}(\bm{\theta},X,T)=
\bm{\Phi}\big(\bm{\theta}+\bm{\theta}_{0}(X,T),{\bf U}(X,T)\big),
\end{gather}
and
\begin{gather*}
S^{\alpha}_{T}(X,T)=\omega^{\alpha}({\bf U}(X,T)),\qquad
S^{\alpha}_{X}(X,T)=k^{\alpha}({\bf U}(X,T)),
\end{gather*}
as follows after the substitution of~\eqref{whithsol} into system
\eqref{insyst}.

In the simplest case the functions
$\bm{\Psi}_{(k)}(\bm{\theta},X,T)$
are determined from the linear systems
\begin{gather}\label{ksyst}
{\hat L}^{i}_{j[{\bf U},\bm{\theta}_{0}]}(X,T) \Psi_{(k)}^{j}(\bm{\theta},X,T)=
f_{(k)}^{i}(\bm{\theta},X,T),
\end{gather}
where ${\hat L}^{i}_{j[{\bf U},\bm{\theta}_{0}]}(X,T)$
is a~linear operator def\/ined by the linearization of system
\eqref{PhaseSyst0} on the solution~\eqref{psi0}.
The resolvability
conditions of systems~\eqref{ksyst} in the space of periodic functions
can be written as the conditions of orthogonality of the functions
${\bf f}_{(k)}(\bm{\theta},X,T)$ to all the ``left eigenvectors''
(the eigenvectors of the adjoint operator) of the operator
${\hat L}^{i}_{j[{\bf U},\bm{\theta}_{0}]}(X,T)$ corresponding
to zero eigenvalue.

We should say, however, that the resolvability conditions of
systems~\eqref{ksyst} can actually be quite complicated in general
multi-phase case, since the eigenspaces of the operators
${\hat L}_{[{\bf U},\bm{\theta}_{0}]}$
and~${\hat L}^{\dagger}_{[{\bf U},\bm{\theta}_{0}]}$
on the space of $2\pi$-periodic functions can be rather nontrivial
in the multi-phase situation.
Thus, even the dimension of the
kernels of ${\hat L}_{[{\bf U},\bm{\theta}_{0}]}$
and ${\hat L}^{\dagger}_{[{\bf U},\bm{\theta}_{0}]}$
can depend in a~highly nontrivial way on the values of ${\bf U}$.
In general, the picture arising in the ${\bf U}$-space can be rather
complicated.
As a~result, the determination of the next corrections from
systems~\eqref{ksyst} is impossible in general multiphase situation and
the corrections to the main approximation~\eqref{psi0} have more
complicated and rather nontrivial form~\cite{dobr1,dobr2,DobrKrichever}.

These dif\/f\/iculties do not arise commonly in the single-phase
situation ($m = 1$) where the behavior of eigenvectors of
${\hat L}_{[{\bf U},\bm{\theta}_{0}]}$
and ${\hat L}^{\dagger}_{[{\bf U},\bm{\theta}_{0}]}$, as a~rule,
is quite regular.
The resolvability conditions of system
\eqref{ksyst} for $k=1$
\begin{gather}\label{1syst}
{\hat L}^{i}_{j[{\bf U},\bm{\theta}_{0}]}(X,T)
\Psi_{(1)}^{j}(\bm{\theta},X,T)=
f_{(1)}^{i}(\bm{\theta},X,T)
\end{gather}
with relations
$k_{T}=\omega_{X}$ def\/ine in this case the Whitham system
for the single-phase solutions of~\eqref{insyst} which plays the
central role in considering the slow modulations.

For the multi-phase solutions the Whitham system is usually given
by the orthogonality conditions of the right-hand part of~\eqref{1syst}
to the maximal set of ``regular'' left eigenvectors corresponding to
zero eigenvalues which are def\/ined for all values of ${\bf U}$
and depend smoothly on ${\bf U}$.
As a~rule, this system is
equivalent to the conditions obtained by the averaging of some complete
set of conservation laws of system~\eqref{insyst} having the local
form
\begin{gather*}
P^{\nu}_{t}(\bm{\varphi},\bm{\varphi}_{t},\bm{\varphi}_{x},\dots)
=
Q^{\nu}_{x}(\bm{\varphi},\bm{\varphi}_{t},\bm{\varphi}_{x},\dots)
,\qquad
\nu=1,\dots,N.
\end{gather*}

The Whitham system is written usually as a~system of hydrodynamic type
\begin{gather}\label{WhithamSystem}
U^{\nu}_{T}=V^{\nu}_{\mu}({\bf U})U^{\mu}_{X}
\end{gather}
and gives the main approximation for the connection between
spatial and time derivatives of the parameters $U^{\nu}(X,T)$.
The variables $T$ and $X$ represent the ``slow'' variables
$T=\epsilon t$, $X=\epsilon x$, connected with the variables
$t$ and $x$ by a~small parameter~$\epsilon$.
Thus, the
Whitham system~\eqref{WhithamSystem} represents a~homogeneous
quasi-linear system of hydrodynamic type connecting the derivatives
of slow modulated parameters.

As mentioned above, the construction of the asymptotic series
\eqref{whithsol} in the multi-phase case is impossible in general
situation (see~\cite{dobr1,dobr2,DobrKrichever}).
Nevertheless, the Whitham system
\eqref{WhithamSystem} and the leading term of the expansion
\eqref{whithsol} play the major role in consideration of modulated
solutions also in this case, representing the main approximation for
corresponding modulated solutions.
The corrections to the main
term have in general more nontrivial form than~\eqref{whithsol},
but they also tend to zero in the limit
$\epsilon\rightarrow0$~\cite{dobr1,dobr2,DobrKrichever}.

Let us give here just some incomplete list of the classical
papers devoted to the foundations of the Whitham method
\cite{AblBenny, dm, DobrMaslMFA, dobr1, dobr2, DobrKrichever,
dn1, dn2, dn3, ffm,Haberman1,Haberman2,
Hayes, krichev1, krichev2, luke, Newell, theorsol,
Nov, whith1, whith2, whith3}.
We will be interested here only in Hamiltonian aspects of
the Whitham method.
In the remaining part of the Introduction we will
give the def\/inition of the ``regular'' Whitham system for
the complete regular family of \mbox{$m$-phase} solutions which will
be used everywhere below.

Let us use for simplicity the notation $\Lambda$ both for the family of
the functions $\bm{\Phi}(\bm{\theta}+\bm{\theta}_{0},{\bf U})$
and the corresponding family of \mbox{$m$-phase} solutions of system
\eqref{insyst}, such that we will denote by $\Lambda$ both the families of
the functions~$\bm{\Phi}(\bm{\theta}+\bm{\theta}_{0},{\bf U})$
in the space of $2\pi$-periodic in all~$\theta^{\alpha}$ functions~$\bm{\varphi}(\bm{\theta})$ and
$\bm{\varphi}_{[{\bf U},\bm{\theta}_{0}]}(x)=\bm{\Phi}
({\bf k}({\bf U})x+\bm{\theta}_{0},{\bf U})$.
We will assume
everywhere below that the family~$\Lambda$ represents a~smooth family
of \mbox{$m$-phase} solutions of system~\eqref{insyst} in the sense
discussed above.

It is generally assumed that the parameters $k^{\alpha}$,
$\omega^{\alpha}$ are independent on the family
$\Lambda$, such that the full family of the \mbox{$m$-phase} solutions
of~\eqref{insyst} depends on $N=2m+s$ ($s\geq0$)
parame\-ters~$U^{\nu}$ and $m$ initials phase shifts
$\theta^{\alpha}_{0}$.
In this case it is convenient to represent
the parameters ${\bf U}$ in the form
${\bf U}=({\bf k},\bm{\omega},{\bf n})$, where ${\bf k}$
represents the wave numbers, $\bm{\omega}$ are the frequencies
of the \mbox{$m$-phase} solutions and ${\bf n}=(n^{1},\dots,n^{s})$ are some additional parameters (if any).

It is easy to see that the functions
$\bm{\Phi}_{\theta^{\alpha}}(\bm{\theta}+\bm{\theta}_{0},
{\bf k},\bm{\omega},{\bf n})$, $\alpha=1,\dots,m$,
$\bm{\Phi}_{n^{l}}(\bm{\theta}+\bm{\theta}_{0},
{\bf k},\bm{\omega},{\bf n})$, $l=1,\dots,s$,
belong to the kernel of the operator
${\hat L}^{i}_{j[{\bf k},\bm{\omega},{\bf n},\bm{\theta}_{0}]}$.
In the regular case it is natural to assume that the set
of the functions ($\bm{\Phi}_{\theta^{\alpha}}$, $\bm{\Phi}_{n^{l}}$)
represents the maximal linearly independent set of the kernel
vectors of the operator ${\hat L}$ regularly depending on
the parameters $({\bf k},\bm{\omega},{\bf n})$.
For the construction
of the ``regular'' Whitham system we have to require the following
property of regularity and completeness of the family of \mbox{$m$-phase}
solutions of system~\eqref{insyst}.

\begin{definition}\label{Definition1.1.}
We call a~family $\Lambda$ a~complete regular family of
\mbox{$m$-phase} solutions of system~\eqref{insyst} if:
\begin{enumerate}\itemsep=0pt
\item[1)] the values ${\bf k}=(k^{1},\dots,k^{m})$,
$\bm{\omega}=(\omega^{1},\dots,\omega^{m})$
represent independent parameters on the family $\Lambda$,
such that the total set of parameters of the \mbox{$m$-phase} solutions
can be represented in the form
$({\bf U},\bm{\theta}_{0})=
({\bf k},\bm{\omega},{\bf n},\bm{\theta}_{0})$;

\item[2)] the functions
$\bm{\Phi}_{\theta^{\alpha}}(\bm{\theta}+\bm{\theta}_{0},
{\bf k},\bm{\omega},{\bf n})$,
$\bm{\Phi}_{n^{l}}(\bm{\theta}+\bm{\theta}_{0},
{\bf k},\bm{\omega},{\bf n})$
are linearly independent and give the maximal linearly independent
set among the kernel vectors of the operator
${\hat L}^{i}_{j[{\bf k},\bm{\omega},{\bf n},\bm{\theta}_{0}]}$,
smoothly depending on the parameters
$({\bf k},\bm{\omega},{\bf n})$ on the whole set of parameters;

\item[3)] the operator
${\hat L}^{i}_{j[{\bf k},\bm{\omega},{\bf n},\bm{\theta}_{0}]}$
has exactly $m+s$ linearly independent left eigenvectors
with zero eigenvalue
\begin{gather*}
\bm{\kappa}^{(q)}_{[{\bf U}]}(\bm{\theta}+\bm{\theta}_{0})
=
\bm{\kappa}^{(q)}_{[{\bf k},\bm{\omega},{\bf n}]}
(\bm{\theta}+\bm{\theta}_{0})
,\qquad
q=1,\dots,m+s,
\end{gather*}
among the vectors smoothly depending on the parameters
$({\bf k},\bm{\omega},{\bf n})$ on the whole set of parameters.
\end{enumerate}
\end{definition}

By def\/inition, we will call the regular Whitham system for a                  
complete regular family of \mbox{$m$-phase} solutions of~\eqref{insyst}
the conditions of orthogonality of the discrepancy
${\bf f}_{(1)}(\bm{\theta},X,T)$ to the functions
$\bm{\kappa}^{(q)}_{[{\bf U}(X,T)]}
(\bm{\theta}+\bm{\theta}_{0}(X,T))$
\begin{gather}\label{ortcond}
\int_{0}^{2\pi} \cdots\int_{0}^{2\pi}\!\!\kappa^{(q)}_{[{\bf U}(X,T)]i}
(\bm{\theta}+\bm{\theta}_{0}(X,T))f^{i}_{(1)}(\bm{\theta},X,T){{\mathrm d}^{m}\theta\over(2\pi)^{m}}=0,\qquad
q=1,\dots,m+s,
\end{gather}
with the compatibility conditions                                          
\begin{gather}\label{comcond}
k^{\alpha}_{T}=\omega^{\alpha}_{X}.
\end{gather}

System~\eqref{ortcond}, \eqref{comcond} gives
$m+(m+s)=2m+s=N$
conditions at every $X$ and $T$ for the parameters of the
zero approximation $\bm{\Psi}_{(0)}(\bm{\theta},X,T)$.

It is well known that the Whitham system does not include
the parameters $\theta^{\alpha}_{0}(X,T)$ and provides
restrictions only to the parameters $U^{\nu}(X,T)$ of the
zero approximation.
Let us prove here a~simple lemma which
conf\/irms this property under the conditions formulated above\footnote{This simple fact was present in the Whitham approach
from the very beginning
(see~\cite{whith1, whith2, whith3, luke}).
In fact, under various assumptions it can be also shown that the
additional phase shifts $\theta^{\alpha}_{0}(X,T)$ can be always
absorbed by the functions $S^{\alpha}(X,T)$ after a~suitable
correction of initial data (see, e.g.,~\cite{Haberman1, Haberman2,
MaltsevJMP, DobrMinenkov}).
It should be noted, however, that
the corresponding phase shift can play rather important role
in the weakly nonlocal case~\cite{Newell} (see also~\cite{MaltsevAmerMath, DobrMinenkov}).}.

\begin{lemma}
Under the regularity conditions formulated above the
orthogonality conditions~\eqref{ortcond} do not contain the functions
$\theta^{\alpha}_{0}(X,T)$ and give constraints only to the
functions $U^{\nu}(X,T)$, having the form
\begin{gather*}
C^{(q)}_{\nu}({\bf U})U^{\nu}_{T}-D^{(q)}_{\nu}({\bf U})U^{\nu}_{X}=0,\qquad q=1,\dots,m+s,
\end{gather*}
with some functions $C^{(q)}_{\nu}({\bf U})$,
$D^{(q)}_{\nu}({\bf U})$.
\end{lemma}

\begin{proof}
Let us write down the part
${\bf f}^{\prime}_{(1)}$ of the function ${\bf f}_{(1)}$,
which contains the derivatives $\theta^{\beta}_{0T}(X,T)$
and $\theta^{\beta}_{0X}(X,T)$.
We have
\begin{gather*}
f^{\prime i}_{(1)}(\bm{\theta},X,T)=
-{\partial F^{i}\over\partial\varphi^{j}_{t}}
\left(\bm{\Psi}_{(0)},\dots\right)
\Psi^{j}_{(0)\theta^{\beta}}\theta^{\beta}_{0T}-
{\partial F^{i}\over\partial\varphi^{j}_{x}}
\left(\bm{\Psi}_{(0)},\dots\right)
\Psi^{j}_{(0)\theta^{\beta}}\theta^{\beta}_{0X}
\\
\phantom{f^{\prime i}_{(1)}(\bm{\theta},X,T)=}
-{\partial F^{i}\over\partial\varphi^{j}_{tt}}
\left(\bm{\Psi}_{(0)},\dots\right)2
\omega^{\alpha}(X,T)
\Psi^{j}_{(0)\theta^{\alpha}\theta^{\beta}}
\theta^{\beta}_{0T}
\\
\phantom{f^{\prime i}_{(1)}(\bm{\theta},X,T)=}
-{\partial F^{i}\over\partial\varphi^{j}_{xx}}
\left(\bm{\Psi}_{(0)},\dots\right)2
k^{\alpha}(X,T)
\Psi^{j}_{(0)\theta^{\alpha}\theta^{\beta}}
\theta^{\beta}_{0X}-\cdots.
\end{gather*}

Let us choose the parameters ${\bf U}$ in the form
\begin{gather*}
{\bf U}=(k^{1},\dots,k^{m},\omega^{1},\dots,\omega^{m}, n^{1},\dots,n^{s}).
\end{gather*}

We can write then
\begin{gather*}
f^{\prime i}_{(1)}(\bm{\theta},X,T)=
\left[-{\partial\over\partial\omega^{\beta}}
F^{i}\left(\bm{\Phi}(\bm{\theta}+\bm{\theta}_{0},{\bf U}),
\dots\right)+{\hat L}^{i}_{j}
{\partial\over\partial\omega^{\beta}}
\Phi^{j}(\bm{\theta}+\bm{\theta}_{0},{\bf U})\right]
\theta^{\beta}_{0T}\\
\phantom{f^{\prime i}_{(1)}(\bm{\theta},X,T)=}
+\left[-{\partial\over\partial k^{\beta}}
F^{i}\left(\bm{\Phi}(\bm{\theta}+\bm{\theta}_{0},{\bf U}),
\dots\right)+{\hat L}^{i}_{j}
{\partial\over\partial k^{\beta}}
\Phi^{j}(\bm{\theta}+\bm{\theta}_{0},{\bf U})\right]
\theta^{\beta}_{0X}.
\end{gather*}

The total derivatives $\partial F^{i}/\partial\omega^{\beta}$
and $\partial F^{i}/\partial k^{\beta}$ are identically equal to
zero on $\Lambda$ according to~\eqref{PhaseSyst0}.
We have then
\begin{gather*}
\int_{0}^{2\pi} \cdots\int_{0}^{2\pi}
\kappa^{(q)}_{[{\bf U}(X,T)]i}
(\bm{\theta}+\bm{\theta}_{0}(X,T))
f^{\prime i}_{(1)}(\bm{\theta},X,T)
{{\mathrm d}^{m}\theta\over(2\pi)^{m}}\equiv0
\end{gather*}
since
$\bm{\kappa}^{(q)}_{[{\bf U}(X,T)]}(\bm{\theta}+\bm{\theta}_{0}(X,T))$
are left eigenvectors of ${\hat L}$ with the zero eigenvalues.

It is not dif\/f\/icult to see also that all the $\bm{\theta}_{0}(X,T)$
in the arguments of $\bm{\Phi}$ and $\bm{\kappa}^{(q)}$
disappear after the integration with respect to $\bm{\theta}$,
so we get the statement of the lemma.
\end{proof}

We can claim then that the regular Whitham system has the following
general form
\begin{gather}
{\partial k^{\alpha}\over\partial U^{\nu}}
U^{\nu}_{T}=
{\partial\omega^{\alpha}\over\partial U^{\nu}}
U^{\nu}_{X},\qquad\alpha=1,\dots,m,
\nonumber\\
C^{(q)}_{\nu}({\bf U})U^{\nu}_{T}=
D^{(q)}_{\nu}({\bf U})U^{\nu}_{X},\qquad
q=1,\dots,m+s.\label{RegWhithSyst}
\end{gather}

Let us note that according to our assumptions we have here
$\operatorname{rank}||\partial k^{\alpha}/\partial U^{\nu}||=m$.
In
generic case the derivatives $U^{\nu}_{T}$ can be expressed in
terms of $U^{\mu}_{X}$ and the Whitham system can be written in
the form~\eqref{WhithamSystem}.

\section{Lagrangian and Hamiltonian formulations\\ of the Whitham
method}

Together with the formulation of Whitham's method the
Lagrangian structure of the equations of slow modulations was proposed~\cite{whith1,whith2,whith3}.
The method of averaging
of Lagrangian function introduced by Whitham can be formulated in
the following way.
We assume that the original system~\eqref{insyst}
is lagrangian with the local action of the form
\begin{gather*}
S=\iint
L\left(\bm{\varphi},\bm{\varphi}_{t},\bm{\varphi}_{x},
\bm{\varphi}_{tt},\bm{\varphi}_{xt},\bm{\varphi}_{xx},\dots\right){\mathrm d}x {\mathrm d}t,
\end{gather*}
such that the functions $F^{i}$ have the form
\begin{gather*}
F^{i}(\bm{\varphi},\bm{\varphi}_{t},\bm{\varphi}_{x},\dots)
={\delta S\over\delta\varphi^{i}(x,t)}
={\partial L\over\partial\varphi^{i}}-
{\partial\over\partial t}
{\partial L\over\partial\varphi^{i}_{t}}-
{\partial\over\partial x}
{\partial L\over\partial\varphi^{i}_{x}}+\cdots.
\end{gather*}

Let us assume here for simplicity that the parameters
$({\bf k},\bm{\omega})=
(k^{1},\dots,k^{m},\omega^{1},\dots,\omega^{m})$
give the complete set of independent parameters on the family
of \mbox{$m$-phase} solutions (excluding the initial phase shifts),
such that the number of parameters $U^{\nu}$ is equal to~$2m$.

The linearized operator
${\hat L}^{i}_{j[{\bf k},\bm{\omega},\bm{\theta}_{0}]}(X,T)$
in~\eqref{ksyst} is given now by the distribution
\begin{gather*}
L^{i}_{j[{\bf k},\bm{\omega},\bm{\theta}_{0}]}
(\bm{\theta},\bm{\theta}^{\prime})=
{\delta^{2}S\over\delta\Phi^{i}(\bm{\theta})
\delta\Phi^{j}(\bm{\theta}^{\prime})},
\end{gather*}
where
\begin{gather*}
S=\int_{0}^{2\pi}\cdots\int_{0}^{2\pi}
L\big(\bm{\Phi},\omega^{\alpha}\bm{\Phi}_{\theta^{\alpha}},
k^{\beta}\bm{\Phi}_{\theta^{\beta}},\dots\big)
{{\mathrm d}^{m}\theta\over(2\pi)^{m}}
\end{gather*}
is a~self-adjoint operator.

Throughout the paper we will always understand the integration
with respect to $\bm{\theta}$ as the averaging procedure.
For this
reason, all the integrals over ${\mathrm d}^{m}\theta$ will be
def\/ined with the factor $1/(2\pi)^{m}$.
In particular, we will
also assume that the variational derivatives of the type
$\delta S/\delta\varphi^{i}(\bm{\theta})$ are def\/ined as
\begin{gather*}
\delta S\equiv
\int_{0}^{2\pi}\cdots\int_{0}^{2\pi}
{\delta S\over\delta\varphi^{i}(\bm{\theta})}
\delta\varphi^{i}(\bm{\theta})
{{\mathrm d}^{m}\theta\over(2\pi)^{m}}
\end{gather*}
on the space of $2\pi$-periodic in $\bm{\theta}$ functions.

We also def\/ine here the delta function
$\delta(\bm{\theta}-\bm{\theta}^{\prime})$
and its higher derivatives
$\delta_{\theta^{\alpha_{1}}\dots\theta^{\alpha_{s}}}
(\bm{\theta}-\bm{\theta}^{\prime})$
on the space of $2\pi$-periodic functions by the formula
\begin{gather*}
\int_{0}^{2\pi}\cdots\int_{0}^{2\pi}
\delta_{\theta^{\alpha_{1}}\cdots\theta^{\alpha_{s}}}
(\bm{\theta}-\bm{\theta}^{\prime})
\psi(\bm{\theta}^{\prime})
{{\mathrm d}^{m}\theta^{\prime}\over(2\pi)^{m}}
\equiv
\psi_{\theta^{\alpha_{1}}\cdots\theta^{\alpha_{s}}}(\bm{\theta}).
\end{gather*}

The functions $\bm{\Phi}_{\theta^{\alpha}}$, $\alpha=1,\dots,m$,
represent both the left and the right eigenfunctions of the operator
${\hat L}^{i}_{j[{\bf k},\bm{\omega},\bm{\theta}_{0}]}(X,T)$,
corresponding to the zero eigenvalue.

Under the assumption that the family of the \mbox{$m$-phase} solutions
$\Lambda$ is a~complete regular family of \mbox{$m$-phase} solutions of~\eqref{insyst} we assume that the functions
$\bm{\Phi}_{\theta^{\alpha}}(\bm{\theta}+\bm{\theta}_{0},
{\bf k},\bm{\omega})$ are linearly independent and give the
maximal linearly independent set among the kernel vectors of the
operator ${\hat L}^{i}_{j[{\bf k},\bm{\omega},\bm{\theta}_{0}]}$
smoothly depending on the parameters $({\bf k},\bm{\omega})$.
The regular Whitham system is given then by the conditions
$k^{\alpha}_{T}=\omega^{\alpha}_{X}$
and $m$ conditions of orthogonality of the function
${\bf f}_{(1)}(\bm{\theta},X,T)$ to the functions
$\bm{\Phi}_{\theta^{\alpha}}(\bm{\theta}+\bm{\theta}_{0},
{\bf k},\bm{\omega})$.

According to the Whitham procedure the Whitham system on the
parameters $({\bf k},\bm{\omega})$ is obtained from the condition
of extremality of the action
\begin{gather}\label{WhithLagr}
\Sigma^{(0)}\left[{\bf S}\right]=
\iiint_{0}^{2\pi}\cdots\int_{0}^{2\pi}
L\big(\bm{\Phi},S^{\alpha}_{T}\bm{\Phi}_{\theta^{\alpha}},
S^{\beta}_{X}\bm{\Phi}_{\theta^{\beta}},\dots\big)
{{\mathrm d}^{m}\theta\over(2\pi)^{m}}{\mathrm d}X {\mathrm d}T
\end{gather}
under the conditions $k^{\alpha}=S^{\alpha}_{X}$,
$\omega^{\alpha}=S^{\alpha}_{T}$.

The conditions $k^{\alpha}_{T}=\omega^{\alpha}_{X}$ and
$\delta\Sigma/\delta S^{\alpha}(X,T)=0$
give a~system of $2m$ equations on the parameters
$({\bf k},\bm{\omega})$.

It is not dif\/f\/icult to see that the system given by the
variation of the ``averaged'' action coincides with the
conditions of orthogonality of the function
${\bf f}_{(1)}(\bm{\theta},X,T)$ to the functions
$\bm{\Phi}_{\theta^{\alpha}}(\bm{\theta}+\bm{\theta}_{0},
{\bf k},\bm{\omega})$.
Indeed, let us consider the action
\begin{gather*}
\Sigma\left[{\bf S},\bm{\varphi},\epsilon\right]
=\int\! L\!\left(\bm{\varphi}\!
\left({{\bf S}(X,T)\over\epsilon}+\bm{\theta},X,T\right)\!,
\epsilon{\partial\over\partial T}\bm{\varphi}\!
\left({{\bf S}(X,T)\over\epsilon}+\bm{\theta},X,T\right),
\dots\right)\!{{\mathrm d}^{m}\theta\over(2\pi)^{m}}{\mathrm d}X {\mathrm d}T
\\
\phantom{\Sigma\left[{\bf S},\bm{\varphi},\epsilon\right]}
=\Sigma^{(0)}\left[{\bf S},\bm{\varphi}\right]
+\epsilon\Sigma^{(1)}\left[{\bf S},\bm{\varphi}\right]
+\epsilon^{2}\Sigma^{(2)}\left[{\bf S},\bm{\varphi}\right]
+\cdots
\end{gather*}
def\/ined on the functions $\bm{\varphi}(\bm{\theta},X,T)$,
$2\pi$-periodic in each $\theta^{\alpha}$.
Taking into
account the relation
\begin{gather*}
{\delta\Sigma\over\delta S^{\alpha}(X,T)}=
\epsilon^{-1}\int_{0}^{2\pi}\cdots\int_{0}^{2\pi}
\varphi^{i}_{\theta^{\alpha}}
\left(\bm{\theta},X,T\right)
{\delta\Sigma\over\delta\varphi^{i}(\bm{\theta},X,T)}
{{\mathrm d}^{m}\theta\over(2\pi)^{m}}
\end{gather*}
and the invariance of the action with respect to the shifts
\begin{gather*}
{\bf S}(X,T)\rightarrow{\bf S}(X,T)
+\Delta{\bf S},
\end{gather*}
it is easy to see that
\begin{gather}
\int_{0}^{2\pi}\cdots\int_{0}^{2\pi}
\varphi^{i}_{\theta^{\alpha}}
\left(\bm{\theta},X,T\right)
{\delta\Sigma^{(0)}\over\delta\varphi^{i}(\bm{\theta},X,T)}
{{\mathrm d}^{m}\theta\over(2\pi)^{m}}\equiv0,
\nonumber\\
 \label{Sigma0Sigma1}
{\delta\Sigma^{(0)}\over\delta S^{\alpha}(X,T)}=
\int_{0}^{2\pi}\cdots\int_{0}^{2\pi}
\varphi^{i}_{\theta^{\alpha}}
\left(\bm{\theta},X,T\right)
{\delta\Sigma^{(1)}\over\delta\varphi^{i}(\bm{\theta},X,T)}
{{\mathrm d}^{m}\theta\over(2\pi)^{m}}
\end{gather}
etc.

Substituting the functions
$\bm{\varphi}(\bm{\theta},X,T)$
in the form $\bm{\varphi}(\bm{\theta},X,T)=
\bm{\Phi}(\bm{\theta}+\bm{\theta}_{0},{\bf S}_{X},{\bf S}_{T})$
in the relations above, we can see that we have to include now the
additional dependence of the functions
$\bm{\varphi}(\bm{\theta},X,T)$ on ${\bf S}_{X}$ and
${\bf S}_{T}$ in relation~\eqref{Sigma0Sigma1}.
However, due to
the relation
\begin{gather*}
{\delta\Sigma^{(0)}\over\delta\varphi^{i}(\bm{\theta},X,T)}
\equiv0
\end{gather*}
on the family $\Lambda$, the relation~\eqref{Sigma0Sigma1} will
not change in this situation.
Taking also into account the equality
\begin{gather*}
f^{i}_{(1)}\left(\bm{\theta},X,T\right)=
{\delta\Sigma^{(1)}\over\delta\varphi^{i}(\bm{\theta},X,T)}
\end{gather*}
we get the required statement.

Under the assumption of the completeness and regularity of the
family $\Lambda$ we can see then that the averaged action
\eqref{WhithLagr} def\/ines a~lagrangian structure of the regular
Whitham system in general multiphase case.
We should say also
that the cases with additional parameters ${\bf n}$, as a~rule,
can be also included into the scheme described above with the aid
of the Whitham ``pseudo-phases''~\cite{whith3}.
Let us note
also that dif\/ferent questions connected with the justif\/ication
of the averaging of Lagrangian functions in dif\/ferent orders
can be found in~\cite{dm}.

Another approach to the construction of the regular Whitham system
is connected with the method of averaging of conservation laws.
According to further consideration of the Hamiltonian structure
of the Whitham equations we will assume now that system~\eqref{insyst}
is written in an evolutionary form
\begin{gather}\label{EvInSyst}
\varphi^{i}_{t}=
F^{i}\left(\bm{\varphi},\bm{\varphi}_{x},\bm{\varphi}_{xx},
\dots\right).
\end{gather}

The families of the \mbox{$m$-phase} solutions of~\eqref{EvInSyst}
are def\/ined then by solutions of the system
\begin{gather}\label{EvPhaseSyst}
\omega^{\alpha}\varphi^{i}_{\theta^{\alpha}}=
F^{i}\big(\bm{\varphi},
k^{\beta}\bm{\varphi}_{\theta^{\beta}},\dots\big)
\end{gather}
on the space of $2\pi$-periodic in each $\theta^{\alpha}$ functions
$\bm{\varphi}(\bm{\theta})$.

We will assume that the conservation laws of system~\eqref{EvInSyst}
have the form
\begin{gather*}
P^{\nu}_{t}\left(\bm{\varphi},\bm{\varphi}_{x},\bm{\varphi}_{xx},
\dots\right)=Q^{\nu}_{x}\left(\bm{\varphi},
\bm{\varphi}_{x},\bm{\varphi}_{xx},\dots\right),
\end{gather*}
such that the values
\begin{gather*}
I^{\nu}=\int_{-\infty}^{+\infty}
P^{\nu}\left(\bm{\varphi},\bm{\varphi}_{x},\bm{\varphi}_{xx},
\dots\right){\mathrm d}x
\end{gather*}
represent translationally invariant conservative quantities
for the system~\eqref{EvInSyst} in the case of
the rapidly decreasing at inf\/inity functions
$\bm{\varphi}(x)$.
We can also def\/ine the conservation laws for
system~\eqref{EvInSyst} in the periodic case with a~f\/ixed period~$K$
\begin{gather*}
I^{\nu}={1\over K}\int_{0}^{K}
P^{\nu}\left(\bm{\varphi},\bm{\varphi}_{x},\bm{\varphi}_{xx},
\dots\right){\mathrm d}x,
\end{gather*}
or in the quasiperiodic case
\begin{gather*}
I^{\nu}=\lim_{K\rightarrow\infty}
{1\over2K}\int_{-K}^{K}
P^{\nu}\left(\bm{\varphi},\bm{\varphi}_{x},\bm{\varphi}_{xx},
\dots\right){\mathrm d}x.
\end{gather*}

It is natural also to def\/ine the variational derivatives of the
functionals $I^{\nu}$ with respect to the variations of
$\bm{\varphi}(x)$ having the same periodic or quasiperiodic
properties as the original functions.
Easy to see then that the
standard Euler--Lagrange expressions for the variational derivatives
can be used in this case.

Let us write the functionals $I^{\nu}$ in the general form
\begin{gather}\label{Integrals}
I^{\nu}=\int
P^{\nu}\left(\bm{\varphi},\bm{\varphi}_{x},\bm{\varphi}_{xx},
\dots\right){\mathrm d}x
\end{gather}
assuming the appropriate def\/inition in the corresponding situations.

Let us def\/ine a~quasiperiodic function $\bm{\varphi}(x)$
with f\/ixed quasiperiods
$(k^{1},\dots,k^{m})$ as a~smooth periodic function
$\bm{\varphi}(\bm{\theta})$ on the torus
$\mathbb{T}^{m}$ which is restricted on the corresponding
straight-line winding
\begin{gather*}
\bm{\varphi}({\bf k}x+\bm{\theta}_{0})
\rightarrow\bm{\varphi}(x).
\end{gather*}

Let us def\/ine the functionals
\begin{gather}\label{Jnu}
J^{\nu}=\int_{0}^{2\pi}\cdots\int_{0}^{2\pi}
P^{\nu}\big(\bm{\varphi},
k^{\beta}\bm{\varphi}_{\theta^{\beta}},
\dots\big){{\mathrm d}^{m}\theta\over(2\pi)^{m}}
\end{gather}
on the space of $2\pi$-periodic in $\bm{\theta}$ functions.

It's not dif\/f\/icult to see that the functions
\begin{gather}\label{VarDer}
\zeta^{(\nu)}_{i[{\bf U}]}(\bm{\theta}+\bm{\theta}_{0})
=\left.
{\delta J^{\nu}
\over\delta\varphi^{i}(\bm{\theta})}
\right|_{\bm{\varphi}(\bm{\theta})=\bm{\Phi}(\bm{\theta}+\bm{\theta}_{0},{\bf U})}
\end{gather}
represent left eigenvectors of the operator
${\hat L}^{i}_{j[{\bf U},\bm{\theta}_{0}]}$ with zero
eigenvalues regularly depending on parameters ${\bf U}$
on a~f\/ixed smooth family $\Lambda$.

Indeed, the operator
${\hat L}^{i}_{j[{\bf U},\bm{\theta}_{0}]}$
is def\/ined in this case by the distribution
\begin{gather*}
L^{i}_{j[{\bf U},\bm{\theta}_{0}]}
(\bm{\theta},\bm{\theta}^{\prime})=
\delta^{i}_{j}\omega^{\alpha}\delta_{\theta^{\alpha}}
(\bm{\theta}-\bm{\theta}^{\prime})-\left.
{\delta F^{i}(\bm{\varphi},
k^{\beta}\bm{\varphi}_{\theta^{\beta}},\dots)\over
\delta\varphi^{j}(\bm{\theta}^{\prime})}
\right|_{\bm{\varphi}(\bm{\theta})=\bm{\Phi}(\bm{\theta}+\bm{\theta}_{0},{\bf U})}.
\end{gather*}

We have
\begin{gather*}
\int_{0}^{2\pi}\cdots\int_{0}^{2\pi}
{\delta J^{\nu}\over\delta\varphi^{i}(\bm{\theta})}
\big(\omega^{\alpha}\varphi^{i}_{\theta^{\alpha}}-
F^{i}(\bm{\varphi},
k^{\beta}\bm{\varphi}_{\theta^{\beta}},\dots)\big)
{{\mathrm d}^{m}\theta\over(2\pi)^{m}}\equiv0
\end{gather*}
for any translationally invariant integral of~\eqref{EvInSyst}.
Taking the variational derivative of this relation with respect to
$\varphi^{j}(\bm{\theta}^{\prime})$
on $\Lambda$ we get the required statement.

Thus, we can write
\begin{gather}\label{SviazZetaKappa}
\zeta^{(\nu)}_{i[{\bf U}]}(\bm{\theta})=
\sum_{q}c^{\nu}_{q}({\bf U})\kappa^{(q)}_{i[{\bf U}]}
(\bm{\theta})
\end{gather}
with some smooth functions $c^{\nu}_{q}({\bf U})$
on a~complete regular family $\Lambda$.

For the construction of the regular Whitham system on a~complete
regular family of \mbox{$m$-phase} solutions of~\eqref{EvInSyst} we need
a suf\/f\/icient number of the f\/irst integrals~\eqref{Integrals} such
that the values of the functionals $J^{\nu}$ on $\Lambda$ represent
the full set of parameters $U^{\nu}=J^{\nu}|_{\Lambda}$.
Besides
that, we should require that the maximal linearly independent
subset of the functions
\eqref{VarDer} give a~complete set of linearly independent left
eigenvectors of the operator
${\hat L}^{i}_{j[{\bf U},\bm{\theta}_{0}]}$
with zero eigenvalues among the vectors regularly depending on the
parameters ${\bf U}$ on the family $\Lambda$.

Coming back to the def\/inition of a~complete regular family of
\mbox{$m$-phase} solutions of system~\eqref{EvInSyst} we can see that in
the case of a~complete regular family $\Lambda$ the number of linearly
independent vectors~\eqref{VarDer} on $\Lambda$ is always f\/inite.
More precisely, if $N=2m+s$ is the number of parameters of
\mbox{$m$-phase} solutions of~\eqref{EvInSyst} (excluding the initial phase
shifts) then for a~complete regular family of \mbox{$m$-phase} solutions
we require the presence of exactly $m+s=N-m$ left eigenvectors
$\bm{\kappa}^{(q)}_{[{\bf U}]}(\bm{\theta}+\bm{\theta}_{0})$
with zero eigenvalues, regularly depending on parameters,
in accordance with the number of the vectors
$\bm{\Phi}_{\theta^{\alpha}}$, $\bm{\Phi}_{n^{l}}$.
Thus, according to Def\/inition~\ref{Definition1.1.},
we assume here that the number of linearly independent vectors
def\/ined by formula~\eqref{VarDer} is exactly equal to
$m+s=N-m$ for a~complete regular family $\Lambda$.

We should note that the conditions on the variational derivatives
of $J^{\nu}$ formulated above do not contradict to the condition
that the values $J^{\nu}$ ($\nu=1,\dots,N$) can be chosen as
parameters~$U^{\nu}$ on the family of \mbox{$m$-phase} solutions.
Indeed, the def\/inition of $J^{\nu}$~\eqref{Jnu} explicitly
includes the additional~$m$ functions~$k^{\alpha}$, which provide
the necessary functional independence of the values of~$J^{\nu}$ on~$\Lambda$.
In other words, we can use the Euler--Lagrange
expressions for the variational derivatives of $I^{\nu}$
only on subspaces with f\/ixed quasiperiods $(k^{1},\dots,k^{m})$.
The variation of the quasiperiods gives linearly growing
variations which do not allow to use the Euler--Lagrange
expressions.

Moreover, under the assumptions formulated above, we can show
that the condition of the completeness of the variational derivatives~\eqref{VarDer} of the functionals $J^{\nu}$ in the space of
regular left eigenvectors of the operator
${\hat L}^{i}_{j[{\bf U},\bm{\theta}_{0}]}$
with zero eigenvalues follows in fact from the condition
that the values $U^{\nu}=J^{\nu}|_{\Lambda}$ can be chosen
as the full set of parameters (excluding the initial phase shifts)
on the family~${\Lambda}$.

Let us make the agreement that we will always assume here that
the Jacobian of the coordinate transformation
\begin{gather*}
\left({\bf k},\bm{\omega},{\bf n}\right)
\rightarrow\left(U^{1},\dots,U^{N}\right)
\end{gather*}
is dif\/ferent from zero on $\Lambda$ whenever we say that the
values $U^{\nu}({\bf k},\bm{\omega},{\bf n})$ represent
a complete set of parameters on ${\Lambda}$
(excluding the initial phase shifts).

Under the conditions formulated above let us
prove here the following proposition.

\begin{proposition}\label{Proposition2.1.}
Let ${\Lambda}$ be a~complete regular family of \mbox{$m$-phase} solutions
of system~\eqref{EvInSyst}.
Let the values $(U^{1},\dots,U^{N})$
of the functionals $(J^{1},\dots,J^{N})$~\eqref{Jnu} give a
complete set of parameters on $\Lambda$ excluding the initial phase
shifts.
Then:
\begin{enumerate}\itemsep=0pt
\item[$1)$] the set of the vectors
\begin{gather*}
\big\{\bm{\Phi}_{\omega^{\alpha}}
(\bm{\theta}+\bm{\theta}_{0},
{\bf k},\bm{\omega},{\bf n}),\;
\bm{\Phi}_{n^{l}}
(\bm{\theta}+\bm{\theta}_{0},
{\bf k},\bm{\omega},{\bf n}),\;
\alpha=1,\dots,m,\;
l=1,\dots,s
\big\}
\end{gather*}
is linearly independent on $\Lambda$;

\item[$2)$] the variation derivatives
$\zeta^{(\nu)}_{i[{\bf U}]}(\bm{\theta}+\bm{\theta}_{0})$,
given by~\eqref{VarDer}, generate the full space of the regular
left eigenvectors of the operator
${\hat L}^{i}_{j[{\bf U},\bm{\theta}_{0}]}$ with zero eigenvalues
on the family~${\Lambda}$.
\end{enumerate}
\end{proposition}

\begin{proof}
Indeed, we require that the rows given by the derivatives
\begin{gather*}
\left({\partial U^{1}\over\partial\omega^{\alpha}},\dots,
{\partial U^{N}\over\partial\omega^{\alpha}}\right),
\qquad
\left({\partial U^{1}\over\partial n^{l}},\dots,
{\partial U^{N}\over\partial n^{l}}\right)
\end{gather*}
are linearly independent on $\Lambda$.
Using the expressions
\begin{gather*}
{\partial U^{\nu}\over\partial\omega^{\alpha}}=
\int_{0}^{2\pi}\cdots\int_{0}^{2\pi}
\zeta^{(\nu)}_{i[{\bf U}]}(\bm{\theta})\,
\Phi^{i}_{\omega^{\alpha}}(\bm{\theta},{\bf U})
{{\mathrm d}^{m}\theta\over(2\pi)^{m}},\qquad
\alpha=1,\dots,m,
\\
{\partial U^{\nu}\over\partial n^{l}}=
\int_{0}^{2\pi}\cdots\int_{0}^{2\pi}
\zeta^{(\nu)}_{i[{\bf U}]}(\bm{\theta})\,
\Phi^{i}_{n^{l}}(\bm{\theta},{\bf U})
{{\mathrm d}^{m}\theta\over(2\pi)^{m}},\qquad
l=1,\dots,s,
\end{gather*}
on $\Lambda$, we get that the set
$\{\bm{\Phi}_{\omega^{\alpha}},\bm{\Phi}_{n^{l}}\}$ is
linearly independent on $\Lambda$ and the number of linearly
independent variation derivatives~\eqref{VarDer} is not less
than $m+s$.

We obtain then that the variation derivatives~\eqref{VarDer}
generate in this case a~space of
regular left eigenvectors of the operator
${\hat L}^{i}_{j[{\bf U},\bm{\theta}_{0}]}$ with zero eigenvalues
of dimension $(m+s)$.
\end{proof}

As a~remark, let us note here some general fact
associated with multiphase solutions of partial dif\/ferential equations.
As is well known, the presence of families of multiphase
quasiperiodic solutions, as a~rule, is connected with integrability
of system~\eqref{insyst} by the inverse scattering methods.

The f\/irst \mbox{$m$-phase} solutions for the KdV equation given by the
Novikov potentials were introduced exactly as the functional families
represented by the extremals of linear combinations of some set
of higher integrals of the system, i.e.
the families where
the variational derivatives of integrals of the set become linearly
dependent.

If we have a~natural hierarchy of the f\/irst integrals and
commuting f\/lows of an integrable system, some of the f\/irst
integrals of the hierarchy $(I^{1},\dots,I^{q})$ are usually used
for the construction of \mbox{$m$-phase} solutions.
So, the functions
given by the conditions
\begin{gather*}
c_{1}\delta I^{1}+\cdots+
c_{q}\delta I^{q}=0
\end{gather*}
for all possible $(c_{1},\dots, c_{q})$ form a~complete family
of \mbox{$m$-phase} solutions of the integrable sys\-tem~\cite{NovikovFuncAn}.

Thus, for Novikov potentials we have $q=m+2$ while the dimensions
of the families of \mbox{$m$-phase} solutions for KdV (excluding initial phase shifts) are equal to $2m+1$.
The f\/irst $2m+1$ of integrals of KdV $(I^{1},\dots,I^{2m+1})$ can be used for the construction
of parameters $(U^{1},\dots,U^{2m+1})$ on the families of
\mbox{$m$-phase} solutions of KdV.
The number of linearly independent
variational derivatives of these functionals on the family of
\mbox{$m$-phase} solutions is exactly equal to $m+1$.
It's not dif\/f\/icult
to show also that the variational derivatives of all the higher
integrals of the KdV hierarchy are given by linear combinations
of the variational derivatives of the set $(I^{1},\dots,I^{q})$
on the families of \mbox{$m$-phase} solutions.

The construction proposed in~\cite{NovikovFuncAn} in fact is used
without substantial changes for many systems that are
integrable by the inverse scattering methods,
and represents the basic scheme for constructing of
\mbox{$m$-phase} solutions of integrable systems.
This circumstance gives
therefore a~convenient method of checking the above relations
for most specif\/ic examples.

Let us prove here the following lemma, which we will need in
further considerations.

\begin{lemma}\label{Lemma2.1.}
Let the values $U^{\nu}$ of the functionals $J^{\nu}$
on a~complete regular family of \mbox{$m$-phase} solutions $\Lambda$
be functionally independent and give a~complete set of parameters
$($excluding initial phase shifts$)$ on $\Lambda$, such that we have
$k^{\alpha}=k^{\alpha}(U^{1},\dots,U^{N})$.
Then the functionals $k^{\alpha}(J^{1},\dots,J^{N})$
have zero variational derivatives on~$\Lambda$.
\end{lemma}

\begin{proof}
As we have seen, the conditions of the lemma imply the existence
of $m$ independent relations{\samepage
\begin{gather}\label{intder}
\sum_{\nu=1}^{N}\lambda^{\alpha}_{\nu}({\bf U})\left.
{\delta J^{\nu}\over\delta\varphi^{i}(\bm{\theta})}
\right|_{\bm{\varphi}(\bm{\theta})=\bm{\Phi}(\bm{\theta}
+\bm{\theta}_{0},{\bf U})}\equiv0
,\qquad
\alpha=1,\dots,m,
\end{gather}
on $\Lambda$.}

For the corresponding coordinates $U^{\nu}$ on $\Lambda$ this
implies the relations
\begin{gather*}
\sum_{\nu=1}^{N}\lambda^{\alpha}_{\nu}({\bf U}){\mathrm d}U^{\nu}
=\sum_{\beta=1}^{m}
\mu^{(\alpha)}_{\beta}({\bf U}){\mathrm d} k^{\beta}({\bf U})
\end{gather*}
for some matrix $\mu^{(\alpha)}_{\beta}({\bf U})$.

Since $U^{\nu}$ provide coordinates on $\Lambda$
the matrix $\mu^{(\alpha)}_{\beta}({\bf U})$ has the full rank,
and therefore invertible.
We can then write
\begin{gather*}
{\mathrm d}k^{\beta}=\sum_{\alpha=1}^{m}
\big({\hat\mu}^{-1}\big)^{\beta}_{(\alpha)}({\bf U})
\sum_{\nu=1}^{N}\lambda^{(\alpha)}_{\nu}({\bf U})
{\mathrm d}U^{\nu}.
\end{gather*}

The assertion of the lemma follows then from~\eqref{intder}.
\end{proof}

The regular Whitham system in the described approach
can be written as
\begin{gather}\label{ConsWhitham}
\langle P^{\nu}\rangle_{T}=\langle Q^{\nu}\rangle_{X},\qquad\nu=1,\dots,N,
\end{gather}
where $\langle\dots\rangle$ denotes the averaging operation
on $\Lambda$ def\/ined by the formula
\begin{gather*}
\langle f(\bm{\varphi},\bm{\varphi}_{x},\dots)\rangle
\equiv\int_{0}^{2\pi}\cdots\int_{0}^{2\pi}
f\left(\bm{\Phi},k^{\beta}\bm{\Phi}_{\theta^{\beta}},
\dots\right){{\mathrm d}^{m}\theta\over(2\pi)^{m}}.
\end{gather*}

Let us prove here the following lemma about the connection between
the systems~\eqref{ConsWhitham} and~\eqref{RegWhithSyst}.

\begin{lemma}\label{Lemma2.2.}
Let the values $U^{\nu}$ of the functionals $J^{\nu}$
on a~complete regular family of \mbox{$m$-phase} solutions $\Lambda$
be functionally independent and give a~complete set of parameters
on $\Lambda$ excluding the initial phase shifts.
Then the system~\eqref{ConsWhitham} is equivalent to~\eqref{RegWhithSyst}.
\end{lemma}

\begin{proof}
Let us introduce the
functions
\begin{gather*}
\Pi^{\nu}_{i(l)}(\bm{\varphi},\bm{\varphi}_{x},\dots)
\equiv
{\partial P^{\nu}(\bm{\varphi},\bm{\varphi}_{x},\dots)
\over\partial\varphi^{i}_{lx}}
\end{gather*}
for $l\geq0$.

Using the expression for the evolution of the densities
$P^{\nu}(\bm{\varphi},\epsilon\bm{\varphi}_{X},\dots)$
we can write the following identities
\begin{gather}\label{PnuQnu}
P^{\nu}_{t}(\bm{\varphi},\epsilon\bm{\varphi}_{X},\dots)
=\!\!\sum_{l\geq0}\epsilon^{l}\Pi^{\nu}_{i(l)}
(\bm{\varphi},\epsilon\bm{\varphi}_{X},\dots)
\!\left(F^{i}(\bm{\varphi},\epsilon\bm{\varphi}_{X},\dots)
\right)_{lX}\equiv\epsilon Q^{\nu}_{X}
(\bm{\varphi},\epsilon\bm{\varphi}_{X},\dots).
\end{gather}

To calculate the values $\epsilon\langle Q^{\nu}\rangle_{X}$ let us put now
\begin{gather}\label{VarphiSubs}
\varphi^{i}(\bm{\theta},X)=\Phi^{i}\left({{\bf S}(X)\over\epsilon}+\bm{\theta},{\bf U}(X)\right),
\end{gather}
where $S^{\alpha}_{X}=k^{\alpha}({\bf U}(X))$.

The operator $\epsilon\partial/\partial X$ acting on the
functions~\eqref{VarphiSubs} can be naturally represented as a~sum
of $k^{\alpha}\partial/\partial\theta^{\alpha}$ and the terms
proportional to $\epsilon$.
So, any expression
$f(\bm{\varphi},\epsilon\bm{\varphi}_{X},\dots)$ on the
submanifold~\eqref{VarphiSubs} can be naturally represented in the
form
\begin{gather*}
f(\bm{\varphi},\epsilon\bm{\varphi}_{X},\dots)=
\sum_{l\geq0}\epsilon^{l}f_{[l]}\left[\bm{\Phi},{\bf U}\right],
\end{gather*}
where $f_{[l]}[\bm{\Phi},{\bf U}]$ are smooth functions of
$(\bm{\Phi},\bm{\Phi}_{\theta^{\alpha}},\bm{\Phi}_{U^{\nu}},\dots)$
and $({\bf U},{\bf U}_{X},{\bf U}_{XX},\dots)$,
polynomial in the derivatives $({\bf U}_{X},{\bf U}_{XX},\dots)$,
and having degree $l$ in terms of the total number of derivations of
${\bf U}$ w.r.t.~$X$.
Note also that the functions
$\bm{\Phi}$ appear in $f_{[l]}$ with the phase shift
${\bf S}(X)/\epsilon$ according to~\eqref{VarphiSubs}.
The common phase shift is not important for the integration with
respect to $\bm{\theta}$, so let us assume below that the phase shift
${\bf S}(X)/\epsilon$ is omitted after taking all the dif\/ferentiations
with respect to~$X$.

According to~\eqref{PnuQnu} and~\eqref{EvPhaseSyst} we can write
\begin{gather*}
\epsilon\langle Q^{\nu}\rangle_{X}=
\epsilon\int_{0}^{2\pi}\cdots\int_{0}^{2\pi} Q^{\nu}_{X[1]}{{\mathrm d}^{m}\theta\over(2\pi)^{m}}
\\
\phantom{\epsilon\langle Q^{\nu}\rangle_{X}}
=\epsilon\int_{0}^{2\pi}\cdots\int_{0}^{2\pi}
\sum_{l\geq0}\left(\Pi^{\nu}_{i(l)[0]}F^{i}_{lX[1]}+\Pi^{\nu}_{i(l)[1]}F^{i}_{lX[0]}\right)
{{\mathrm d}^{m}\theta\over(2\pi)^{m}}
\\
\phantom{\epsilon\langle Q^{\nu}\rangle_{X}}
=\epsilon\int_{0}^{2\pi}\cdots\int_{0}^{2\pi}
\sum_{l\geq0}\Bigg(\Pi^{\nu}_{i(l)[0]}k^{\gamma_{1}}\cdots k^{\gamma_{l}}
F^{i}_{[1]\theta^{\gamma_{1}}\dots\theta^{\gamma_{l}}}
\\
\phantom{\epsilon\langle Q^{\nu}\rangle_{X}=}
+\Pi^{\nu}_{i(l)[0]}l k^{\gamma_{1}}\cdots k^{\gamma_{l-1}}
\big(\omega^{\beta}
\Phi^{i}_{\theta^{\beta}\theta^{\gamma_{1}}\cdots\theta^{\gamma_{l-1}}}
\big)_{X[1]}
\\
\phantom{\epsilon\langle Q^{\nu}\rangle_{X}=}
+\Pi^{\nu}_{i(l)[0]}{l(l-1)\over2}
k^{\gamma_{1}}_{X}k^{\gamma_{2}}
\cdots k^{\gamma_{l-1}}\omega^{\beta}
\Phi^{i}_{\theta^{\beta}\theta^{\gamma_{1}}\cdots\theta^{\gamma_{l-1}}}
\\
\phantom{\epsilon\langle Q^{\nu}\rangle_{X}=}
+\Pi^{\nu}_{i(l)[1]}k^{\gamma_{1}}\cdots k^{\gamma_{l}}
\omega^{\beta}
\Phi^{i}_{\theta^{\beta}\theta^{\gamma_{1}}\cdots\theta^{\gamma_{l}}}
\Bigg){{\mathrm d}^{m}\theta\over(2\pi)^{m}}.
\end{gather*}

It is not dif\/f\/icult to see also that for arbitrary dependence of
parameters~${\bf U}$ of~$T$, the derivative of the average
$\langle P^{\nu}\rangle$ w.r.t.~$T$ can be written as
\begin{gather*}
\langle P^{\nu}\rangle_{T}=
\int_{0}^{2\pi}\cdots\int_{0}^{2\pi} \sum_{l\geq0}
\Pi^{\nu}_{i(l)[0]}\left(k^{\gamma_{1}}\cdots k^{\gamma_{l}}
\Phi^{i}_{\theta^{\gamma_{1}}\cdots\theta^{\gamma_{l}}}\right)_{T}
{{\mathrm d}^{m}\theta\over(2\pi)^{m}}.
\end{gather*}

Now, we can write the relations
$\langle P^{\nu}\rangle_{T}=\langle Q^{\nu}\rangle_{X}$
as
\begin{gather*}
\int_{0}^{2\pi}\cdots\int_{0}^{2\pi} \sum_{l\geq0}
\left(\Pi^{\nu}_{i(l)[0]}k^{\gamma_{1}}\cdots k^{\gamma_{l}}
\Phi^{i}_{\theta^{\gamma_{1}}\cdots\theta^{\gamma_{l}}T}
+\Pi^{\nu}_{i(l)[0]}
l k^{\gamma_{1}}\cdots k^{\gamma_{l-1}}k^{\gamma_{l}}_{T}
\Phi^{i}_{\theta^{\gamma_{1}}\cdots\theta^{\gamma_{l}}}\right)
{{\mathrm d}^{m}\theta\over(2\pi)^{m}}
\\
{}=\int_{0}^{2\pi}\cdots\int_{0}^{2\pi}
\sum_{l\geq0}\Bigg(
\Pi^{\nu}_{i(l)[0]}k^{\gamma_{1}}\cdots k^{\gamma_{l}}
F^{i}_{[1]\theta^{\gamma_{1}}\cdots\theta^{\gamma_{l}}}
\\
\quad
{}+\Pi^{\nu}_{i(l)[0]}l k^{\gamma_{1}}\cdots k^{\gamma_{l-1}}\omega^{\beta}_{X}
\Phi^{i}_{\theta^{\beta}\theta^{\gamma_{1}}\cdots \theta^{\gamma_{l-1}}}
+\Pi^{\nu}_{i(l)[0]}l k^{\gamma_{1}}\cdots k^{\gamma_{l-1}}
\omega^{\beta}
\Phi^{i}_{\theta^{\beta}\theta^{\gamma_{1}}\cdots\theta^{\gamma_{l-1}}X[1]}
\\
\quad
{}+\Pi^{\nu}_{i(l)[0]}{l(l-1)\over2}
k^{\gamma_{1}}_{X}k^{\gamma_{2}}\cdots k^{\gamma_{l-1}}\omega^{\beta}
\Phi^{i}_{\theta^{\beta}\theta^{\gamma_{1}}\cdots \theta^{\gamma_{l-1}}}
+\Pi^{\nu}_{i(l)[1]}k^{\gamma_{1}}\cdots k^{\gamma_{l}}
\omega^{\beta}
\Phi^{i}_{\theta^{\beta}\theta^{\gamma_{1}}\cdots\theta^{\gamma_{l}}}
\Bigg){{\mathrm d}^{m}\theta\over(2\pi)^{m}}.
\end{gather*}

The last three terms in the right-hand part represent the integral
of the value
\begin{gather*}
\sum_{l\geq0}\omega^{\beta}\left(\Pi^{\nu}_{i(l)[0]}
\Phi^{i}_{\theta^{\beta},lX[1]}+\Pi^{\nu}_{i(l)[1]}
\Phi^{i}_{\theta^{\beta},lX[0]}\right)=
\omega^{\beta}\partial P^{\nu}_{[1]}/\partial\theta^{\beta}
\end{gather*}
and are equal to zero.
The remaining terms after integration by
parts can be written in the form{\samepage
\begin{gather*}
\int_{0}^{2\pi}\cdots\int_{0}^{2\pi} \Bigg(
\zeta^{(\nu)}_{i[{\bf U}(X)]} (\bm{\theta}) \big[
\Phi^{i}_{T} (\bm{\theta}, {\bf U}(X)) -
F^{i}_{[1]} (\bm{\theta}, X) \big]
\\
\hphantom{\int_{0}^{2\pi}\cdots\int_{0}^{2\pi} \Bigg(}
{}+\big(k^{\beta}_{T} - \omega^{\beta}_{X}\big)
\sum_{l \geq 0} \Pi^{\nu}_{i(l)[0]}lk^{\gamma_{1}}\cdots k^{\gamma_{l-1}}
\Phi^{i}_{\theta^{\beta}\theta^{\gamma_{1}}\cdots\theta^{\gamma_{l-1}}}
\Bigg){{\mathrm d}^{m}\theta\over(2\pi)^{m}}=0,
\end{gather*}
where the values $\zeta^{(\nu)}_{i[{\bf U}(X)]}(\bm{\theta})$
are given by~\eqref{VarDer}.}

Consider the convolution (in $\nu$) of the above expression
with the values $\partial k^{\alpha}/\partial U^{\nu}$.
The expressions
\begin{gather*}
{\partial k^{\alpha}\over\partial U^{\nu}}({\bf U}(X))
\zeta^{(\nu)}_{i[{\bf U}(X)]}(\bm{\theta})
\end{gather*}
are identically equal to zero according to Lemma~\ref{Lemma2.1.}.

From the other hand we have
\begin{gather}
{\partial k^{\alpha}\over\partial U^{\nu}}
\int_{0}^{2\pi}\cdots\int_{0}^{2\pi} \sum_{l\geq1}l
k^{\beta_{1}}\cdots k^{\beta_{l-1}}
\Phi^{i}_{\theta^{\beta}\theta^{\beta_{1}}\cdots\theta^{\beta_{l-1}}}
\Pi^{\nu}_{i(l)}\left(\bm{\Phi},
k^{\gamma}\bm{\Phi}_{\theta^{\gamma}},
\dots\right){{\mathrm d}^{m}\theta\over(2\pi)^{m}}
\nonumber\\
\left.\qquad
{}=
\left({\partial k^{\alpha}\over\partial U^{\nu}}
{\partial\over\partial k^{\beta}}
J^{\nu}[\bm{\varphi},{\bf k}]\right)
\right|_{\bm{\varphi}(\bm{\theta})=\bm{\Phi}(\bm{\theta},{\bf U})}
=\delta^{\alpha}_{\beta},\label{kalphakbetaUnu}
\end{gather}
since the variations of the functions $\bm{\Phi}$
are insignif\/icant for the values of $k^{\alpha}$ according to
Lemma~\ref{Lemma2.1.}.

We get then that conditions~\eqref{ConsWhitham} imply
the relations $k^{\alpha}_{T}=\omega^{\alpha}_{X}$, which are the
f\/irst part of system~\eqref{RegWhithSyst}.

Now the conditions
\begin{gather*}
\int_{0}^{2\pi}\cdots\int_{0}^{2\pi}
\zeta^{(\nu)}_{i[{\bf U}(X)]}(\bm{\theta})\big[
\Phi^{i}_{T}(\bm{\theta},{\bf U}(X))-
F^{i}_{[1]}(\bm{\theta},X)\big]
{{\mathrm d}^{m}\theta\over(2\pi)^{m}}=0
\end{gather*}
express the conditions of orthogonality of the vectors
\eqref{VarDer} to the function
$-\bm{\Phi}_{T}+{\bf F}_{[1]}$, which coincides exactly with
the right-hand part of the equation~\eqref{1syst} in our case.
Since
the linear span of the vectors~\eqref{VarDer} coincides with the
linear span of the complete set of the regular left eigenvectors
of the operator ${\hat L}^{i}_{j[{\bf U},\bm{\theta}_{0}]}(X,T)$
with zero eigenvalues, we get that system~\eqref{ConsWhitham}
is equivalent to system~\eqref{RegWhithSyst}.
\end{proof}

Let us note here that it follows from Lemma~\ref{Lemma2.2.} that systems
\eqref{ConsWhitham}, obtained from dif\/ferent sets of conservation laws
are equivalent to each other.
In other words, if system
\eqref{EvInSyst} has additional conservation laws then their averaging
gives relations following from system~\eqref{ConsWhitham}.

Let us note also that the justif\/ication questions discussed above, as a~rule, can be considered
in a~simpler way under additional assumptions about the next
corrections to the main approximation~\eqref{psi0} (see, e.g.~\cite{dm}).
We note again that here we don't make any additional assumptions of this
kind and consider the regular Whitham system as an independent object
that is asso\-cia\-ted only with description of the main approximation~\eqref{psi0}.
As we have said, we will follow this approach
everywhere in the paper.

The Hamiltonian properties of systems~\eqref{ConsWhitham} and more
general systems~\eqref{WhithamSystem} play very important role in their
consideration.
The general theory of systems~\eqref{WhithamSystem},
which are Hamiltonian with respect to local Poisson brackets of
hydrodynamic type (Dubrovin--Novikov brackets) was constructed by
B.A.~Dubrovin and S.P.~Novikov.
Let us give here a~brief description
of the Dubrovin--Novikov Hamiltonian structures and of the properties
of the corresponding systems~\eqref{WhithamSystem}.

The Dubrovin--Novikov bracket on the space of f\/ields
$(U^{1}(X),\dots,U^{N}(X))$ has the form
\begin{gather}\label{DNbr}
\{U^{\nu}(X),U^{\mu}(Y)\}=g^{\nu\mu}({\bf U})
\delta^{\prime}(X - Y)+
b^{\nu\mu}_{\gamma}({\bf U})U^{\gamma}_{X}
\delta(X - Y),\qquad\nu,\mu=1,\dots,N.
\end{gather}

The Hamiltonian operator corresponding to~\eqref{DNbr} can be
written in the form
\begin{gather*}
{\hat J}^{\nu\mu}=
g^{\nu\mu}({\bf U}){{\rm d}\over {\rm d} X}+b^{\nu\mu}_{\gamma}({\bf U})U^{\gamma}_{X}.
\end{gather*}

As was shown by B.A.~Dubrovin and S.P.~Novikov
 \cite{dn1,dn2,dn3}, expression~\eqref{DNbr} with non-degenerate
tensor $g^{\nu\mu}({\bf U})$ def\/ines a~Poisson bracket on the space
of f\/ields ${\bf U}(X)$ if and only~if:
\begin{enumerate}\itemsep=0pt
\item[1)] tensor $g^{\nu\mu}({\bf U})$ gives a~symmetric f\/lat pseudo-Riemannian
metric with upper indexes on the space of parameters
$(U^{1},\dots,U^{N})$;

\item[2)] the values
\begin{gather*}
\Gamma^{\nu}_{\mu\gamma}=-g_{\mu\lambda}b^{\lambda\nu}_{\gamma},
\end{gather*}
where
$g^{\nu\lambda}({\bf U})g_{\lambda\mu}({\bf U})=\delta^{\nu}_{\mu}$,
represent the Christof\/fel symbols for the corresponding metric~$g_{\nu\mu}({\bf U})$.
\end{enumerate}

As follows from the statements above, every Dubrovin--Novikov
bracket with non-degenerate tensor $g^{\nu\mu}({\bf U})$ can be written
in the canonical form~\cite{dn1,dn2,dn3}
\begin{gather*}
\{n^{\nu}(X),n^{\mu}(Y)\}=\epsilon^{\nu}\delta^{\nu\mu}\delta^{\prime}(X - Y),\qquad \epsilon^{\nu}=\pm1,
\end{gather*}
after the transition to the f\/lat coordinates
$n^{\nu}=n^{\nu}({\bf U})$ for the metric~$g_{\nu\mu}({\bf U})$.

The functionals
\begin{gather*}
N^{\nu}=\int_{-\infty}^{+\infty}
n^{\nu}(X){\rm d}X
\end{gather*}
represent the annihilators of the Dubrovin--Novikov bracket
while the functional
\begin{gather*}
P=\int_{-\infty}^{+\infty}\frac12
\sum_{\nu=1}^{N}\epsilon^{\nu}(n^{\nu})^{2}(X){\rm d}X
\end{gather*}
represents the momentum functional for the bracket~\eqref{DNbr}.

The Hamiltonian functions in the theory of brackets~\eqref{DNbr}
are represented by the functionals of hydrodynamic type, i.e.\
\begin{gather*}
H=\int_{-\infty}^{+\infty}
h({\bf U}){\rm d}X.
\end{gather*}

The bracket~\eqref{DNbr} has also two other important forms
on the space of ${\bf U}(X)$.
One of them is the ``Liouville''
form~\cite{dn1,dn2,dn3} having the form
\begin{gather*}
\{U^{\nu}(X),U^{\mu}(Y)\}=(\gamma^{\nu\mu}({\bf U})
+\gamma^{\mu\nu}({\bf U}))\delta^{\prime}(X - Y)+
{\partial\gamma^{\nu\mu}\over\partial U^{\lambda}}
U^{\lambda}_{X}\delta(X - Y)
\end{gather*}
for some functions $\gamma^{\nu\mu}({\bf U})$.

The ``Liouville'' form of the Dubrovin--Novikov bracket is called
also the physical form and corresponds to the case when the integrals
of coordinates $U^{\nu}$
\begin{gather*}
I^{\nu}=\int_{-\infty}^{+\infty}
U^{\nu}(X){\rm d}X
\end{gather*}
commute with each other.

Another important form of the Dubrovin--Novikov bracket is the
diagonal form.
It corresponds to the case when the coordinates
$U^{\nu}$ represent the diagonal coordinates for the metric~$g_{\nu\mu}({\bf U})$ and the tensor $g^{\nu\mu}({\bf U})$
in~\eqref{DNbr} has a~diagonal form.
This form of the
Dubrovin--Novikov bracket is closely connected with the integration
theory of systems of hydrodynamic type
\begin{gather*}
U^{\nu}_{T}=
V^{\nu}_{\mu}({\bf U})U^{\mu}_{X}
\end{gather*}
which can be written in the diagonal form{\samepage
\begin{gather}\label{DiagSystHT}
U^{\nu}_{T}=
V^{\nu}({\bf U})U^{\nu}_{X}
\end{gather}
(no summation) and are Hamiltonian with respect to bracket~\eqref{DNbr}.}

It was conjectured by S.P.~Novikov that all the systems of
hydrodynamic type having the form~\eqref{DiagSystHT} and Hamiltonian
with respect to any bracket~\eqref{DNbr} are integrable.
This
conjecture was proved by S.P.~Tsarev in~\cite{tsarev} where the
method (the ``generalized hodograph method'') of integration of
these systems was suggested.
However, the method of Tsarev proved to
be appli\-cab\-le to a~wider class of diagonalizable systems of hydrodynamic
type which was called by Tsarev ``semi-Hamiltonian''.
As it turned out
later in the class of ``semi-Hamiltonian systems'' fall also the
systems Hamiltonian with respect to generalizations of the
Dubrovin--Novikov bracket~--- the weakly nonlocal Mokhov--Ferapontov bracket~\cite{mohfer1} and the Ferapontov bracket~\cite{fer1, fer2}.
Various aspects of the weakly nonlocal brackets of hydrodynamic
type are discussed in~\cite{mohfer1, fer1, fer2, fer3, fer4, Pavlov1, PhysD}.
Let us note also that the generalization of the Dubrovin--Novikov
procedure for weakly nonlocal case was proposed in~\cite{malnloc2}.

Let us describe now the procedure for constructing the
Dubrovin--Novikov bracket for the Whitham system in the case when the original system~\eqref{EvInSyst}
is Hamiltonian with respect to a~local f\/ield-theoretic bracket
\begin{gather}\label{LocTheorFieldBr}
\{\varphi^{i}(x),\varphi^{j}(y)\}=\sum_{k\geq0}
B^{ij}_{(k)}(\bm{\varphi},\bm{\varphi}_{x},\dots)
\delta^{(k)}(x-y)
\end{gather}
with the local Hamiltonian of the form
\begin{gather}\label{hamfun}
H=\int P_{H}
(\bm{\varphi},\bm{\varphi}_{x},\dots){\mathrm d}x   ,                     
\end{gather}
which was suggested by B.A.~Dubrovin and S.P.~Novikov~\cite{dn1,dn2,dn3}.

Method of B.A.~Dubrovin and S.P.~Novikov is based on
the existence of $N$ (equal to the number of parameters
$U^{\nu}$ of the family of \mbox{$m$-phase} solutions of~\eqref{EvInSyst})
local integrals
\begin{gather}\label{integ}
I^{\nu}=\int
P^{\nu}(\bm{\varphi},\bm{\varphi}_{x},\dots){\mathrm d}x,
\end{gather}
which commute with the Hamiltonian~\eqref{hamfun} and with each other
\begin{gather}\label{invv}
\{I^{\nu},H\}=0,\qquad
\{I^{\nu},I^{\mu}\}=0
\end{gather}
and can be described as follows.

We calculate the pairwise Poisson brackets of the densities
$P^{\nu}$ having the form
\begin{gather}\label{PnuPmuskob}
\{P^{\nu}(x),P^{\mu}(y)\}=\sum_{k\geq0}
A^{\nu\mu}_{k}(\bm{\varphi},\bm{\varphi}_{x},\dots)
\delta^{(k)}(x-y),
\end{gather}
where
\begin{gather*}
A^{\nu\mu}_{0}(\bm{\varphi},\bm{\varphi}_{x},\dots)\equiv
\partial_{x}Q^{\nu\mu}(\bm{\varphi},\bm{\varphi}_{x},\dots)
\end{gather*}
according to~\eqref{invv}.

The corresponding Dubrovin--Novikov bracket on the space of functions
${\bf U}(X)$ has the form
\begin{gather}\label{dubrnovb}
\{U^{\nu}(X),U^{\mu}(Y)\}=\langle A^{\nu\mu}_{1}\rangle({\bf U})\delta^{\prime}(X - Y)+
{\partial\langle Q^{\nu\mu}\rangle\over\partial U^{\gamma}} U^{\gamma}_{X}\delta(X - Y).
\end{gather}

Let us remind that we assume that the parameters~$U^{\nu}$
coincide with the values of the functionals~$I^{\nu}$
def\/ined on the corresponding quasiperiodic solutions of the
family~$\Lambda$
\begin{gather*}
U^{\nu}=\langle P^{\nu}(x)\rangle.
\end{gather*}

The Whitham system~\eqref{ConsWhitham} is Hamiltonian with respect
to the Dubrovin--Novikov brac\-ket~\eqref{dubrnovb} with the
Hamiltonian
\begin{gather}\label{UsrHamFunc}
H^{av}=\int_{-\infty}^{+\infty}
\langle P_{H}\rangle\left({\bf U}(X)\right){\mathrm d}X.
\end{gather}

The proof of the Jacobi identity for the bracket
\eqref{dubrnovb} was suggested in~\cite{izvestia}
under certain assumptions about the family of \mbox{$m$-phase} solutions
of~\eqref{EvInSyst}.
Besides that, it was
shown in~\cite{MalPav} that the Dubrovin--Novikov procedure is
compatible with the procedure of averaging of local Lagrangian
functions when carrying out of both the procedures is possible.

Let us note also that in~\cite{Pavlov2} all the local brackets
\eqref{DNbr} for the Whitham equations for KdV, NLS, and SG
equations were found.
Besides that, in \cite{Alekseev} 
the hierarchies of the weakly nonlocal Hamiltonian structures
for the Whitham systems for KdV were represented.

In this paper we give the most detailed discussion and justif\/ication
of the Dubrovin--Novikov procedure separately for the single-phase and
the multiphase cases.
In the multiphase case we will show also that
the justif\/ication of the Dubrovin--Novikov procedure can be done under
weaker assumptions than in previous reviews.
In particular, we
will show that the justif\/ication of the procedure is in fact insensitive
to the appearance of ``resonances'' which can arise in the multi-phase
case, which is the basis for its widespread use in the multiphase
situation.
In all cases we will consider the regular Whitham system
def\/ined above without any assumptions about the form of higher
corrections to the main approximation~\eqref{psi0} of a~slowly
modulated solution.

In this chapter we consider general constructions connected with the
Dubrovin--Novikov procedure and prove some technical lemmas needed
for its justif\/ication.
At the end of the chapter, we give a~detailed
justif\/ication of the procedure in the single-phase case.
In the next
chapter we will consider multiphase case, where the justif\/ication of
the Dubrovin--Novikov procedure will have a~more complicated form.
In conclusion, we consider a~simple but typical example of the
justif\/ication of the Dubrovin--Novikov procedure using the
Gardner--Zakharov--Faddeev bracket for KdV.

As we said above, we will now consider the general structures
connected with the construction of the Hamiltonian structure for
the regular Whitham system.

{\sloppy Let us note that according to~\eqref{invv} the f\/lows
\begin{gather}\label{Snuflows}
\varphi^{i}_{t^{\nu}}=
S^{i\nu}(\bm{\varphi},\bm{\varphi}_{x},\dots)=
\{\varphi^{i}(x),I^{\nu}\}
\end{gather}
generated by the functionals $I^{\nu}$ according to the bracket~\eqref{LocTheorFieldBr} commute with the initial f\/low~\eqref{EvInSyst}.
The f\/lows~\eqref{Snuflows} leave invariant the full families of
\mbox{$m$-phase} solutions of~\eqref{EvInSyst} as well as the values
$U^{\nu}=I^{\nu}$ of the functionals $I^{\nu}$ on them.
For
a complete regular family of \mbox{$m$-phase} solutions with independent
parameters
$(k^{1},\dots,k^{m})$ it's not dif\/f\/icult to show that the f\/lows~\eqref{Snuflows} generate linear (in time) shifts of the phases
$\theta_{0}^{\alpha}$ with some constant frequen\-cies~$\omega^{\alpha\nu}({\bf U})$, such that
\begin{gather}\label{LinPotSnu}
S^{i\nu}\big(\bm{\Phi},k^{\beta}\bm{\Phi}_{\theta^{\beta}},
\dots\big)=\omega^{\alpha\nu}({\bf U})\Phi^{i}_{\theta^{\alpha}}(\bm{\theta},{\bf U}).
\end{gather}}

According to Lemma~\ref{Lemma2.1.} we have also that the functionals
$k^{\alpha}({\bf I})$ should generate the zero f\/lows on the
corresponding family of \mbox{$m$-phase} solutions of~\eqref{EvInSyst}.
For $U^{\nu}$ coinciding with the values of $I^{\nu}$ on
$\Lambda$ we get then the relations
\begin{gather}\label{komeganulrel}
{\partial k^{\alpha}({\bf U})\over\partial U^{\nu}}\omega^{\beta\nu}({\bf U})\equiv0,\qquad\alpha,\beta=1,\dots,m.
\end{gather}

Let us give here the def\/inition of a~regular Hamiltonian
family of \mbox{$m$-phase} solutions of system~\eqref{EvInSyst} and
of a~complete Hamiltonian set of the functionals~\eqref{integ}.

\begin{definition}\label{Definition2.1.}
We call family $\Lambda$ of \mbox{$m$-phase} solutions of system
\eqref{EvInSyst} a~regular Hamiltonian family if:
\begin{enumerate}\itemsep=0pt
\item[1)] it represents a~complete regular family of \mbox{$m$-phase} solutions
of~\eqref{EvInSyst} in the sense of Def\/inition~\ref{Definition1.1.};

\item[2)] the corresponding bracket~\eqref{LocTheorFieldBr} has on $\Lambda$
constant number of annihilators $N^{1}, \dots,N^{s}$
with linearly independent variational derivatives
$\delta N^{l}/\delta\varphi^{i}(x)$ which
coincides with the number of independent annihilators in the
neighborhood of $\Lambda$.
\end{enumerate}
\end{definition}

Let us say that according to the generalized Darboux theorem
we can identify the number of the variational derivatives
$\delta N^{l}/\delta\varphi^{i}(x)$ on $\Lambda$ with the
number of linearly independent quasiperiodic solutions
$v^{(l)}_{i}(x)$ of the equation
\begin{gather*}
\sum_{k\geq0}\left.
B^{ij}_{(k)}(\bm{\varphi},\bm{\varphi}_{x},\dots)\right|_{\Lambda}v^{(l)}_{j,kx}=0,
\end{gather*}
where $v^{(l)}_{i}(x)$ have the same quasiperiods as the corresponding
functions $\bm{\varphi}(x)$ on $\Lambda$.

\begin{definition}\label{Definition2.2.}
We call a~set $(I^{1},\dots,I^{N})$ of commuting functionals
\eqref{integ} a~complete Hamiltonian set on a~regular Hamiltonian family
$\Lambda$ of \mbox{$m$-phase} solutions of system~\eqref{EvInSyst}
if:
\begin{enumerate}\itemsep=0pt
\item[1)] the restriction of the functionals $(I^{1},\dots,I^{N})$
on the quasiperiodic solutions of the family $\Lambda$ gives a~complete
set of parameters $(U^{1},\dots,U^{N})$ on this family;

\item[2)] the Hamiltonian f\/lows generated by $(I^{1},\dots,I^{N})$
generate on $\Lambda$ linear phase shifts of $\bm{\theta}_{0}$
with frequencies $\bm{\omega}^{\nu}({\bf U})$, such that
\begin{gather*}
\operatorname{rank}||\omega^{\alpha\nu}({\bf U})||=m;
\end{gather*}

\item[3)] the linear space generated by the variational derivatives
$\delta I^{\nu}/\delta\varphi^{i}(x)$ on $\Lambda$ contains
the variational derivatives of all the annihilators $N^{q}$ of the
bracket~\eqref{LocTheorFieldBr}, such that
\begin{gather*}
\left.
{\delta N^{l}\over\delta\varphi^{i}(x)}
\right|_{\Lambda}=
\sum_{\nu}\gamma^{l}_{\nu}({\bf U})\left.
{\delta I^{\nu}\over\delta\varphi^{i}(x)}\right|_{\Lambda}
\end{gather*}
for some smooth functions $\gamma^{l}_{\nu}({\bf U})$ on the
family $\Lambda$.
\end{enumerate}
\end{definition}

Let us note that it follows from Def\/inition~\ref{Definition2.2.}
that if a~complete
Hamiltonian set of integrals $(I^{1},\dots,I^{N})$ exists for
a regular Hamiltonian family $\Lambda$ then the number of the additional
para\-me\-ters $(n^{1},\dots,n^{s})$ discussed above is equal to the
number of annihilators of the bracket~\eqref{LocTheorFieldBr}.
Indeed, according to Def\/initions~\ref{Definition2.1.} and
\ref{Definition2.2.},
the number of the functionals
$I^{\nu}$ having linearly independent variational derivatives on
$\Lambda$ exactly equals to $m+s$, where $s$ is the
number of annihilators of the bracket~\eqref{LocTheorFieldBr}.
The total number of independent parameters $U^{\nu}$ on $\Lambda$
is then equal to $2m+s$ due to the wave
vectors $k^{\alpha}$, $\alpha=1,\dots,m$,
which implies the above assertion.

As follows from the condition (2) of
Def\/inition~\ref{Definition2.2.} and
from the invariance of the functio\-nals~$N^{l}$ and~$I^{\nu}$
with respect to the f\/lows~\eqref{Snuflows}, the values
$\gamma^{l}_{\nu}({\bf U})$ can always be chosen independent
on the initial phase shifts on the family~$\Lambda$.
The values
$\delta I^{\nu}/\delta\varphi^{i}(x)|_{\Lambda}$ are linearly
dependent on $\Lambda$, so it's natural to choose a
complete linearly independent subsystem.
Remembering that the
variational derivatives of $J^{\nu}$~\eqref{VarDer} are linear
combinations of the regular left eigenvectors
$\kappa^{(q)}_{i[{\bf U}]}(\bm{\theta}+\bm{\theta}_{0})$,
we can write
\begin{gather}\label{RazlozhAnn}
\left.
{\delta N^{l}\over\delta\varphi^{i}(x)}
\right|_{\bm{\varphi}(x)=\bm{\Phi}({\bf k}({\bf U})x+\bm{\theta}_{0},{\bf U})}
=\sum_{q}n^{l}_{q}({\bf U})
\kappa^{(q)}_{i[{\bf U}]}\left({\bf k}({\bf U})x+\bm{\theta}_{0}\right)
\end{gather}
for some smooth functions $n^{l}_{q}({\bf U})$.
The functions
$n^{l}_{q}({\bf U})$ are then uniquely determined on
$\Lambda$ and we have
$\text{rank}||n^{l}_{q}({\bf U})||=s$ by
Def\/inition~\ref{Definition2.1.}.

We will need now the construction of the Dirac restriction
of a~Poisson bracket on a~sub\-ma\-ni\-fold.
We describe here brief\/ly
this construction.
Using the terminology of f\/inite-dimensional spaces
we can say that the Dirac restriction of a~Poisson bracket on a
submanifold ${\cal N}^{k}\subset{\cal M}^{n}$ is associated with
a special choice of coordinates in the neighborhood of the submanifold
${\cal N}^{k}$.
The coordinates in the neighborhood of the submanifold
${\cal N}^{k}$ are divided into the ``coordinates on the submanifold''
$(U^{1},\dots,U^{k})$ and the constraints
$(g^{1},\dots,g^{n-k})$ which def\/ine the sub\-ma\-ni\-fold~${\cal N}^{k}$.
It is assumed that the submanifold ${\cal N}^{k}$
is def\/ined by the conditions
\begin{gather*}
g^{i}({\bf x})=0,\qquad i=1,\dots,n-k,
\end{gather*}
while the functions $U^{1}({\bf x}), \dots,U^{k}({\bf x})$ on ${\cal M}^{n}$ play the role of a~coordinate
system on ${\cal N}^{k}$ after the restriction to this submanifold.

If the Hamiltonian f\/lows generated by the functions
$U^{j}({\bf x})$ leave the submanifold ${\cal N}^{k}$
invariant, i.e.\
we have
\begin{gather*}
\{U^{j}({\bf x}),g^{i}({\bf x})\}=0
\qquad\text{for}\quad{\bf g}({\bf x})=0,
\end{gather*}
then the pairwise Poisson brackets of the functions $U^{j}({\bf x})$
def\/ine the Poisson tensor in the coordinates $(U^{1},\dots,U^{k})$
after restriction on ${\cal N}^{k}$ which is called the Dirac
restriction of the Poisson bracket $\{\cdot,\cdot\}$
on the submanifold ${\cal N}^{k}\subset{\cal M}^{n}$.

In general, according to the procedure of Dirac, if we have some
constraints $g^{i}({\bf x})$ which def\/ine the submanifold
${\cal N}^{k}$ and some functions $U^{j}({\bf x})$ which give
a coordinate system on ${\cal N}^{k}$ we must f\/ind $k$
linear combinations
$\beta^{j}_{s}({\bf U})g^{s}({\bf x})$ at every point of
${\cal N}^{k}$, such that for the functions
\begin{gather*}
{\tilde{U}}^{j}({\bf x})=U^{j}({\bf x})+\beta^{j}_{s}({\bf U})g^{s}({\bf x}),\qquad
j=1,\dots,k,
\end{gather*}
we have the relations
\begin{gather*}
\big\{{\tilde{U}}^{j}({\bf x}),g^{i}({\bf x})\big\}=0
\end{gather*}
for ${\bf g}({\bf x})=0$.

The functions ${\tilde{U}}^{j}({\bf x})$ take the same values
that $U^{j}({\bf x})$ at the points of ${\cal N}^{k}$
and we can def\/ine the Dirac bracket
$\{\cdot,\cdot\}_{D}$ on ${\cal N}^{k}$ by the formula
\begin{gather*}
\{U^{i},U^{j}\}_{D}=
\big\{{\tilde{U}}^{i}({\bf x}),{\tilde{U}}^{j}({\bf x})\big\}\big|_{{\cal N}^{k}}({\bf U}).
\end{gather*}

The functions $\beta^{j}_{s}({\bf U})$ are determined from the
linear systems
\begin{gather*}
\big\{g^{i}({\bf x}),g^{s}({\bf x})\big\}\big|_{{\cal N}^{k}}
\beta^{j}_{s}({\bf U})+\big\{g^{i}({\bf x}),U^{j}({\bf x})\big\}\big|_{{\cal N}^{k}}=0,\qquad i=1,\dots,n-k,
\end{gather*}
and we can also write{\samepage
\begin{gather*}
\big\{U^{i},U^{j}\big\}_{D}=\big\{U^{i}({\bf x}),U^{j}({\bf x})\big\}\big|_{{\cal N}^{k}}
-\beta^{i}_{s}({\bf U})
\big\{g^{s}({\bf x}),g^{q}({\bf x})\big\}\big|_{{\cal N}^{k}}\beta^{j}_{q}({\bf U})
\end{gather*}
for the Dirac bracket on ${\cal N}^{k}$.}

We now describe the procedure of the Dirac restriction,
which will be needed in our situation.
We consider now system~\eqref{EvInSyst} which is
Hamiltonian with respect to some local bracket~\eqref{LocTheorFieldBr}
with a~local Hamiltonian function of the form~\eqref{hamfun}.
We f\/irst introduce the extended space of f\/ields
\begin{gather*}
\bm{\varphi}(x)\rightarrow\bm{\varphi}(\bm{\theta},x),
\end{gather*}
where the functions $\bm{\varphi}(\bm{\theta},x)$
are $2\pi$-periodic with respect to each $\theta^{\alpha}$,
and def\/ine the extended Poisson bracket
\begin{gather*}
\big\{\varphi^{i}(\bm{\theta},x),\varphi^{j}(\bm{\theta}^{\prime},y)\big\}
=\sum_{k\geq0}
B^{ij}_{(k)}(\bm{\varphi},\bm{\varphi}_{x},\dots)
\delta^{(k)}(x-y)\delta(\bm{\theta}-\bm{\theta}^{\prime}).
\end{gather*}

We also make the replacement $x\rightarrow X=\epsilon x$ and
def\/ine the Poisson bracket
\begin{gather}\label{EpsExtBracket}
\big\{\varphi^{i}(\bm{\theta},X),\varphi^{j}(\bm{\theta}^{\prime},Y)\big\}
=\sum_{k\geq0}\epsilon^{k}
B^{ij}_{(k)}(\bm{\varphi},\epsilon\bm{\varphi}_{X},\dots)
\delta^{(k)}(X - Y)\delta(\bm{\theta}-\bm{\theta}^{\prime})
\end{gather}
on the space of f\/ields $\bm{\varphi}(\bm{\theta},X)$.

Let us def\/ine again the submanifold ${\cal K}$ by the formula
\eqref{VarphiSubs} in the extended space of f\/ields.
We thus assume
that the functions $\bm{\varphi}(\bm{\theta},X)$ represent
functions from the family $\Lambda$
of the \mbox{$m$-phase} solutions of~\eqref{EvInSyst}
with some parameters ${\bf U}(X)$ at every $X$.

It is convenient to choose the boundary conditions in the form
${\bf U}(X)\rightarrow{\bf U}_{0}$, $X\rightarrow\pm\infty$,
such that $k^{\alpha}({\bf U}_{0})=0$ and def\/ine the functions
$S^{\alpha}(X)$ by the formula
\begin{gather*}
S^{\alpha}(X)=\frac12 \int_{-\infty}^{+\infty}\operatorname{sgn}(X - Y)k^{\alpha}(Y){\mathrm d}Y.
\end{gather*}

We can write then
\begin{gather}\label{FormaPhinaK}
\varphi^{i}(\bm{\theta},X)=\Phi^{i}\left({{\bf S}[{\bf U}](X)\over\epsilon}+\bm{\theta},{\bf U}(X)\right),
\end{gather}
where $S^{\alpha}[{\bf U}]$ is a~single-valued functional of
coordinates ${\bf U}(X)$ on the submanifold~${\cal K}$.\footnote{The
case when the values $k^{\alpha}=0$ are absent on the space of the
parameters~${\bf U}$ requires just a~simple modif\/ication of the
def\/inition of~${\bf S}(X)$ which we do not consider here.}

Let us introduce the functionals $J^{\nu}(X)$ on the functions
$\bm{\varphi}(\bm{\theta},X)$ by the formula
\begin{gather}\label{FuncJnuX}
J^{\nu}(X)=\int_{0}^{2\pi}\cdots\int_{0}^{2\pi}
P^{\nu}\big(\bm{\varphi},\epsilon\bm{\varphi}_{X},\epsilon^{2}
\bm{\varphi}_{XX},\dots\big){{\mathrm d}^{m}\theta\over(2\pi)^{m}}
\end{gather}
and consider their values on the functions of the family~${\cal K}$.

We can write on ${\cal K}$,
\begin{gather}\label{JUtransform}
J^{\nu}(X)=U^{\nu}(X)+\sum_{l\geq1}\epsilon^{l}J^{\nu}_{(l)}(X),
\end{gather}
where $J^{\nu}_{(l)}(X)$ are polynomials in the derivatives
${\bf U}_{X}, {\bf U}_{XX}, \dots$
with coef\/f\/icients depending on~${\bf U}$ and have grading degree~$l$ in terms of total
number of derivations with respect to~$X$.

The transformation~\eqref{JUtransform} can be inverted as a~formal
series in $\epsilon$, such that we can write
\begin{gather}\label{UJtransform}
U^{\nu}(X)=J^{\nu}(X)+\sum_{l\geq1}
\epsilon^{l}U^{\nu}_{(l)}(X)
\end{gather}
on the functions of the submanifold ${\cal K}$.
In the formula~\eqref{UJtransform} the functions $U^{\nu}_{(l)}$
are functions of
${\bf J}, {\bf J}_{X}, {\bf J}_{XX}, \dots$,
polynomial in the derivatives
${\bf J}_{X}, {\bf J}_{XX}, \dots$, and
having degree $l$ in terms of the number of derivations w.r.t.~$X$.

The values of the functionals $J^{\nu}(X)$ or $U^{\nu}(X)$
on the functions from ${\cal K}$ can thus be chosen as coordinates on
${\cal K}$.
We can def\/ine also the functionals ${\bf U}(X)$ on the
whole functional space using the def\/inition of the functionals
${\bf J}(X)$ and relations~\eqref{UJtransform}.
Let us introduce also the constraints
$g^{i}(\bm{\theta},X)$ def\/ining the submanifold ${\cal K}$
by the conditions $g^{i}(\bm{\theta},X)=0$
and numbered by the values of $\bm{\theta}$ and $X$
\begin{gather}\label{gconstr}
g^{i}(\bm{\theta},X)=\varphi^{i}(\bm{\theta},X)
-\Phi^{i}\left({{\bf S}[{\bf U}[{\bf J}]](X)\over\epsilon}
+\bm{\theta},{\bf U}[{\bf J}](X)\right).
\end{gather}

The constraints $g^{i}(\bm{\theta},X)$
are functionals on the whole extended space of f\/ields
$\varphi^{i}(\bm{\theta},X)$ by virtue of the
corresponding def\/inition of the functionals
$J^{\nu}(X)$.

The constraints~\eqref{gconstr} are not independent since
the following relations hold identically for the ``gradients''
$\delta g^{i}(\bm{\theta},X)/\delta\varphi^{j}
(\bm{\theta}^{\prime},Y)$ on the submanifold ${\cal K}$
\begin{gather}\label{dependence}
\int_{-\infty}^{+\infty}\int_{0}^{2\pi}\cdots\int_{0}^{2\pi}
\left.
{\delta J^{\nu}(Z)\over
\delta\varphi^{i}(\bm{\theta},X)}\right|_{\cal K}
\left.
{\delta g^{i}(\bm{\theta},X)\over
\delta\varphi^{j}(\bm{\theta}^{\prime},Y)}\right|_{\cal K}
{{\mathrm d}^{m}\theta\over(2\pi)^{m}}{\mathrm d}X\equiv0.
\end{gather}

Nevertheless, it will be convenient here not to choose an independent
subsystem from~\eqref{gconstr} and keep the system of constraints
in the form~\eqref{gconstr} keeping in mind the existence of identi\-ties~\eqref{dependence}.

Thus, we consider now the values of the functionals
$[{\bf U}(X),{\bf g}(\bm{\theta},X)]$ as a~coordinate system
in the neighborhood of ${\cal K}$ with the relations
\eqref{dependence}.
The values of the functionals ${\bf U}(X)$ will
be considered as a~coordinate system on ${\cal K}$.

Using the explicit expression for the quantities
$\delta J^{\nu}(Z)/\delta\varphi^{i}(\bm{\theta},X)$ on
${\cal K}$
\begin{gather*}
\left.
{\delta J^{\nu}(Z)\over\delta\varphi^{i}(\bm{\theta},X)}\right|_{\cal K}\\
\qquad{} =
\sum_{l\geq0}\Pi^{\nu}_{i(l)}
\left( \bm{\Phi}\left( {{\bf S}(Z)\over\epsilon}+\bm{\theta},
{\bf U}(Z) \right) ,\epsilon{\partial\over\partial Z}
\bm{\Phi}\left( {{\bf S}(Z)\over\epsilon}+\bm{\theta},
{\bf U}(Z) \right) ,\dots \right) \epsilon^{l}
\delta^{(l)}(Z - X),
\end{gather*}
we can write the convolution~\eqref{dependence} in the form of
action of the operator
\begin{gather}\label{DeistvSvert}
\!\int_{0}^{2\pi}\!\!\cdots\!\!\int_{0}^{2\pi}\!\!
{{\mathrm d}^{m}\theta\over(2\pi)^{m}}\!\sum_{l\geq0}\Pi^{\nu}_{i(l)}\!\!
\left(\!\bm{\Phi}\!\left(\!{{\bf S}(Z)\over\epsilon}+\bm{\theta},
{\bf U}(Z)\!\right)\!,\epsilon{\partial\over\partial Z}
\bm{\Phi}\!\left(\!{{\bf S}(Z)\over\epsilon}+\bm{\theta},
{\bf U}(Z)\!\right)\!,\dots\!\right)\!\epsilon^{l}
{{\mathrm d}^{l}\over{\mathrm d}Z^{l}}\!\!
\end{gather}
on the distributions
$\delta g^{i}(\bm{\theta},Z)/\delta\varphi^{j}
(\bm{\theta}^{\prime},Y)|_{\cal K}$.

Pairwise Poisson brackets of the functionals
$J^{\nu}(X)$, $J^{\mu}(Y)$
have the form
\begin{gather*}
\left\{J^{\nu}(X),J^{\mu}(Y)\right\}=
\sum_{k\geq0}\int_{0}^{2\pi}\cdots\int_{0}^{2\pi}
A^{\nu\mu}_{k}\big(\bm{\varphi}(\bm{\theta},X),
\epsilon\bm{\varphi}_{X}(\bm{\theta},X),\dots\big)
{{\mathrm d}^{m}\theta\over(2\pi)^{m}}\epsilon^{k}
\delta^{(k)}(X - Y),
\end{gather*}
where
\begin{gather}\label{epsAQRel}
A^{\nu\mu}_{0}(\bm{\varphi},\epsilon\bm{\varphi}_{X},\dots)
\equiv\epsilon\partial_{X}Q^{\nu\mu}(\bm{\varphi},\epsilon\bm{\varphi}_{X},\dots).
\end{gather}

Substituting the functions $\varphi^{i}(\bm{\theta},X)$ in the form~\eqref{FormaPhinaK} it is easy to see that the Dubrovin--Novikov
bracket is the main part
(in $\epsilon$) of the pairwise Poisson brackets
$\{J^{\nu}(X),J^{\mu}(Y)\}$ on the submanifold ${\cal K}$
in the coordinates $[{\bf U}(Z)]$.

The Poisson brackets of the f\/ields $\varphi^{i}(\bm{\theta},X)$
with the functionals $J^{\mu}(Y)$ can be written as
\begin{gather}\label{phiJmuexp}
\{\varphi^{i}(\bm{\theta},X),J^{\mu}(Y)\}=
\sum_{l\geq0}\epsilon^{l}C^{i\mu}_{(l)}
\left(\bm{\varphi}(\bm{\theta},X),\epsilon
\bm{\varphi}_{X}(\bm{\theta},X),\dots\right)
\delta^{(l)}(X - Y)
\end{gather}
with some smooth functions
$C^{i\mu}_{(l)}(\bm{\varphi},\epsilon\bm{\varphi}_{X},\dots)$.

We also have in this case{\samepage
\begin{gather}\label{C0SmuRel}
C^{i\mu}_{(0)}\left(\bm{\varphi}(\bm{\theta},X),\epsilon
\bm{\varphi}_{X}(\bm{\theta},X),\dots\right)\equiv
S^{i\mu}\left(\bm{\varphi}(\bm{\theta},X),\epsilon
\bm{\varphi}_{X}(\bm{\theta},X),\dots\right)
\end{gather}
by virtue of~\eqref{Snuflows}.}

For any function of the slow variable $q(Y)$ we can write
\begin{gather}\label{phiJq}
\left\{\varphi^{i}(\bm{\theta},X),
\int q(Y)J^{\mu}(Y){\mathrm d}Y\right\}=
\sum_{l\geq0}\epsilon^{l}C^{i\mu}_{(l)}
\left(\bm{\varphi}(\bm{\theta},X),\epsilon
\bm{\varphi}_{X}(\bm{\theta},X),\dots\right) q_{lX}.
\end{gather}

The leading term in the expression~\eqref{phiJq} on ${\cal K}$
has the form
\begin{gather*}
\left.
\left\{\varphi^{i}(\bm{\theta},X),
\int q(Y)J^{\mu}(Y){\mathrm d}Y\right\}\right|_{{\cal K}[0]}
\\
\qquad{}=C^{i\mu}_{(0)}\left(\bm{\Phi}\left(
{{\bf S}(X)\over\epsilon}+\bm{\theta},X\right),
k^{\beta}\bm{\Phi}_{\theta^{\beta}}\left(
{{\bf S}(X)\over\epsilon}+\bm{\theta},X\right),\dots
\right)q(X)
\\
\qquad{}=q(X) S^{i\mu}\left(\bm{\Phi}\left(
{{\bf S}(X)\over\epsilon}+\bm{\theta},X\right),
k^{\beta}\bm{\Phi}_{\theta^{\beta}}\left(
{{\bf S}(X)\over\epsilon}+\bm{\theta},X\right),\dots
\right),
\end{gather*}
where $S^{i\mu}(\bm{\varphi},\bm{\varphi}_{x},\dots)$ is the f\/low
\eqref{Snuflows} generated by the functional $I^{\mu}$.
According to~\eqref{LinPotSnu} we can write then
\begin{gather}\label{phiJbr}
\left.
\left\{
\varphi^{i}(\bm{\theta},X),\int q(Y)J^{\mu}(Y){\mathrm d}Y
\right\}\right|_{\cal K}
=\omega^{\alpha\mu}(X)
\Phi^{i}_{\theta^{\alpha}}\left({{\bf S}(X)\over\epsilon}+\bm{\theta},{\bf U}(X)\right)
q(X)+O(\epsilon).
\end{gather}

Similarly, for any smooth function $Q(\bm{\theta},X)$,
$2\pi$-periodic in each $\theta^{\alpha}$,
we can write on the basis of~\eqref{phiJmuexp}            
\begin{gather}
\int_{0}^{2\pi}\cdots\int_{0}^{2\pi}
Q\left({{\bf S}(X)\over\epsilon}+\bm{\theta},X\right)
\left.
\left\{\varphi^{i}(\bm{\theta},X),J^{\mu}(Y)\right\}
\right|_{\cal K}{{\mathrm d}^{m}\theta\over(2\pi)^{m}}
\nonumber\\
\qquad{}=\omega^{\alpha\mu}(X)
\int_{0}^{2\pi}\cdots\int_{0}^{2\pi} Q(\bm{\theta},X)
\Phi^{i}_{\theta^{\alpha}}(\bm{\theta},{\bf U}(X))
{{\mathrm d}^{m}\theta\over(2\pi)^{m}}\delta(X - Y)
+O(\epsilon).\label{QphiJbr}
\end{gather}

In view of relations~\eqref{JUtransform}, \eqref{UJtransform} for
the functionals $J^{\mu}(Y)$, $U^{\mu}(Y)$ we can also write
\begin{gather}
\left.
\left\{
\varphi^{i}(\bm{\theta},X),\int q(Y)U^{\mu}(Y){\mathrm d}Y
\right\}\right|_{\cal K}
=\omega^{\alpha\mu}(X)
\Phi^{i}_{\theta^{\alpha}}\left(
{{\bf S}(X)\over\epsilon}+\bm{\theta},{\bf U}(X)\right)
q(X)+O(\epsilon),\label{phiUbr}
\\
\int_{0}^{2\pi}\cdots\int_{0}^{2\pi}
Q\left({{\bf S}(X)\over\epsilon}+\bm{\theta},X\right)
\left.
\left\{\varphi^{i}(\bm{\theta},X),U^{\mu}(Y)\right\}
\right|_{\cal K}{{\mathrm d}^{m}\theta\over(2\pi)^{m}}
\nonumber\\
\qquad{} =\omega^{\alpha\mu}(X)
\int_{0}^{2\pi}\cdots\int_{0}^{2\pi} Q(\bm{\theta},X)
\Phi^{i}_{\theta^{\alpha}}(\bm{\theta},{\bf U}(X))
{{\mathrm d}^{m}\theta\over(2\pi)^{m}}\delta(X - Y)
+O(\epsilon).\label{QphiUbr}
\end{gather}

In this case, by virtue of~\eqref{komeganulrel} we have
\begin{gather}
\left.
\left\{
\varphi^{i}(\bm{\theta},X),\int q(Y)k^{\alpha}({\bf U}(Y))
{\mathrm d}Y\right\}\right|_{\cal K}
=O(\epsilon),\label{phiknulbr}
\\
\int_{0}^{2\pi}\cdots\int_{0}^{2\pi}
Q\left({{\bf S}(X)\over\epsilon}+\bm{\theta},X\right)
\left.
\left\{\varphi^{i}(\bm{\theta},X),
k^{\alpha}({\bf U}(Y))\right\}\right|_{\cal K}
{{\mathrm d}^{m}\theta\over(2\pi)^{m}}=O(\epsilon)\label{QvarphikalphaKzerorel}
\end{gather}
for f\/ixed values of coordinates $[{\bf U}(Z)]$.

Let us note here that since we consider the change of coordinates,
which depends explicitly on~$\epsilon$, we shall assume now,
that all relations are written in the coordinate system
$[{\bf U}(X),{\bf g}(\bm{\theta},X)]$ near ${\cal K}$.
In particular, all the values on ${\cal K}$ will be functionals of
${\bf U}(X)$, represented usually in the form of graded expansions
in $\epsilon$ with coef\/f\/icients depending on
${\bf U}, {\bf U}_{X}, {\bf U}_{XX}, \dots$.

Let us prove here some lemmas about the structure of Poisson
brackets on the submani\-fold~${\cal K}$ which we will need in the
further consideration.

\begin{lemma}\label{Lemma2.3.}
Let the values $U^{\nu}$ of the functionals $I^{\nu}$ on a
complete regular family $\Lambda$ of \mbox{$m$-phase} solutions of
\eqref{EvInSyst} be functionally independent
and give a~complete set of parameters on $\Lambda$,
excluding the initial phases.
Then for the Poisson brackets of the
functionals $k^{\alpha}({\bf U}(X))$ and~$J^{\mu}(Y)$ on ${\cal K}$
we have the following relation
\begin{gather}\label{kJmuskob}
\big\{k^{\alpha}({\bf U}(X)),J^{\mu}(Y)\big\}\big|_{\cal K}=
\epsilon\big[\omega^{\alpha\mu}({\bf U}(X))\delta(X - Y)\big]_{X}+O\big(\epsilon^{2}\big).
\end{gather}
\end{lemma}

\begin{proof}
The conditions of Lemma~\ref{Lemma2.3.}
coincide with the conditions of
Lemmas~\ref{Lemma2.1.} and~\ref{Lemma2.2.}.
Consider the Hamiltonian f\/low generated by the
functional $\int q(Y)J^{\mu}(Y){\mathrm d}Y$ according to the bracket~\eqref{EpsExtBracket} with a~compactly supported function
$q(Y)$ of the slow variable $Y=\epsilon y$.

According to~\eqref{phiJq} we can write at the ``points''
of the submanifold ${\cal K}$
\begin{gather}\label{varphiqXev}
\varphi^{i}_{t}=q(X)\omega^{\beta\mu}({\bf U}(X))
\Phi^{i}_{\theta^{\beta}}\left(
{{\bf S}(X)\over\epsilon}+\bm{\theta},{\bf U}(X)\right)
+\epsilon\eta^{i}_{[q]}\left(
{{\bf S}(X)\over\epsilon}+\bm{\theta},X\right)+
O\big(\epsilon^{2}\big)
\end{gather}
with some ($2\pi$-periodic in each $\theta^{\alpha}$)
functions $\eta^{i}_{[q]}(\bm{\theta},X)$.

Let us introduce the functionals
$k^{\alpha}({\bf J}(X))=k^{\alpha}(J^{1}(X),\dots,J^{N}(X))$
using the same functions~$k^{\alpha}({\bf U})$.
According
to~\eqref{JUtransform}, \eqref{UJtransform}
we can write
\begin{gather*}
\big\{k^{\alpha}({\bf U}(X)),J^{\mu}(Y)\big\}\big|_{\cal K}=
\big\{k^{\alpha}({\bf J}(X)),J^{\mu}(Y)\big\}\big|_{\cal K}+O\big(\epsilon^{2}\big).
\end{gather*}

Consider the evolution of the functionals
$k^{\alpha}({\bf J}(X))$ according to the f\/low generated by
\linebreak                                                                                 %
$\int q(Y)J^{\mu}(Y){\mathrm d}Y$ on the submanifold ${\cal K}$.
Using relations~\eqref{varphiqXev} we can write
\begin{gather*}
k^{\alpha}_{t}({\bf J}(X))=
{\partial k^{\alpha}\over\partial U^{\nu}}\left({\bf J}(X)\right)
J^{\nu}_{t}(X)
={\partial k^{\alpha} \over \partial U^{\nu}} \left( {\bf J}(X) \right)
q (X) \omega^{\beta\mu}({\bf U}(X))
\\
{}
\times \int_{0}^{2\pi} \cdots \int_{0}^{2\pi}
 \sum_{l\geq0} \Pi^{\nu}_{i(l)}  \left( \bm{\Phi} \left(
{{\bf S}(X) \over \epsilon} +\bm{\theta}, {\bf U}(X) \right) ,
\epsilon {\partial \over \partial X} \bm{\Phi}\!\left(
{{\bf S}(X) \over \epsilon} +\bm{\theta}, {\bf U}(X) \right) ,
\dots \right)
\\
\phantom{\times \int_{0}^{2\pi} \cdots \int_{0}^{2\pi}}
\times\epsilon^{l} {\partial^{l} \over \partial X^{l}}
\Phi^{i}_{\theta^{\beta}} \left(
{{\bf S}(X) \over \epsilon} +\bm{\theta}, {\bf U}(X) \right)
{{\mathrm d}^{m} \theta \over (2\pi)^{m}}
\\
{}
+\epsilon
{\partial k^{\alpha} \over \partial U^{\nu}} \left( {\bf U}(X) \right)
\big( q (X) \omega^{\beta\mu}({\bf U}(X)) \big)_{X}
\\
\qquad{}
\times \int_{0}^{2\pi} \cdots\int_{0}^{2\pi}
\sum_{l\geq1} l \Pi^{\nu}_{i(l)} \left( \bm{\Phi} \left(
\bm{\theta}, {\bf U}(X) \right), k^{\gamma}
\bm{\Phi}_{\theta^{\gamma}} \left( \bm{\theta}, {\bf U}(X) \right),
\dots \right)
\\
\qquad\phantom{\times \int_{0}^{2\pi} \cdots\int_{0}^{2\pi}}
{}\times k^{\alpha_{1}}(X) \cdots k^{\alpha_{l-1}}(X)
\Phi^{i}_{\theta^{\beta}\theta^{\alpha_{1}}\cdots\theta^{\alpha_{l-1}}}
\left( \bm{\theta}, {\bf U}(X) \right)
{{\mathrm d}^{m} \theta \over (2\pi)^{m}}
\\
{}
+\epsilon
{\partial k^{\alpha} \over \partial U^{\nu}} \left( {\bf U}(X) \right)
\int_{0}^{2\pi} \cdots\int_{0}^{2\pi}  \sum_{l\geq0}
\Pi^{\nu}_{i(l)} \left( \bm{\Phi} \left( \bm{\theta}, {\bf U}(X) \right),
k^{\gamma} \bm{\Phi}_{\theta^{\gamma}} \left( \bm{\theta},
{\bf U}(X) \right), \dots \right)
\\
\phantom{+\epsilon
{\partial k^{\alpha} \over \partial U^{\nu}} \left( {\bf U}(X) \right)
\int_{0}^{2\pi} \cdots\int_{0}^{2\pi}}
{}\times k^{\alpha_{1}}(X) \cdots k^{\alpha_{l}}(X)
\eta^{i}_{[q]\theta^{\alpha_{1}}\cdots\theta^{\alpha_{l}}}
(\bm{\theta}, X)
{{\mathrm d}^{m} \theta \over (2\pi)^{m}}+ O\big(\epsilon^{2}\big).
\end{gather*}

It's not dif\/f\/icult to see that the f\/irst part of the above expression
contains the integrals over~$\bm{\theta}$ of the expressions
$P^{\nu}_{\theta^{\beta}}(\bm{\theta},X)$ and is equal to zero.
It is easy to see also after integration by parts that the third part
of the above expression represents the value
\begin{gather*}
\epsilon\int_{0}^{2\pi}\cdots\int_{0}^{2\pi}
{\partial k^{\alpha}\over\partial U^{\nu}}\left({\bf U}(X)\right)
\zeta^{(\nu)}_{i[{\bf U}(X)]}(\bm{\theta})
\eta^{i}_{[q]}(\bm{\theta},X)
{{\mathrm d}^{m}\theta\over(2\pi)^{m}}
\end{gather*}
and is equal to zero according to Lemma~\ref{Lemma2.1.}.

Thus, we can write in the main order the expressions
for the evolution of the functionals $k^{\alpha}({\bf J}(X))$ on
the submanifold ${\cal K}$ in the form
\begin{gather*}
k^{\alpha}_{t} ({\bf J}(X))=
\epsilon \big(q (X) \omega^{\beta\mu}({\bf U}(X))\big)_{X}
\\
\qquad{}\times{\partial k^{\alpha} \over \partial U^{\nu}}
\int_{0}^{2\pi}\cdots \int_{0}^{2\pi}
\sum_{l\geq1} l k^{\beta_{1}} \cdots k^{\beta_{l-1}}
\Phi^{i}_{\theta^{\beta}\theta^{\beta_{1}}\cdots\theta^{\beta_{l-1}}}
\Pi^{\nu}_{i(l)} \left( \bm{\Phi},
k^{\gamma} \bm{\Phi}_{\theta^{\gamma}},
\dots \right) {{\mathrm d}^{m} \theta \over (2\pi)^{m}}+O\big(\epsilon^{2}\big).
\end{gather*}

Using relations~\eqref{kalphakbetaUnu} we get then
\begin{gather*}
k^{\alpha}_{t}({\bf J}(X))=\epsilon\left[q(X)
\omega^{\alpha\mu}({\bf U}(X))\right]_{X}+O\big(\epsilon^{2}\big),
\end{gather*}
i.e.
\begin{gather*}
\{k^{\alpha}({\bf J}(X)),J^{\mu}(Y)\}|_{\cal K}=
\epsilon\omega^{\alpha\mu}({\bf U}(X))
\delta^{\prime}(X - Y)+\epsilon
\omega^{\alpha\mu}_{X}\delta(X - Y)+O\big(\epsilon^{2}\big),
\end{gather*}
which implies~\eqref{kJmuskob}.
\end{proof}

Easy to see that according to Lemma~\ref{Lemma2.3.} and relations
\eqref{JUtransform}, \eqref{UJtransform} we can write also
\begin{gather}\label{kUmuskob}
\{k^{\alpha}({\bf U}(X)),U^{\mu}(Y)\}|_{\cal K}=
\epsilon\left[\omega^{\alpha\mu}({\bf U}(X))
\delta(X - Y)\right]_{X}+O\big(\epsilon^{2}\big)
\end{gather}
on the submanifold ${\cal K}$.

\begin{corollary}
Under the conditions of Lemma~{\rm \ref{Lemma2.3.}}, we have also
\begin{gather}\label{kknulrel}
\big\{k^{\alpha}(X),k^{\beta}(Y)\big\}\big|_{\cal K}=O\big(\epsilon^{2}\big)
\end{gather}
for the functionals ${\bf k}({\bf U}(X))$.
\end{corollary}

Indeed, by virtue of~\eqref{komeganulrel} we have
\begin{gather*}
\big\{k^{\alpha}(X),k^{\beta}(Y)\big\}\big|_{\cal K}=
\big\{k^{\alpha}(X),U^{\mu}(Y)\big\}\big|_{\cal K}
{\partial k^{\beta}\over\partial U^{\mu}}({\bf U}(Y))
\\
\phantom{\{k^{\alpha}(X),k^{\beta}(Y)\}|_{\cal K}}
=\epsilon\omega^{\alpha\mu}(X)k^{\beta}_{U^{\mu}}(X)
\delta^{\prime}(X - Y)
+\epsilon\omega^{\alpha\mu}(X)
\big(k^{\beta}_{U^{\mu}}(X)\big)_{X}\delta(X - Y)
\\
\phantom{\{k^{\alpha}(X),k^{\beta}(Y)\}|_{\cal K}=}
+\epsilon\left(
\omega^{\alpha\mu}(X)\right)_{X}k^{\beta}_{U^{\mu}}(X)
\delta(X - Y)+O\big(\epsilon^{2}\big)=O\big(\epsilon^{2}\big).
\end{gather*}

Let us note now that in the proof of Lemma~\ref{Lemma2.3.},
we have not used
the fact that the functionals~$J^{\mu}(Y)$ belong to our special
set of functionals~\eqref{FuncJnuX} and used only the fact
that the f\/low generated by the functional $I^{\mu}$
leaves invariant the family $\Lambda$ generating linear shifts of
$\theta^{\alpha}_{0}$ with constant frequencies
$\omega^{\alpha\mu}({\bf U})$.
We can therefore formulate
here the following lemma.

\smallskip

\noindent
{\bf Lemma 2.3$^\prime$.} {\it Let the values $U^{\nu}$                        
of the functionals $I^{\nu}$ on a
complete regular family $\Lambda$ of \mbox{$m$-phase} solutions of system
\eqref{EvInSyst} be functionally independent
and give a~complete set of parameters on~$\Lambda$, excluding
the initial phases.
Let the flow generated by the functional{\samepage
\begin{gather}\label{ItildeFunc}
{\tilde I}=\int{\tilde P}
(\bm{\varphi},\bm{\varphi}_{x},\dots){\rm d} x
\end{gather}
leave invariant the family $\Lambda$ generating a~linear shift of
$\theta^{\alpha}_{0}$ with constant frequencies
${\tilde\omega}^{\alpha}({\bf U})$.}

Consider the functionals
\begin{gather}\label{JXtildeFunc}
{\tilde J}(X)=
\int_{0}^{2\pi}\cdots\int_{0}^{2\pi}
{\tilde P}\big(\bm{\varphi},\epsilon\bm{\varphi}_{X},\epsilon^{2}
\bm{\varphi}_{XX},\dots\big){{\mathrm d}^{m}\theta\over(2\pi)^{m}}.
\end{gather}
Then for the Poisson brackets of the functionals
$k^{\alpha}({\bf U}(X))$ and
${\tilde J}(Y)$ on ${\cal K}$ we have the relation}
\begin{gather*}
\big\{k^{\alpha}({\bf U}(X)),{\tilde J}(Y)\big\}\big|_{\cal K}=
\epsilon\left[{\tilde\omega}^{\alpha}({\bf U}(X))
\delta(X - Y)\right]_{X}+O\big(\epsilon^{2}\big).
\end{gather*}

Proof of Lemma 2.3$^\prime$ completely repeats the proof of
Lemma~\ref{Lemma2.3.}.

\begin{lemma}\label{Lemma2.4.}
Let the values $U^{\nu}$ of the functionals $I^{\nu}$ on a
complete regular family $\Lambda$ of \mbox{$m$-phase} solutions of system~\eqref{EvInSyst} be functionally independent
and give a~complete set of parameters on~$\Lambda$,
excluding the initial phases.
Then for the constraints
$g^{i}(\bm{\theta},X)$ imposed by~\eqref{gconstr}
and smooth compactly supported function
$q(X)$ as well as smooth $2\pi$-periodic
in each $\theta^{\alpha}$ function $Q(\bm{\theta},X)$ we have
the following relations on the submanifold~${\cal K}$
\begin{gather}
\left.
\left\{
g^{i}(\bm{\theta},X),\int q(Y)J^{\mu}(Y){\mathrm d}Y
\right\}\right|_{\cal K}=O(\epsilon),
\nonumber\\
\left.
\left[\int_{0}^{2\pi}\cdots\int_{0}^{2\pi}
Q\left({{\bf S}(X)\over\epsilon}+\bm{\theta},X\right)
\left\{g^{i}(\bm{\theta},X),J^{\mu}(Y)\right\}
{{\mathrm d}^{m}\theta\over(2\pi)^{m}}
\right]\right|_{\cal K}=O(\epsilon).\label{QRelLemma24}
\end{gather}
\end{lemma}

\begin{proof}
Indeed, by (\ref{phiJbr}),~\eqref{QphiJbr}, and
Lemma~\ref{Lemma2.3.} we have
\begin{gather*}
\left.
\left\{
g^{i}(\bm{\theta},X),\int q(Y)J^{\mu}(Y){\mathrm d}Y
\right\}\right|_{{\cal K}[0]}
=\Phi^{i}_{\theta^{\alpha}}\left(
{{\bf S}(X)\over\epsilon}+\bm{\theta},{\bf U}(X)\right)
\omega^{\alpha\mu}(X) q(X)
\\
 {}-\frac12 \Phi^{i}_{\theta^{\alpha}}\left(
{{\bf S}(X)\over\epsilon}+\bm{\theta},{\bf U}(X)\right)
\int\operatorname{sgn}(X - Z)\left(\omega^{\alpha\mu}(Z)
q(Z)\right)_{Z}{\rm d}Z\equiv0,
\\
\left.
\left[\int_{0}^{2\pi}\cdots\int_{0}^{2\pi}
Q\left({{\bf S}(X)\over\epsilon}+\bm{\theta},X\right)
\left\{g^{i}(\bm{\theta},X),J^{\mu}(Y)\right\}
{{\mathrm d}^{m}\theta\over(2\pi)^{m}}
\right]\right|_{{\cal K}[0]}
\\
 =\omega^{\alpha\mu}(X)
\int_{0}^{2\pi}\cdots\int_{0}^{2\pi}
Q(\bm{\theta},X)
\Phi^{i}_{\theta^{\alpha}}(\bm{\theta},{\bf U}(X))
{{\mathrm d}^{m}\theta\over(2\pi)^{m}}\delta(X - Y)
\\
{}- \int_{0}^{2\pi}\cdots \int_{0}^{2\pi}
Q(\bm{\theta},X)\Phi^{i}_{\theta^{\alpha}}(\bm{\theta},{\bf U}(X))
{{\mathrm d}^{m}\theta\over(2\pi)^{m}}
\frac12\operatorname{sgn}(X - Z)\left[\omega^{\alpha\mu}(Z)
\delta(Z - Y)\right]_{Z}{\mathrm d}Z\equiv0.\!\!\!\tag*{\qed}
\end{gather*}
  \renewcommand{\qed}{}
\end{proof}                                                    

Similarly to the previous case, we can formulate also the following lemma.

\smallskip

\noindent                         
{\bf Lemma 2.4$^\prime$.} {\it Let the values $U^{\nu}$ of
the functionals $I^{\nu}$ on a
complete regular family $\Lambda$ of \mbox{$m$-phase} solutions of system~\eqref{EvInSyst} be functionally independent
and give a~complete set of parameters on~$\Lambda$,
excluding the initial phases.
Let the flow generated by
the functional~\eqref{ItildeFunc}
leave invariant the family $\Lambda$ generating a~linear shift of
$\theta^{\alpha}_{0}$ with constant frequencies
${\tilde\omega}^{\alpha}({\bf U})$.
Then for the constraints
$g^{i}(\bm{\theta},X)$ and the functionals~${\tilde J}(X)$
imposed by~\eqref{JXtildeFunc} we have the following relations
on the submanifold ${\cal K}$}
\begin{gather*}
\left.
\left\{
g^{i}(\bm{\theta},X),\int q(Y){\tilde J}(Y){\mathrm d}Y
\right\}\right|_{\cal K}=O(\epsilon),
\\
\left.
\left[\int_{0}^{2\pi}\cdots\int_{0}^{2\pi}
Q\left({{\bf S}(X)\over\epsilon}+\bm{\theta},X\right)
\big\{g^{i}(\bm{\theta},X),{\tilde J}(Y)\big\}
{{\mathrm d}^{m}\theta\over(2\pi)^{m}}
\right]\right|_{\cal K}=O(\epsilon).
\end{gather*}

Let us consider now the Dirac restriction of the bracket~\eqref{EpsExtBracket} on the submanifold~${\cal K}$.

For the Dirac restriction of the bracket~\eqref{EpsExtBracket}
on the submanifold ${\cal K}$ we have to modify the functionals
$U^{\mu}(Y)$ by linear combinations of the constraints
$g^{i}(\bm{\theta},X)$
\begin{gather*}
{\tilde U}^{\mu}(Y)=U^{\mu}(Y)+\int_{-\infty}^{+\infty}\int_{0}^{2\pi}\cdots\int_{0}^{2\pi}
g^{i}(\bm{\theta},X)\, \alpha^{\mu}_{i}
\left({{\bf S}[{\bf U}](X)\over\epsilon}+\bm{\theta},
[{\bf U}],X,Y\right){{\mathrm d}^{m}\theta\over(2\pi)^{m}}{\mathrm d}X,
\end{gather*}
so that the f\/lows generated by the functionals~${\tilde U}^{\mu}(Y)$ leave ${\cal K}$ invariant and then use
the functionals~${\tilde U}^{\mu}(Y)$
to construct the Dirac bracket on~${\cal K}$.

Since the functionals $U^{\mu}(Y)$ are def\/ined with the aid of the
functionals ${\bf J}(Z)$, technically it is more convenient to modify the functionals~$J^{\mu}(Y)$
\begin{gather*}
{\tilde J}^{\mu}(Y)=J^{\mu}(Y)+\int_{-\infty}^{+\infty}\!\!
 \int_{0}^{2\pi}\!\cdots \int_{0}^{2\pi}\!
g^{i}(\bm{\theta},X)\,\beta^{\mu}_{i}
\!\left({{\bf S}[{\bf U}[{\bf J}]](X)\over\epsilon}+\bm{\theta},
[{\bf U}[{\bf J}]],X,Y\!\right){{\mathrm d}^{m}\theta\over(2\pi)^{m}}{\mathrm d}X,
\end{gather*}
so that the f\/lows generated by the functionals
${\tilde J}^{\mu}(Y)$ leave ${\cal K}$ invariant and then put
\begin{gather*}
{\tilde U}^{\mu}(Y)={\tilde J}^{\mu}(Y)+\sum_{l\geq1}\epsilon^{l}{\tilde U}^{\mu}_{(l)}(Y)
\end{gather*}
just using the functionals ${\tilde{\bf J}}(Y)$ instead of
${\bf J}(Y)$ in~\eqref{UJtransform}.
Both the approaches give, certainly, the same result for the Dirac
restriction of the brac\-ket~\eqref{EpsExtBracket} on ${\cal K}$.

The distributions
$\beta^{\mu}_{i}({\bf S}(X)/\epsilon+\bm{\theta},X,Y)$
must satisfy the relation
\begin{gather}
\int_{-\infty}^{+\infty}\int_{0}^{2\pi}\cdots\int_{0}^{2\pi}\!\!
\left.
\big\{ g^{i} (\bm{\theta}, X),g^{j} (\bm{\theta}^{\prime}, Z)\big\}\right|_{\cal K}
\beta^{\mu}_{j} \left( {{\bf S} (Z) \over \epsilon} +\bm{\theta}^{\prime}, Z,
Y \right) {{\mathrm d}^{m} \theta^{\prime}\over (2\pi)^{m}}{\mathrm d}Z
\nonumber\\
\hphantom{\int_{-\infty}^{+\infty}\int_{0}^{2\pi}\cdots\int_{0}^{2\pi}}{}
+\left.\big\{g^{i}(\bm{\theta},X),J^{\mu}(Y)\big\}\right|_{\cal K}=0\label{betasystem}
\end{gather}
on ${\cal K}$ and are def\/ined at each ``point'' of ${\cal K}$
up to linear combinations of the distributions \linebreak
$\delta J^{\zeta}(W)/\delta\varphi^{i}(\bm{\theta},X)$
by virtue of the original dependence~\eqref{dependence}
of the constraints~\eqref{gconstr}.

To obtain a~local Poisson bracket on ${\cal K}$ we are
trying to f\/ind the functions $\beta^{\mu}_{i}(\bm{\theta},X,Y)$
in the form
\begin{gather}\label{beta1def}
\beta^{\mu}_{i}(\bm{\theta},X,Y)=\sum_{l\geq1}\epsilon^{l}\beta_{i(l)}^{\mu}(\bm{\theta},X,Y),
\end{gather}
where the functions $\beta_{i(l)}^{\mu}(\bm{\theta},X,Y)$
represent the local distributions
\begin{gather*}
\beta_{i(l)}^{\mu}(\bm{\theta},X,Y)=\sum_{p=0}^{l}
\beta_{i(l),p}^{\mu}(\bm{\theta},X)\,\delta^{(p)}(X - Y)
\end{gather*}
with the total grading $l$ in $\epsilon$, considering that the
derivatives of the delta function $\delta^{(p)}(X - Y)$ have
by def\/inition the degree~$p$.

Thus, we assume that all the functions
$\beta_{i(l),p}^{\mu}(\bm{\theta},X)$ on ${\cal K}$
are local functionals of
$({\bf U}(X),{\bf U}_{X}$, $\dots)$
at every $\bm{\theta}$, polynomial in the derivatives
$({\bf U}_{X},\dots)$ and having degree
$l-p$ under the previous def\/inition.
Such a~structure of $\beta^{\mu}_{i}(\bm{\theta},X,Y)$
is obviously equivalent to saying that the functionals
\begin{gather*}
J_{[{\bf q}]}=
\int_{-\infty}^{+\infty}J^{\mu}(Y) q_{\mu}(Y){\mathrm d}Y
\end{gather*}
can be modif\/ied by linear combinations of constraints~\eqref{gconstr}
with the coef\/f\/icients
\begin{gather*}
B_{i[{\bf q}]}(\bm{\theta},X)=
\sum_{l\geq1}\epsilon^{l}B_{i[{\bf q}](l)}(\bm{\theta},X)
=\sum_{l\geq1}
\epsilon^{l}\sum_{p=0}^{l}\beta_{i(l),p}^{\mu}(\bm{\theta},X)
{{\mathrm d}^{p}q_{\mu}(X)\over{\mathrm d}X^{p}},
\end{gather*}
so that the corresponding f\/lows leave invariant
the submanifold ${\cal K}$.
Under this scheme, the derivatives
${\mathrm d}^{p}q_{\mu}/{\mathrm d}X^{p}$ of slowly varying functions
$q_{\mu}(X)$ also have grading $p$ as the derivatives of the
functionals ${\bf U}(X)$.

The pairwise Poisson brackets of the constraints
$g^{i}(\bm{\theta},X)$, $g^{j}(\bm{\theta}^{\prime},Y)$ on ${\cal K}$ can be written in the form
\begin{gather}
\{g^{i}(\bm{\theta},X),
g^{j}(\bm{\theta}^{\prime},Y)\}|_{\cal K}
\nonumber\\
\qquad
=\{\varphi^{i}(\bm{\theta},X),
\varphi^{j}(\bm{\theta}^{\prime},Y)\}|_{\cal K}-
\{\varphi^{i}(\bm{\theta},X),U^{\lambda}(Y)\}|_{\cal K}\,
\Phi^{j}_{U^{\lambda}}\left({{\bf S}(Y)\over\epsilon}+\bm{\theta}^{\prime},{\bf
U}(Y)\right)
\nonumber\\
\qquad
{}-\Phi^{i}_{U^{\nu}}\left({{\bf S}(X)\over\epsilon}+\bm{\theta},{\bf U}(X)\right)\{U^{\nu}(X),
\varphi^{j}(\bm{\theta}^{\prime},Y)\}|_{\cal K}
\nonumber\\
\qquad
{}+\Phi^{i}_{U^{\nu}}\left({{\bf S}(X)\over\epsilon}+\bm{\theta},{\bf U}(X)\right)
\{U^{\nu}(X),U^{\lambda}(Y)\}|_{\cal K}\,
\Phi^{j}_{U^{\lambda}}\left({{\bf S}(Y)\over\epsilon}+\bm{\theta}^{\prime},{\bf
U}(Y)\right)
\nonumber\\
\qquad
{} -{1\over2\epsilon}\int
\{\varphi^{i}(\bm{\theta},X),k^{\beta}(W)\}|_{\cal K}
\operatorname{sgn}(Y - W){\mathrm d}W\,
\Phi^{j}_{\theta^{\prime\beta}}\left({{\bf S}(Y)\over\epsilon}+\bm{\theta}^{\prime},{\bf
U}(Y)\right)
\nonumber\\
\qquad
{} -{1\over2\epsilon}\Phi^{i}_{\theta^{\alpha}}\left(
{{\bf S}(X)\over\epsilon}+\bm{\theta},{\bf U}(X)\right)\int\operatorname{sgn}(X - Z)
\{k^{\alpha}(Z),\varphi^{j}(\bm{\theta}^{\prime},Y)\}|_{\cal K}
{\mathrm d}Z
\nonumber\\
\qquad
{} +{1\over2\epsilon}\Phi^{i}_{\theta^{\alpha}}\left(
{{\bf S}(X)\over\epsilon}+\bm{\theta},{\bf U}(X)\right)
\nonumber\\
\qquad\quad
{}\times \int
\operatorname{sgn}(X - Z)\{k^{\alpha}(Z),U^{\lambda}(Y)\}|_{\cal K}
{\mathrm d}Z\,\Phi^{j}_{U^{\lambda}}\left({{\bf S}(Y)\over\epsilon}+\bm{\theta}^{\prime},{\bf U}(Y)\right)
\nonumber\\
\qquad
{} +{1\over2\epsilon}
\Phi^{i}_{U^{\nu}}\left({{\bf S}(X)\over\epsilon}+\bm{\theta},{\bf
U}(X)\right)
\nonumber\\
\qquad\quad
{} \times \int
\{U^{\nu}(X),k^{\beta}(W)\}|_{\cal K}\operatorname{sgn}(Y - W){\mathrm d}W\,
\Phi^{j}_{\theta^{\prime\beta}}\left({{\bf S}(Y)\over\epsilon}
+\bm{\theta}^{\prime},{\bf U}(Y)\right)
\nonumber\\
\qquad
{} +{1\over4\epsilon^{2}}\Phi^{i}_{\theta^{\alpha}}\left(
{{\bf S}(X)\over\epsilon}+\bm{\theta},{\bf U}(X)\right)
\label{SkobSv}\\
\qquad\quad
{} \times \int
\operatorname{sgn}(X - Z)\{k^{\alpha}(Z),k^{\beta}(W)\}|_{\cal K}
\operatorname{sgn}(Y - W){\mathrm d}Z{\mathrm d}W\,\Phi^{j}_{\theta^{\prime\beta}}
\left({{\bf S}(Y)\over\epsilon}+\bm{\theta}^{\prime},
{\bf U}(Y)\right). \nonumber
\end{gather}

For the purposes of this chapter, we need only
``approximate'' restriction of the bracket~\eqref{EpsExtBracket}
on the submanifold ${\cal K}$.
Specif\/ically, we will need to prove the
existence of only the f\/irst term $\beta^{\mu}_{i(1)}(\bm{\theta},X,Y)$
in the expansion~\eqref{beta1def} having the form
\begin{gather}\label{VidBeta}
\beta^{\mu}_{i(1)}(\bm{\theta},X,Y)=
\beta^{\mu}_{i(1),1}(\bm{\theta},{\bf U}(X))
\delta^{\prime}(X - Y)+\beta^{\mu}_{i(1),0\lambda}(\bm{\theta},{\bf U}(X))
U^{\lambda}_{X}\delta(X - Y),
\end{gather}
which is equivalent to the approximation
\begin{gather}\label{B1ApprSol}
B_{i[{\bf q}](1)}(\bm{\theta},X)=
\beta^{\mu}_{i(1),1}(\bm{\theta},{\bf U}(X))
q_{\mu X}(X)+\beta^{\mu}_{i(1),0\lambda}(\bm{\theta},{\bf U}(X))
U^{\lambda}_{X} q_{\mu}(X)
\end{gather}
for the functions $B_{i[{\bf q}]}(\bm{\theta},X)$.

To f\/ind it we need only the leading term of the convolution
of the Poisson brackets of constraints on ${\cal K}$ with
$B_{j[{\bf q}](1)}({\bf S}(Y)/\epsilon+\bm{\theta}^{\prime},Y)$
and the f\/irst (in $\epsilon$) non-vanishing term of the brackets
of constraints and the functional
$J_{[{\bf q}]}$.

Note that according to the relations~\eqref{phiknulbr},
\eqref{QvarphikalphaKzerorel},~\eqref{kUmuskob}, and~\eqref{kknulrel}
all the terms of the bracket~\eqref{SkobSv} behave as the values of
the order of at most $O(1)$ for $\epsilon\rightarrow0$ in the
convolution with
$B_{j[{\bf q}](1)}({\bf S}(Y)/\epsilon+\bm{\theta}^{\prime},Y)$.

We impose additional conditions
\begin{gather}\label{BetaCond1}
\int_{0}^{2\pi}\cdots\int_{0}^{2\pi}
\Phi^{j}_{\theta^{\alpha}}(\bm{\theta},{\bf U}(X))\,
\beta^{\mu}_{j(1)}(\bm{\theta},X,Y)
{{\mathrm d}^{m}\theta\over(2\pi)^{m}}\equiv0,\qquad\alpha=1,\dots,m,
\end{gather}
or equivalently
\begin{gather}\label{BetaCond}
\int_{0}^{2\pi}\cdots\int_{0}^{2\pi}
\Phi^{j}_{\theta^{\alpha}}(\bm{\theta},{\bf U}(X))\,
B_{j[{\bf q}](1)}(\bm{\theta},X)
{{\mathrm d}^{m}\theta\over(2\pi)^{m}}\equiv0,\qquad\alpha=1,\dots,m,
\end{gather}
to be conf\/irmed {\it aposteriori} for the functions
$\beta^{\mu}_{j(1)}(\bm{\theta},X,Y)$
and $B_{j[{\bf q}](1)}(\bm{\theta},X)$.
The Poisson brackets of constraints~\eqref{SkobSv} on ${\cal K}$
can then be reduced to the form
\begin{gather}
\{g^{i}(\bm{\theta},X),
g^{j}(\bm{\theta}^{\prime},Y)\}|_{\cal K}\rightarrow
\{g^{i}(\bm{\theta},X),
g^{j}(\bm{\theta}^{\prime},Y)\}|^{\rm ef\/f}_{\cal K}
\nonumber\\
\qquad
{} =\{\varphi^{i}(\bm{\theta},X),
\varphi^{j}(\bm{\theta}^{\prime},Y)\}|_{\cal K}-
\{\varphi^{i}(\bm{\theta},X),U^{\lambda}(Y)\}|_{\cal K}\,
\Phi^{j}_{U^{\lambda}}\left({{\bf S}(Y)\over\epsilon}+\bm{\theta}^{\prime},{\bf
U}(Y)\right)
\nonumber\\
\qquad
{}-\Phi^{i}_{U^{\nu}}\left({{\bf S}(X)\over\epsilon}+\bm{\theta},{\bf U}(X)\right)\{U^{\nu}(X),
\varphi^{j}(\bm{\theta}^{\prime},Y)\}|_{\cal K}
\nonumber\\
\qquad
{} +\Phi^{i}_{U^{\nu}}\left({{\bf S}(X)\over\epsilon}+\bm{\theta},{\bf U}(X)\right)
\{U^{\nu}(X),U^{\lambda}(Y)\}|_{\cal K}\,
\Phi^{j}_{U^{\lambda}}\left({{\bf S}(Y)\over\epsilon}+\bm{\theta}^{\prime},{\bf
U}(Y)\right)
\nonumber\\
\qquad
{} -{1\over2\epsilon}\Phi^{i}_{\theta^{\alpha}}\left(
{{\bf S}(X)\over\epsilon}+\bm{\theta},{\bf U}(X)\right)\int\operatorname{sgn}(X - Z)
\{k^{\alpha}(Z),\varphi^{j}(\bm{\theta}^{\prime},Y)
\}|_{\cal K}{\mathrm d}Z
\nonumber\\
\qquad
{} +{1\over2\epsilon}\Phi^{i}_{\theta^{\alpha}}\left(
{{\bf S}(X)\over\epsilon}+\bm{\theta},{\bf U}(X)\right)\int
\operatorname{sgn}(X - Z)\{k^{\alpha}(Z),U^{\lambda}(Y)\}|_{\cal K}{\mathrm d}Z
\nonumber\\
\qquad\quad
{} \times\Phi^{j}_{U^{\lambda}}\left({{\bf S}(Y)\over\epsilon}+\bm{\theta}^{\prime},{\bf
U}(Y)\right)\label{Reduction}
\end{gather}
in f\/inding
$B_{i[{\bf q}](1)}(\bm{\theta},X)$.

To f\/ind the leading term of the convolution of the Poisson brackets
of constraints on ${\cal K}$ and
$B_{j[{\bf q}](1)}({\bf S}(Y)/\epsilon+\bm{\theta}^{\prime},Y)$
we can simplify further the relation~\eqref{Reduction}.
Namely, taking into account relations~\eqref{phiUbr},~\eqref{kUmuskob}
we can omit in the leading order the second and the last terms
of~\eqref{Reduction} in the convolution with
$B_{j[{\bf q}](1)}({\bf S}(Y)/\epsilon+\bm{\theta}^{\prime},Y)$.
By~\eqref{QphiUbr} and~\eqref{BetaCond}, we can omit
in the leading order also the third term of~\eqref{Reduction}
in the convolution with
$B_{j[{\bf q}](1)}({\bf S}(Y)/\epsilon+\bm{\theta}^{\prime},Y)$.
The fourth term of~\eqref{Reduction}, obviously, is of the order
$O(\epsilon)$ and also can be omitted in the leading order.

As a~result, the principal term of the convolution of~\eqref{SkobSv}
with the functions \linebreak $\epsilon
B_{j[{\bf q}](1)}({\bf S}(Y)/\epsilon+\bm{\theta}^{\prime},Y)$
can be written in the form
\begin{gather}
\left[ \int \{ g^{i} (\bm{\theta}, X),
g^{j} (\bm{\theta}^{\prime}, Y) \}|_{\cal K} \epsilon
B_{j[{\bf q}](1)} \left(
{{\bf S}(Y) \over \epsilon} +\bm{\theta}^{\prime}, Y \right)
{{\mathrm d}^{m} \theta^{\prime} \over (2\pi)^{m}}{\mathrm d}Y
\right]_{[1]}
\nonumber\\
\qquad
{} \equiv \sum_{s\geq 0} B^{ij}_{(s)} \left( \bm{\Phi}
\left( {{\bf S}(X) \over \epsilon} +\bm{\theta}, {\bf U}(X) \right),
k^{\gamma}(X) \bm{\Phi}_{\theta^{\gamma}}
\left( {{\bf S}(X) \over \epsilon} +\bm{\theta}, {\bf U}(X) \right),
\dots \right)
\nonumber\\
\qquad\quad
{}\times
k^{\alpha_{1}}(X) \cdots k^{\alpha_{s}}(X)
B_{j[{\bf q}](1)\theta^{\alpha_{1}}\cdots\theta^{\alpha_{s}}}
\left( {{\bf S}(X) \over \epsilon} +\bm{\theta}, X \right)
\nonumber\\
\qquad
{}-{1 \over 2} \Phi^{i}_{\theta^{\alpha}} \left(
{{\bf S}(X) \over \epsilon} +\bm{\theta}, {\bf U}(X) \right) \int \operatorname{sgn} (X - Z)
\nonumber\\
\qquad\quad
{}\times\left[ \int \left\{
k^{\alpha}(Z), \varphi^{j} (\bm{\theta}^{\prime}, Y)
\right\}|_{\cal K} \epsilon B_{j[{\bf q}](1)} \left(
{{\bf S}(Y) \over \epsilon} +\bm{\theta}^{\prime}, Y\right)
{{\mathrm d}^{m} \theta^{\prime} \over (2\pi)^{m}}{\mathrm d}Y\right]_{[2]}{\mathrm d}Z.\label{FinalReduction}
\end{gather}

Let us remind here that the indexes $[1]$, $[2]$ mean as before the
terms of the corresponding graded expansion on ${\cal K}$ having
degree $1$ and $2$ respectively.

We have also
\begin{gather}
\{ g^{i}(\bm{\theta}, X), J_{[{\bf q}]} \}|_{{\cal K}[1]}=
\{ \varphi^{i}(\bm{\theta}, X), J_{[{\bf q}]} \}|_{{\cal K}[1]} -
\Phi^{i}_{U^{\nu}} \left(
{{\bf S}(X) \over \epsilon} +\bm{\theta}, {\bf U}(X) \right)
\{ U^{\nu}(X), J_{[{\bf q}]} \}|_{{\cal K}[1]}
\nonumber\\
\qquad
{}- {1 \over 2} \Phi^{i}_{\theta^{\alpha}} \left(
{{\bf S}(X) \over \epsilon} +\bm{\theta}, {\bf U}(X) \right)
\int \operatorname{sgn} (X - Z)
\{ k^{\alpha}(Z), J_{[{\bf q}]} \}|_{{\cal K}[2]}{\mathrm d}Z.\label{giJmu1skob}
\end{gather}

Let us prove here the following lemma which allows to investigate
the further construction of the functions
$B_{i[{\bf q}](1)}(\bm{\theta},X)$.

\begin{lemma}
Let the functions $B_{i[{\bf q}](1)}(\bm{\theta},X)$
satisfy the conditions~\eqref{BetaCond}, i.e.
\begin{gather*}
\int_{0}^{2\pi}\cdots\int_{0}^{2\pi}
\Phi^{j}_{\theta^{\alpha}}(\bm{\theta},{\bf U}(X))
B_{j[{\bf q}](1)}(\bm{\theta},X)
{{\mathrm d}^{m}\theta\over(2\pi)^{m}}\equiv0
\end{gather*}
and the relation
\begin{gather}\label{BAcond}
{\hat B}^{ij}_{[0]}(X)
B_{j[{\bf q}](1)}(\bm{\theta},X)
+A^{i}_{[1][{\bf q}]}(\bm{\theta},X)=0,
\end{gather}
where
\begin{gather}\label{OperatorB}
{\hat B}^{ij}_{[0]}(X)\\
\qquad{}=\sum_{s\geq0}
B^{ij}_{(s)}\left(\bm{\Phi}(\bm{\theta},{\bf U}(X)),
k^{\alpha}(X)\bm{\Phi}_{\theta^{\alpha}}(\bm{\theta},{\bf U}(X)),
\dots\right)k^{{\alpha}_{1}}(X)\cdots k^{{\alpha}_{s}}(X)
{\partial^{s}\over\partial\theta^{{\alpha}_{1}}\cdots
\partial\theta^{{\alpha}_{s}}},\nonumber
\\
\label{FunktsiiA}
A^{i}_{[1][{\bf q}]}\left({{\bf S}(X)\over\epsilon}+\bm{\theta},X\right)\\
\qquad{}=
\{\varphi^{i}(\bm{\theta},X),J_{[{\bf q}]}\}|_{{\cal K}[1]}-
\Phi^{i}_{U^{\nu}}\left(
{{\bf S}(X)\over\epsilon}+\bm{\theta},{\bf U}(X)\right)
\{U^{\nu}(X),J_{[{\bf q}]}\}|_{{\cal K}[1]}.\nonumber
\end{gather}

Then the functions $B_{i[{\bf q}](1)}(\bm{\theta},X)$
satisfy the relation
\begin{gather}
\left[\int
\{g^{i}(\bm{\theta},X),g^{j}(\bm{\theta}^{\prime},Z)
\}|_{{\cal K}}\, \epsilon B_{j[{\bf q}](1)}
\left({{\bf S}(Z)\over\epsilon}+\bm{\theta}^{\prime},Z
\right){{\mathrm d}^{m}\theta^{\prime}\over(2\pi)^{m}}{\mathrm d}Z\right]_{[1]}\nonumber\\
\qquad{}
+\{g^{i}(\bm{\theta},X),
J_{[{\bf q}]}\}|_{{\cal K}[1]}=0.\label{EffCond}
\end{gather}
\end{lemma}

\begin{proof}
The fulf\/illment of condition~\eqref{BetaCond} allows to
reduce the expression
\begin{gather*}
\int\{g^{i}(\bm{\theta},X),g^{j}(\bm{\theta}^{\prime},Z)
\}|_{{\cal K}}\, \epsilon B_{j[{\bf q}](1)}
\left({{\bf S}(Z)\over\epsilon}+\bm{\theta}^{\prime},Z
\right){{\mathrm d}^{m}\theta^{\prime}\over(2\pi)^{m}}{\mathrm d}Z
\end{gather*}
according to formula~\eqref{FinalReduction} up to the terms
$O(\epsilon^{2})$.
From~\eqref{BAcond} we can get then
\begin{gather}
\int \{ g^{i} (\bm{\theta}, X), g^{j} (\bm{\theta}^{\prime}, Z)
\}|_{{\cal K}}\,  \epsilon B_{j[{\bf q}](1)}
\left( {{\bf S}(Z) \over \epsilon} +\bm{\theta}^{\prime}, Z
\right) {{\mathrm d}^{m} \theta^{\prime} \over (2\pi)^{m}}{\mathrm d}Z+
\{ g^{i} (\bm{\theta}, X),
J_{[{\bf q}]} \}|_{\cal K}
\nonumber\\
\qquad=- {\epsilon \over 2} \Phi^{i}_{\theta^{\alpha}} \left(
{{\bf S}(X) \over \epsilon} +\bm{\theta}, {\bf U}(X) \right) \int
\operatorname{sgn} (X - Z)
\nonumber\\
\qquad\quad{} \times \left[ \int
\{k^{\alpha}(Z), \varphi^{j} (\bm{\theta}^{\prime}, W) \}|_{\cal K}\,
\epsilon B_{j[{\bf q}](1)} \left(
{{\bf S}(W) \over \epsilon} +\bm{\theta}^{\prime}, W \right)
{{\mathrm d}^{m} \theta^{\prime} \over (2\pi)^{m}}{\mathrm d}W\right]_{[2]}{\mathrm d}Z
\nonumber\\
\qquad{} -{\epsilon \over 2} \Phi^{i}_{\theta^{\alpha}} \left(
{{\bf S}(X) \over \epsilon} +\bm{\theta}, {\bf U}(X) \right) \int
\operatorname{sgn} (X - Z)
\{k^{\alpha}(Z), J_{[{\bf q}]} \}|_{{\cal K}[2]}{\mathrm d}Z+
O\big(\epsilon^{2}\big).\label{RedggBgJ}
\end{gather}

The proof of the Lemma follows then from the orthogonality
conditions of the left-hand part of~\eqref{RedggBgJ} to the
variational derivatives
\begin{gather}\label{kbetavarder}
\left.
{\delta k^{\beta}({\bf J}(X^{\prime}))\over
\delta\varphi^{i}(\bm{\theta},X)}\right|_{\cal K}=
{\partial k^{\beta}\over\partial U^{\lambda}}({\bf J}(X^{\prime}))
\left.
{\delta J^{\lambda}(X^{\prime})\over
\delta\varphi^{i}(\bm{\theta},X)}\right|_{\cal K}
,\qquad\beta=1,\dots,m
\end{gather}
according to~\eqref{dependence} in the order $\epsilon^{2}$ on
${\cal K}$.
Here we use the functions $k^{\beta}({\bf U})$ to def\/ine
the functionals $k^{\beta}({\bf J}(X^{\prime}))$.
Let us remind also
that the functionals ${\bf U}(X)$ and ${\bf J}(X)$ are connected
by the transformations~\eqref{JUtransform} and~\eqref{UJtransform}.

Indeed, the right-hand part of~\eqref{RedggBgJ} has the form
\begin{gather}\label{PhiThetaOst}
{\epsilon\over2}\Phi^{i}_{\theta^{\alpha}}\left(
{{\bf S}(X)\over\epsilon}+\bm{\theta},{\bf U}(X)\right)\int
\operatorname{sgn}(X - Z)Q^{\alpha}_{[{\bf q}]}(Z){\mathrm d}Z+
O\big(\epsilon^{2}\big)
\end{gather}
with some functions $Q^{\alpha}_{[{\bf q}]}(Z)$.

The convolution with the quantities~\eqref{kbetavarder} can be
represented as the action of the operator
\begin{gather}
\int_{0}^{2\pi}\cdots \int_{0}^{2\pi}
{{\mathrm d}^{m} \theta \over (2\pi)^{m}}
{\partial k^{\beta} \over \partial U^{\lambda}}
\left( {\bf J}(X^{\prime}) \right)
\label{kbetaVarDerOper}\\
\qquad{}\times \sum_{l\geq0}
\Pi^{\lambda}_{i(l)} \left( \bm{\Phi}
\left( {{\bf S}(X^{\prime}) \over \epsilon} +\bm{\theta},
{\bf U}(X^{\prime}) \right),
\epsilon {\partial \over \partial X^{\prime}}
\bm{\Phi} \left( {{\bf S}(X^{\prime}) \over \epsilon} +\bm{\theta},
{\bf U}(X^{\prime}) \right), \dots \right) \epsilon^{l}
{{\mathrm d}^{l} \over{\mathrm d}X^{\prime l}}\nonumber
\end{gather}
on the functions of ${\bf S}(X^{\prime})/\epsilon+\bm{\theta}$
and $X^{\prime}$.

Expanding the action of the operator~\eqref{kbetaVarDerOper}
in powers of $\epsilon$ we can write it's main part in the form
\begin{gather*}
\int_{0}^{2\pi}\cdots\int_{0}^{2\pi}
{{\mathrm d}^{m} \theta \over (2\pi)^{m}}
{\partial k^{\beta} \over \partial U^{\lambda}}
\left( {\bf U}(X^{\prime}) \right)\\
\qquad\times\sum_{l\geq0} \Pi^{\lambda}_{i(l)}
\left( \bm{\Phi} ( \dots ), k^{\alpha}
\bm{\Phi}_{\theta^{\alpha}} ( \dots ), \dots \right)
k^{\alpha_{1}}(X^{\prime}) \cdots k^{\alpha_{l}}(X^{\prime})
{\partial^{l} \over
\partial \theta^{\alpha_{1}} \cdots \partial \theta^{\alpha_{l}}}.
\end{gather*}

We can see then after the integration by parts that the main part
of the operator~\eqref{kbetaVarDerOper} reduces to the convolution
(w.r.t.~$\bm{\theta}$) with the vectors
\begin{gather*}
{\partial k^{\beta}\over\partial U^{\lambda}}
\left({\bf U}(X^{\prime})\right)
\zeta^{(\lambda)}_{i[{\bf U}(X^{\prime})]}
\left({{\bf S}(X^{\prime})\over\epsilon}+\bm{\theta}
\right),
\end{gather*}
which are identically equal to zero according to Lemma~\ref{Lemma2.1.}.

We can conclude, therefore, that terms of the order of
$O(\epsilon^{2})$ in the right-hand part of~\eqref{RedggBgJ}
do not play any role in the convolution with the
values~\eqref{kbetavarder} in the order $\epsilon^{2}$.

Let us consider now the action of the operator
\eqref{kbetaVarDerOper} on the terms of the order of
$\epsilon$ in the expression~\eqref{PhiThetaOst}.
We can see that the terms in which
the operators ${\mathrm d}^{s}/{\mathrm d}X^{\prime s}$
apply only to the functions
$\bm{\Phi}_{\theta^{\alpha}}({\bf S}(X^{\prime})/\epsilon+\bm{\theta},
{\bf U}(X^{\prime}))$ yield the quantity
\begin{gather*}
{\epsilon\over2}{\partial k^{\beta}\over\partial U^{\lambda}}
\left({\bf J}(X^{\prime})\right)
\int_{0}^{2\pi}\cdots\int_{0}^{2\pi}
{\partial\over\partial\theta^{\alpha}}P^{\lambda}
(\bm{\varphi},\epsilon\bm{\varphi}_{X^{\prime}},\dots)
\Big|_{\cal K}{{\mathrm d}^{m}\theta\over(2\pi)^{m}}
\int\operatorname{sgn}(X^{\prime}-Z)Q^{\alpha}_{[{\bf q}]}(Z){\mathrm d}Z
\end{gather*}
identically equal to zero.

The terms of the order of $\epsilon^{2}$ occur at the same time with
a single dif\/ferentiation of the function
$\operatorname{sgn}(X^{\prime}-Z)$
and application of the remaining derivations
$\partial/\partial X^{\prime}$ to
$\bm{\Phi}_{\theta^{\alpha}}({\bf S}(X^{\prime})/\epsilon+\bm{\theta},
{\bf U}(X^{\prime}))$ in the leading order.
In the required approximation
the functionals ${\bf J}(X)$ should be also replaced by ${\bf U}(X)$
in the argument of $\partial k^{\beta}/\partial U^{\lambda}$.

The corresponding values are then equal to
\begin{gather*}
\epsilon^{2}{\partial k^{\beta}\over\partial U^{\lambda}}
\left({\bf U}(X^{\prime})\right)\\
\qquad{}\times
\int_{0}^{2\pi}\cdots\int_{0}^{2\pi}
\sum_{l\geq1}\Pi^{\lambda}_{i(l)}\left(\bm{\Phi},
k^{\gamma}\bm{\Phi}_{\theta^{\gamma}},\dots\right)
l k^{\alpha_{1}}\cdots k^{\alpha_{l-1}}
\Phi^{i}_{\theta^{\alpha_{1}}\cdots\theta^{\alpha_{l-1}}\theta^{\alpha}}
{{\mathrm d}^{m}\theta\over(2\pi)^{m}} Q^{\alpha}_{[{\bf q}]}
(X^{\prime}).
\end{gather*}

Using again the relations~\eqref{kalphakbetaUnu}, we obtain
that the orthogonality of the right-hand part of~\eqref{RedggBgJ}
to the values~\eqref{kbetavarder} in the order $\epsilon^{2}$
is equivalent to the relations
$Q^{\beta}_{[{\bf q}]}(X^{\prime})\equiv0$.
Given that the main part of~\eqref{PhiThetaOst}
represents the dif\/ference between
the left-hand sides of~\eqref{EffCond} and~\eqref{BAcond}
we obtain the assertion of the lemma.
\end{proof}

For a~regular Hamiltonian family $\Lambda$ and a~complete
Hamiltonian set of integrals $(I^{1},\dots,I^{N})$ we can also
prove the following lemma.

\begin{lemma}
Let the functions $B_{j[{\bf q}](1)}(\bm{\theta},X)$
satisfy conditions~\eqref{BAcond}.
Then the functions
$B_{j[{\bf q}](1)}(\bm{\theta},X)$
automatically satisfy~\eqref{BetaCond}.
\end{lemma}

\begin{proof}
We f\/irst prove the following statement.

The values $\zeta^{(\nu)}_{i[{\bf U}(X)]}(\bm{\theta})$
are orthogonal (for any $X$) to the values
$A^{i}_{[1][{\bf q}]}(\bm{\theta},X)$, i.e.
\begin{gather}\label{kappaAort}
\int_{0}^{2\pi}\cdots\int_{0}^{2\pi}
\zeta^{(\nu)}_{i[{\bf U}(X)]}(\bm{\theta})
A^{i}_{[1][{\bf q}]}(\bm{\theta},X)
{{\mathrm d}^{m}\theta\over(2\pi)^{m}}\equiv0.
\end{gather}

Indeed, as we know, the values
$\{g^{i}(\bm{\theta},X),J_{[{\bf q}]}\}|_{\cal K}$
identically vanish under action of the operator~\eqref{DeistvSvert}
(with the replacement of~$Z$ to~$X$).
In the leading order
($\sim\epsilon$) it's action is given by the convolution
w.r.t.~$\bm{\theta}$ of the values
$\{g^{i}(\bm{\theta},X),J_{[{\bf q}]}\}|_{{\cal K}[1]}$,
imposed by~\eqref{giJmu1skob}, with the values
$\zeta^{(\nu)}_{i[{\bf U}(X)]}({\bf S}(X)/\epsilon+\bm{\theta})$.
We know also that the values
$\zeta^{(\nu)}_{i[{\bf U}(X)]}({\bf S}(X)/\epsilon+\bm{\theta})$
are automatically orthogonal to the functions
$\Phi^{i}_{\theta^{\alpha}}({\bf S}(X)/\epsilon+\bm{\theta},
{\bf U}(X))$, so we get the relation~\eqref{kappaAort}.

Thus, the implementation of~\eqref{BAcond} implies the conditions
\begin{gather*}
\int_{0}^{2\pi}\cdots\int_{0}^{2\pi}
\zeta^{(\nu)}_{i[{\bf U}(X)]}(\bm{\theta})
{\hat B}^{ij}_{[0]}(X)B_{j[{\bf q}](1)}(\bm{\theta},X)
{{\mathrm d}^{m}\theta\over(2\pi)^{m}}=0,
\end{gather*}
which is equivalent to
\begin{gather*}
\omega^{\alpha\nu}({\bf U}(X))
\int_{0}^{2\pi}\cdots\int_{0}^{2\pi}
\Phi^{j}_{\theta^{\alpha}}(\bm{\theta},{\bf U}(X))
B_{j[{\bf q}](1)}(\bm{\theta},X)
{{\mathrm d}^{m}\theta\over(2\pi)^{m}}=0
\end{gather*}
view the skew-symmetry of ${\hat B}^{ij}_{[0]}(X)$.

From condition (2) of the def\/inition of a~complete Hamiltonian family
of commuting functionals we immediately obtain now conditions
\eqref{BetaCond}.
\end{proof}

It is not dif\/f\/icult to see that from the relation~\eqref{kappaAort}
follow also the conditions
\begin{gather}\label{LeftVecAort}
\int_{0}^{2\pi}\cdots\int_{0}^{2\pi}
\kappa^{(q)}_{i[{\bf U}(X)]}(\bm{\theta})
A^{i}_{[1][{\bf q}]}(\bm{\theta},X)
{{\mathrm d}^{m}\theta\over(2\pi)^{m}}\equiv0
\end{gather}
by Proposition~\ref{Proposition2.1.}.

We come, therefore, to investigation of system~\eqref{BAcond}
to construct an ``approximate'' restriction of the Poisson bracket~\eqref{EpsExtBracket} on the submanifold~${\cal K}$.
In the remaining
part of the article the investigation of system~\eqref{BAcond}
will play the basic role for the justif\/ication of the main results.

In what follows we consider separately the single-phase $(m=1)$
and the multiphase $(m\geq2)$ cases.

The following lemma can be formulated for the single-phase case
$m=1$.

\begin{lemma}
Let $\Lambda$ be a~regular Hamiltonian family of single-phase
solutions of~\eqref{EvInSyst} and $(I^{1},\dots,I^{N})$ be a~complete
Hamiltonian set of the first integrals of the form~\eqref{integ}.
Then
the functions $B_{j[{\bf q}](1)}(\bm{\theta},X)$
can be found from system~\eqref{BAcond} and can be written in
the form~\eqref{B1ApprSol} with smooth dependence on parameters
${\bf U}(X)$.
\end{lemma}

\begin{proof}
System~\eqref{BAcond} in the single-phase case is a~system of ordinary
dif\/ferential equations in $\theta$ with skew-symmetric operator
${\hat B}^{ij}_{[0]}(X)$.
It is easy to see also that the
right-hand side of system~\eqref{BAcond} has the form
\begin{gather*}
\xi^{i\mu}\left(\theta,{\bf U}(X)\right)
q_{\mu,X}(X)+\eta^{i\mu}_{\lambda}
\left(\theta,{\bf U}(X)\right)U^{\lambda}_{X}
q_{\mu}(X)
\end{gather*}
with periodic in $\theta$ functions
$\xi^{i\mu}(\theta,{\bf U}(X))$,
$\eta^{i\mu}_{\lambda}(\theta,{\bf U}(X))$.

The orthogonality conditions~\eqref{LeftVecAort} imply the
orthogonality of both the sets of
functions \linebreak $\xi^{i\mu}(\theta,{\bf U}(X))$
and $\eta^{i\mu}_{\lambda}(\theta,{\bf U}(X))$ to the functions
$\kappa^{(q)}_{i[{\bf U}(X)]}(\theta)$, such that system
\eqref{BAcond} can be split into independent inhomogeneous
systems
\begin{gather}\label{BbetaZeta}
{\hat B}^{ij}_{[0]}(X)\beta^{\mu}_{j(1),1}\left(
\theta,{\bf U}(X)\right)=
\xi^{i\mu}\left(\theta,{\bf U}(X)\right),
\\
\label{BbetaEta}
{\hat B}^{ij}_{[0]}(X)\beta^{\mu}_{j(1),0\lambda}
\left(\theta,{\bf U}(X)\right)=
\eta^{i\mu}_{\lambda}\left(\theta,{\bf U}(X)\right)
\end{gather}
def\/ining functions~\eqref{VidBeta}.

Both the systems~\eqref{BbetaZeta},~\eqref{BbetaEta}
are systems of ordinary linear dif\/ferential equations
with periodic coef\/f\/icients and the skew-symmetric opera\-tor~${\hat B}^{ij}_{[0]}(X)$.
The zero modes of the opera\-tor~${\hat B}^{ij}_{[0]}(X)$ are given by the variational
derivatives of annihilators of the bracket~\eqref{LocTheorFieldBr}
represented as functions of~$\theta$
on the manifold of single-phase solutions and are orthogonal to the
right-hand parts of~\eqref{BbetaZeta} and~\eqref{BbetaEta} according
to~\eqref{RazlozhAnn} and~\eqref{LeftVecAort}.
Eigenfunctions of~${\hat B}^{ij}_{[0]}(X)$ form a~basis in the space of
$2\pi$-periodic functions $\bm{\varphi}(\theta)$.
Besides that,
the nonzero eigenvalues of~${\hat B}^{ij}_{[0]}(X)$ are separated from
zero in this case.
Thus, the $2\pi$-periodic functions
$\beta^{\mu}_{j(1),1}(\theta,{\bf U}(X))$,
$\beta^{\mu}_{j(1),0\lambda}(\theta,{\bf U}(X))$ can be found
from systems~\eqref{BbetaZeta},~\eqref{BbetaEta} up to
the variational derivatives of the annihilators of the bracket~\eqref{LocTheorFieldBr}.
If we impose additional conditions of
orthogonality of $\beta^{\mu}_{j(1)}(\theta,X,Y)$ to the
variational derivatives of the annihilators of the bracket~\eqref{LocTheorFieldBr} on the manifold of single-phase solutions
we can suggest a~unique procedure of construction of the functions
$\beta^{\mu}_{j(1)}(\theta,X,Y)$.
The functions
$\beta^{\mu}_{j(1)}(\theta,X,Y)$ satisfy~\eqref{betasystem} in the order of $O(\epsilon)$
and conditions~\eqref{BetaCond1}
and smoothly depend on the parameters ${\bf U}(X)$, which implies
the required properties for the functions
$B_{j[{\bf q}](1)}(\theta,X)$.
\end{proof}

Let us formulate now the theorem that gives the justif\/ication
of the Dubrovin--Novikov procedure for a~regular Hamiltonian
family of single-phase solutions of system~\eqref{EvInSyst}.

\begin{theorem}\label{Theorem2.1.}
Let $\Lambda$ be a~regular Hamiltonian family of
single-phase solutions of system~\eqref{EvInSyst}.
Let
$(I^{1},\dots,I^{N})$ be a~complete Hamiltonian set of
commuting first integrals of system~\eqref{EvInSyst} on~$\Lambda$
having the form~\eqref{integ}.
Then the Dubrovin--Novikov procedure
gives a~Poisson bracket of hydrodynamic type on the space
of parameters $\{U^{\nu}(X),\; \nu=1,\dots,N\}$,
where $U^{\nu}\equiv\langle P^{\nu}\rangle$.
\end{theorem}

\begin{proof}
The antisymmetry of the Dubrovin--Novikov bracket follows immediately
from the antisymmetry of the original bracket.
Let us prove the
existence of the Jacobi identity under the conditions of the theorem.

Consider the Poisson bracket~\eqref{EpsExtBracket} and the functionals
\begin{gather*}
{\tilde J}^{\nu}_{(1)}(X)=J^{\nu}(X)+\epsilon\iint_{0}^{2\pi}
g^{j}(\theta,Z)\beta^{\nu}_{j(1)}\left(
{S(Z)\over\epsilon}+\theta,Z,X\right)
{{\mathrm d}\theta\over2\pi}{\mathrm d}Z
\end{gather*}
in the neighborhood of the submanifold ${\cal K}$.
(Note that the functions
$\beta^{\nu}_{j(1)}$ depending on the values ${\bf U}(X)$ on
${\cal K}$, can be considered as functionals
in the neighborhood of ${\cal K}$
by the def\/inition of the functionals ${\bf U}(X)$).

The pairwise brackets of the functionals
${\tilde J}^{\nu}_{(1)}(X)$,
${\tilde J}^{\mu}_{(1)}(Y)$ can be written as
\begin{gather}
\big\{{\tilde J}^{\nu}_{(1)}(X),
{\tilde J}^{\mu}_{(1)}(Y)\big\}
=\big\{J^{\nu}(X),J^{\mu}(Y)\big\}
\nonumber\\
{}+\epsilon\iint_{0}^{2\pi}
\left\{\beta^{\nu}_{i(1)}
\left({S(Z)\over\epsilon}+\theta,Z,X\right)
g^{i}(\theta,Z),J^{\mu}(Y)\right\}
{{\mathrm d}\theta\over2\pi}{\mathrm d}Z
\nonumber\\
{}+\epsilon\iint_{0}^{2\pi}
\left\{J^{\nu}(X),g^{j}(\theta,W)\beta^{\mu}_{j(1)}
\left({S(W)\over\epsilon}+\theta,W,Y\right)\right\}
{{\mathrm d}\theta\over2\pi}{\mathrm d}W
\label{JbracketnearK}\\
{}+\epsilon^{2}\!\int\!
\left\{\beta^{\nu}_{i(1)}\!
\left({S(Z)\over\epsilon}+\theta,Z,X\!\right)
g^{i}(\theta,Z),g^{j}(\theta^{\prime},W)
\beta^{\mu}_{j(1)} \!
\left({S(W)\over\epsilon}+\theta^{\prime},W,Y\!\right)\right\}\!
{{\mathrm d}\theta\over2\pi}{{\mathrm d}\theta^{\prime}\over2\pi}
{\mathrm d}Z{\mathrm d}W.\nonumber
\end{gather}

We note here that despite the presence of non-local terms in the
brackets of constraints \linebreak
$\{g^{i}(\theta,Z),g^{j}(\theta^{\prime},W)\}$,
and
$\{J^{\nu}(X),g^{j}(\theta,W)\}$,
$\{g^{i}(\theta,Z),J^{\mu}(Y)\}$, the bracket of the
functionals \linebreak
$\{{\tilde J}^{\nu}_{(1)}(X),{\tilde J}^{\mu}_{(1)}(Y)\}$
on ${\cal K}$ has a~purely local form because of the conditions~\eqref{BetaCond1}.
We can also see here that expression~\eqref{JbracketnearK} is a~regular at
$\epsilon\rightarrow0$ distribution after the restriction on~${\cal K}$.

From the form of the constraints and condition~\eqref{BetaCond1}
it is easy to obtain that the pairwise Poisson brackets of the
functionals ${\tilde J}^{\nu}_{(1)}(X)$,
${\tilde J}^{\mu}_{(1)}(Y)$ on ${\cal K}$ can be written as
\begin{gather}\label{Jnu1Jmu1genbr}
\big\{{\tilde J}^{\nu}_{(1)}(X),
{\tilde J}^{\mu}_{(1)}(Y)\big\}\big|_{\cal K}
=\sum_{l\geq1}\epsilon^{l}\sum_{q=0}^{l}
a^{\nu\mu}_{[l-q]}({\bf U},{\bf U}_{X},\dots)
\delta^{(q)}(X - Y),
\end{gather}
where $a^{\nu\mu}_{[l-q]}$ are some functions of
${\bf U},{\bf U}_{X},\dots$,
polynomial in the derivatives and having deg\-ree~\mbox{$l-q$}.
By virtue of relations~\eqref{QRelLemma24}
we can state also that the leading term in $\epsilon$
in the expression~\eqref{Jnu1Jmu1genbr} is still the same
Dubrovin--Novikov bracket on ${\cal K}$, since the corrections given
by the last three terms of~\eqref{JbracketnearK} are of the order
of $O(\epsilon^{2})$ on ${\cal K}$.
We thus have
\begin{gather*}
\big\{{\tilde J}^{\nu}_{(1)}(X),
{\tilde J}^{\mu}_{(1)}(Y)\big\}\big|_{\cal K}
=\epsilon\left\{U^{\nu}(X),U^{\mu}(Y)
\right\}_{\rm DN}+\sum_{l\geq2}\epsilon^{l}\sum_{q=0}^{l}
a^{\nu\mu}_{[l-q]}({\bf U},{\bf U}_{X},\dots)
\delta^{(q)}(X - Y),
\end{gather*}
where
\begin{gather*}
\left\{U^{\nu}(X),U^{\mu}(Y)\right\}_{\rm DN}=
\langle A^{\nu\mu}_{1}\rangle\left({\bf U}(X)\right)
\delta^{\prime}(X - Y)+
{\partial\langle Q^{\nu\mu}\rangle\over\partial U^{\gamma}}
U^{\gamma}_{X}\delta(X - Y)
\end{gather*}
is the Dubrovin--Novikov bracket on the space of f\/ields
${\bf U}(X)$.
Thus, we can write
\begin{gather}
{\delta\over\delta U^{\gamma}(W)}
\big\{{\tilde J}^{\nu}_{(1)}(X),
{\tilde J}^{\mu}_{(1)}(Y)\big\}
\big|_{\cal K}=\epsilon
{\delta\left\{U^{\nu}(X),U^{\mu}(Y)\right\}_{\rm DN}
\over\delta U^{\gamma}(W)}\nonumber\\
\label{VarPrSkobJ}
\qquad{}+
{\delta\over\delta U^{\gamma}(W)}
\sum_{l\geq2}\epsilon^{l}\sum_{q=0}^{l}
a^{\nu\mu}_{[l-q]}({\bf U},{\bf U}_{X},\dots)
\delta^{(q)}(X - Y)
\end{gather}
on the submanifold ${\cal K}$.

For further discussion it is convenient to introduce ``regularized''
functionals.
Namely, for any smooth compactly supported vector-valued
function
${\bf q}(X)=(q_{1}(X),\dots,q_{N}(X))$
we introduce the functionals
\begin{gather*}
J_{[{\bf q}]}=\int q_{\nu}(X)J^{\nu}(X){\mathrm d}X
,\!\qquad {\tilde J}_{(1)[{\bf q}]}=
\int q_{\nu}(X){\tilde J}^{\nu}_{(1)}(X){\mathrm d}X,\!\qquad
U_{[{\bf q}]}=\int q_{\nu}(X)U^{\nu}(X){\mathrm d}X.
\end{gather*}

We can write then
\begin{gather}
\big\{{\tilde J}_{(1)[{\bf q}]},
{\tilde J}_{(1)[{\bf p}]}\big\}
=\left\{J_{[{\bf q}]},J_{[{\bf p}]}\right\}
+\epsilon\int_{-\infty}^{+\infty}\int_{0}^{2\pi}
\left\{B_{i[{\bf q}](1)}
\left({S(Z)\over\epsilon}+\theta,Z\right)
g^{i}(\theta,Z),J_{[{\bf p}]}\right\}
{{\mathrm d}\theta\over2\pi}{\mathrm d}Z
\nonumber\\
{} +\epsilon\int_{-\infty}^{+\infty}\int_{0}^{2\pi}
\left\{J_{[{\bf q}]},g^{j}(\theta,W)B_{j[{\bf p}](1)}
\left({S(W)\over\epsilon}+\theta,W\right)\right\}
{{\mathrm d}\theta\over2\pi}{\mathrm d}W
\label{JbracketnearK1}\\
{}+\epsilon^{2}\!\int\!
\left\{B_{i[{\bf q}](1)}\!
\left({S(Z)\over\epsilon}+\theta,Z\!\right)g^{i}(\theta,Z)
,g^{j}(\theta^{\prime},W)B_{j[{\bf p}](1)} \!
 \left({S(W)\over\epsilon}+\theta^{\prime},W\!\right)\right\}\!
{{\mathrm d}\theta\over2\pi}{{\mathrm d}\theta^{\prime}\over2\pi}
{\mathrm d}Z{\mathrm d}W.\nonumber
\end{gather}

We can see again that the last three terms of~\eqref{JbracketnearK1}
give corrections of the order of $\epsilon^{2}$ on the submanifold
${\cal K}$.

For the expansion of the brackets $\eqref{JbracketnearK1}$
in the neighborhood of the submanifold
${\cal K}$ it is convenient to use the expressions
\begin{gather}\label{PhigRel}
\varphi^{i}(\theta,X)=g^{i}(\theta,X)
+\Phi^{i}\left({S(X)\over\epsilon}+\theta,
{\bf U}(X)\right)
\end{gather}
making it possible to easily expand any functionals of
$\bm{\varphi}(\theta,X)$ with respect to the values of~${\bf g}$.
For example, using the expression~\eqref{PnuPmuskob},
we can write for the brackets $\{J_{[{\bf q}]},J_{[{\bf p}]}\}$
in the neighborhood of ${\cal K}$
\begin{gather*}
\left\{J_{[{\bf q}]},J_{[{\bf p}]}\right\}=
\left.\left\{J_{[{\bf q}]},J_{[{\bf p}]}\right\}\right|_{\cal K}
\\
{}+\iint_{0}^{2\pi}\! \sum_{s\geq0}
\left[
{\delta\over\delta\varphi^{j}(\theta,W)}
\left.\iint_{0}^{2\pi}\! \epsilon^{s}
A^{\nu\mu}_{s}\left(\bm{\varphi}(\theta^{\prime},X),
\epsilon\bm{\varphi}_{X}(\theta^{\prime},X),\dots\right)
q_{\nu}(X)p_{\mu,sX}(X)
{{\mathrm d}\theta^{\prime}\over2\pi}{\mathrm d}X\right]\right|_{\cal K}
\\
\qquad {}
\times g^{j}(\theta,W){{\mathrm d}\theta\over2\pi}{\mathrm d}W+O\big({\bf g}^{2}\big),
\end{gather*}
where the values $\{J_{[{\bf q}]},J_{[{\bf p}]}\}|_{\cal K}$
as well as all the other values on ${\cal K}$
are calculated on the functions
${\tilde\varphi}^{i}(\theta,W)\in{\cal K}$ corresponding to the
same values of the functionals $[{\bf U}(Z)]$ as the original functions
$\varphi^{i}(\theta,W)$.
We can see that the f\/irst term
of the expansion of $\{J_{[{\bf q}]},J_{[{\bf p}]}\}$
near ${\cal K}$ also has the order $O(\epsilon)$ as well as the
values $\{J_{[{\bf q}]},J_{[{\bf p}]}\}$ on ${\cal K}$
view the condition~\eqref{epsAQRel}.

Considering relations~\eqref{phiJmuexp},
we can write a~similar representation for the remaining terms of
the expression~\eqref{JbracketnearK1}.

We can see that the last term of the right-hand side of
\eqref{JbracketnearK1} has the order $O(\epsilon^{2})$ in the
whole space which gives the same order in $\epsilon$ for the
terms of its expansion after the substitution of~\eqref{PhigRel}.

By Lemma~\ref{Lemma2.4.},
the second and the third terms of~\eqref{JbracketnearK1}
are of the order $O(\epsilon^{2})$ on the submanifold ${\cal K}$.
However, this can not be said about the next terms of the expansion
of these brackets in the values of ${\bf g}$.
To f\/ind the values of
these terms near ${\cal K}$ we write e.g.
\begin{gather*}
\iint_{0}^{2\pi}\!\!B_{i[{\bf q}](1)}\left(
{S(Z)\over\epsilon}+\theta,Z\right)
g^{i}(\theta,Z){{\mathrm d}\theta\over2\pi}{\mathrm d}Z
\\
{}
=\iint_{0}^{2\pi} B_{i[{\bf q}](1)}\left(
{S(Z)\over\epsilon}+\theta,Z\right)
\varphi^{i}(\theta,Z){{\mathrm d}\theta\over2\pi}{\mathrm d}Z-
\iint_{0}^{2\pi} B_{i[{\bf q}](1)}(\theta,Z)
\Phi^{i}(\theta,{\bf U}(Z)){{\mathrm d}\theta\over2\pi}{\mathrm d}Z.
\end{gather*}

After rather simple calculations it's not dif\/f\/icult to get then
for the second term of~\eqref{JbracketnearK1} near ${\cal K}$
\begin{gather*}
\iint_{0}^{2\pi} \left\{B_{i[{\bf q}](1)}\left(
{S(Z)\over\epsilon}+\theta,Z\right)g^{i}(\theta,Z)
,J_{[{\bf p}]}\right\}{{\rm d}\theta\over2\pi}{\mathrm d}Z
\\
{}=\iint_{0}^{2\pi} \left.
\left\{B_{i[{\bf q}](1)}\left(
{S(Z)\over\epsilon}+\theta,Z\right)g^{i}(\theta,Z)
,J_{[{\bf p}]}\right\}\right|_{\cal K}
{{\rm d}\theta\over2\pi}{\mathrm d}Z
\\
{}+\int\Bigg[\int B_{i[{\bf q}](1)}\left(
{S(Z)\over\epsilon}+\theta,Z\right)\left.
{\delta C^{i\mu}_{(0)}(\bm{\varphi}(\theta,Z),\epsilon
\bm{\varphi}_{Z}(\theta,Z),\dots)\over
\delta\varphi^{j}(\theta^{\prime},W)}\right|_{\cal K}
p_{\mu}(Z){{\mathrm d}\theta\over2\pi}{\mathrm d}Z+O(\epsilon)
\Bigg]
\\
\qquad{} \times g^{j}(\theta^{\prime},W)
{{\mathrm d}\theta^{\prime}\over2\pi}{\mathrm d}W
\\
{}+\int\Bigg[\int B_{i[{\bf q}](1)\theta}\left(
{S(Z)\over\epsilon}+\theta,Z\right)\!{1\over2\epsilon}
\operatorname{sgn}(Z - W)\!\left.
\left\{k(W),J_{[{\bf p}]}
\right\}\right|_{\cal K}{\mathrm d}W+O(\epsilon)\Bigg]
\\
\qquad{}\times
g^{i}(\theta,Z){{\mathrm d}\theta\over2\pi}{\mathrm d}Z+O\big({\bf g}^{2}\big)
\\
=\iint_{0}^{2\pi} \left.
\left\{B_{i[{\bf q}](1)}\left(
{S(Z)\over\epsilon}+\theta,Z\right)g^{i}(\theta,Z)
,J_{[{\bf p}]}\right\}\right|_{\cal K}
{{\mathrm d}\theta\over2\pi}{\mathrm d}Z
\\
{}+\int\Bigg[\int B_{j[{\bf q}](1)}\left(
{S(Z)\over\epsilon}+\theta^{\prime},Z\right)\left.
{\delta S^{j\mu}(\bm{\varphi}(\theta^{\prime},Z),\epsilon
\bm{\varphi}_{Z}(\theta^{\prime},Z),\dots)\over
\delta\varphi^{i}(\theta,W)}\right|_{\cal K}
p_{\mu}(Z){{\mathrm d}\theta^{\prime}\over2\pi}{\mathrm d}Z+
O(\epsilon)\Bigg]
\\
\qquad
{}\times g^{i}(\theta,W){{\mathrm d}\theta\over2\pi}{\mathrm d}W
\\
{}+\int\left[\int B_{i[{\bf q}](1)\theta}\left(
{S(Z)\over\epsilon}+\theta,Z\right)\omega^{\mu}(Z)
p_{\mu}(Z)+O(\epsilon)\right]
g^{i}(\theta,Z){{\mathrm d}\theta\over2\pi}{\mathrm d}Z+
O\big({\bf g}^{2}\big).
\end{gather*}

In general, considering the expansion of the brackets
\eqref{JbracketnearK1} in the neighborhood of ${\cal K}$,
we can write the relations
\begin{gather}
\big\{{\tilde J}_{(1)[{\bf q}]},
{\tilde J}_{(1)[{\bf p}]}\big\}=
\big\{{\tilde J}_{(1)[{\bf q}]},
{\tilde J}_{(1)[{\bf p}]}\big\}\big|_{\cal K}
 \nonumber\\
\qquad
{}+\epsilon\iint_{0}^{2\pi}
{\tilde T}_{i[{\bf q},{\bf p}]}\left({S(W)\over\epsilon}+\theta,
W,\epsilon\right)g^{i}(\theta,W){{\mathrm d}\theta\over2\pi}
{\mathrm d}W+O\big({\bf g}^{2}\big)\label{JbracketExpnearK}
\end{gather}
with some regular at $\epsilon\rightarrow0$,
$2\pi$-periodic in $\theta$, functions
${\tilde T}_{i[{\bf q},{\bf p}]}(\theta,W,\epsilon)$.

Now for the bracket~\eqref{EpsExtBracket}
we consider at the ``points'' of ${\cal K}$
the Jacobi identity of the form
\begin{gather}\label{YacobiJtozhd}
\big\{\big\{{\tilde J}_{(1)[{\bf q}]},
{\tilde J}_{(1)[{\bf p}]}\big\},
{\tilde J}_{(1)[{\bf r}]}\big\}+{\rm c.p.}
\equiv0
\end{gather}
with some smooth compactly supported ${\bf q}(X)$, ${\bf p}(X)$,
${\bf r}(X)$.

We have on ${\cal K}$ the relations
\begin{gather*}
\big\{g^{i}(\theta,W),
{\tilde J}_{(1)[{\bf r}]}\big\}\big|_{\cal K}
=O\big(\epsilon^{2}\big)
\end{gather*}
for the bracket of the functionals $g^{i}(\theta,W)$ and
${\tilde J}_{(1)[{\bf r}]}$.

It's not dif\/f\/icult to get also
\begin{gather*}
\big\{U^{\gamma}(W),{\tilde J}_{(1)[{\bf r}]}
\big\}\big|_{\cal K}=
\big\{J^{\gamma}(W),{\tilde J}_{(1)[{\bf r}]}
\big\}\big|_{\cal K}+O\big(\epsilon^{2}\big)
\\
\hphantom{\big\{U^{\gamma}(W),{\tilde J}_{(1)[{\bf r}]}
\big\}\big|_{\cal K}}{}
=
\big\{J^{\gamma}(W),J_{[{\bf r}]}
\big\}\big|_{\cal K}+O\big(\epsilon^{2}\big)
=\epsilon
\big\{U^{\gamma}(W),U_{[{\bf r}]}\big\}_{\rm DN}
+O\big(\epsilon^{2}\big).
\end{gather*}

Now using relations~\eqref{VarPrSkobJ} and expansion
\eqref{JbracketExpnearK}, we can see that the Jacobi identities
\eqref{YacobiJtozhd} on ${\cal K}$ coincide in the leading order in
$\epsilon$ $(\epsilon^{2})$ with the similar Jacobi identities for
the Dubrovin--Novikov bracket
\begin{gather*}
\int{\delta\{U_{[{\bf q}]},U_{[{\bf p}]}\}_{\rm DN}\over
\delta U^{\gamma}(W)}\{U^{\gamma}(W),U_{[{\bf r}]}
\}_{\rm DN}\, {\mathrm d}W+\text{c.p.} \equiv 0
\end{gather*}
on the space of f\/ields ${\bf U}(X)$.
\end{proof}

Let us formulate here also the Theorems justifying the invariance
of the Dubrovin--Novikov procedure and the Hamiltonian properties
of the Whitham system~\eqref{ConsWhitham} with respect to the
averaged bracket in the single-phase case.

\begin{theorem}\label{Theorem2.2.}
Let $\Lambda$ be a~regular Hamiltonian family of single-phase
solutions of~\eqref{EvInSyst}.
Let $(I^{1},\dots,I^{N})$ and
$(I^{\prime1},\dots,I^{\prime N})$ be two different complete
Hamiltonian sets of commuting first integrals of
\eqref{EvInSyst} having the form~\eqref{integ}.
Then the
Dubrovin--Novikov brackets obtained using the sets
$(I^{1},\dots,I^{N})$ and $(I^{\prime1},\dots,I^{\prime N})$
coincide with each other.
\end{theorem}

\begin{theorem}\label{Theorem2.3.}
Let $\Lambda$ be a~regular Hamiltonian family of single-phase
solutions of~\eqref{EvInSyst}.
Let $(I^{1},\dots,I^{N})$ be a
complete Hamiltonian set of commuting first integrals of
\eqref{EvInSyst} having the form~\eqref{integ} and $H$ be the
Hamiltonian function for the system~\eqref{EvInSyst}
having the form~\eqref{hamfun}.
Then the Whitham system
\eqref{ConsWhitham} is Hamiltonian with respect to the
corresponding Dubrovin--Novikov bracket~\eqref{dubrnovb}
with the Hamiltonian function~\eqref{UsrHamFunc}
\begin{gather*}
H^{av}=\int_{-\infty}^{+\infty}\langle P_{H}\rangle
\left({\bf U}(X)\right){\mathrm d}X.
\end{gather*}
\end{theorem}

Proofs of Theorems~\ref{Theorem2.2.} and~\ref{Theorem2.3.}
are identical for the single-phase
and the multiphase cases, so we prove them in general \mbox{$m$-phase}
case in the next chapter.

\section[Dubrovin-Novikov bracket and the multiphase case]{Dubrovin--Novikov bracket and the multiphase case}

We turn now to the general multiphase situation.
We note
that the dif\/ference with the single-phase case occur only in
Theorem~\ref{Theorem2.1.}, since f\/inding the functions
$B_{j[{\bf q}](1)}(\bm{\theta},X)$
here is a~more complicated problem.
Let us return to the system~\eqref{BAcond}, i.e.
\begin{gather}\label{BAcondNextChapt}
{\hat B}^{ij}_{[0]}(X)B_{j[{\bf q}](1)}(\bm{\theta},X)
+A^{i}_{[1][{\bf q}]}(\bm{\theta},X)=0,
\end{gather}
where the operator ${\hat B}^{ij}_{[0]}(X)$ and the functions
$A^{i}_{[1][{\bf q}]}(\bm{\theta},X)$ are given by~\eqref{OperatorB}
and~\eqref{FunktsiiA} respectively.
It is easy to see that the operator
${\hat B}^{ij}_{[0]}(X)$ is a~dif\/ferential operator on the torus,
which acts along a~constant direction
$k^{\alpha}\partial/\partial\theta^{\alpha}$.
The corresponding
vector f\/ield generates one-dimensional trajectories which are everywhere
dense in the torus $\mathbb{T}^{m}$ for generic vectors ${\bf k}(X)$.
For special values of ${\bf k}(X)$ the closures of the trajectories may
be lower-dimensional tori, in particular, the closed one-dimensional
trajectories in $\mathbb{T}^{m}$.
Note that the operator
${\hat B}^{ij}_{[0]}(X)\equiv{\hat B}^{ij}_{[0]}({\bf U}(X))$
is exactly the Hamiltonian operator for the bracket
\eqref{LocTheorFieldBr} on the manifold of \mbox{$m$-phase} solutions
$\Lambda$.

The operator ${\hat B}^{ij}_{[0]}({\bf U})$ has in general
f\/inite number of ``regular'' eigenvectors with zero eigenvalues
def\/ined for all values of the parameters ${\bf U}$ and smoothly
depending on the parameters.
However, for special values of
${\bf U}$ the set of eigenvectors with zero eigenvalues is inf\/inite
and determined by the dimension of closures of the straight-line
trajectories in $\mathbb{T}^{m}$, def\/ined by the vector
${\bf k}({\bf U})$.\footnote{For some special brackets
\eqref{LocTheorFieldBr} the dif\/ferential part can be absent in the
opera\-tor~${\hat B}^{ij}_{[0]}({\bf U})$.
The opera\-tor~${\hat B}^{ij}_{[0]}({\bf U})$ reduces then to an ultralocal
operator acting independently at every point of~$\mathbb{T}^{m}$.
As a~rule, the mat\-rix~$B^{ij}_{[0]}({\bf U})$ is non-degenerate
in this case.
For example, for the ultralocal Poisson bracket
\begin{gather*}
\{\psi(x),{\bar\psi}(y)\}=i
\delta(x-y)
\end{gather*}
for the NLS equation
\begin{gather*}
i\psi_{t}=\psi_{xx}+\nu|\psi|^{2}\psi
\end{gather*}
we have exactly this situation.
Easy to see that system
\eqref{BAcondNextChapt} represents a~simple algebraic system in this
case and is trivially solvable.
The multiphase situation is not
dif\/ferent here from the single-phase one, so all the arguments of the
previous chapter can be used also for the multiphase case.
However,
for arbitrary brackets~\eqref{LocTheorFieldBr} the operators
${\hat B}^{ij}_{[0]}({\bf U})$ have more general form described above.}

Let us def\/ine in the space of the parameters ${\bf U}$ the set
${\cal M}$, such that for all ${\bf U}\in{\cal M}$ the trajectories
of the vector f\/ield $(k^{1}({\bf U}),\dots,k^{m}({\bf U}))$
are everywhere dense in $\mathbb{T}^{m}$.
From the condition
\begin{gather*}
\operatorname{rank}||\partial k^{\alpha}/\partial U^{\nu}||
=m
\end{gather*}
it follows that the set ${\cal M}$ is everywhere dense in the
parameter space ${\bf U}$ and, moreover, has the full measure.

In the study of the solubility of~\eqref{BAcondNextChapt}
we must f\/irst require the orthogonality of the functions
$A^{i}_{[1][{\bf q}]}(\bm{\theta},X)$ to the ``regular''
eigenvectors of ${\hat B}^{ij}_{[0]}(X)$ with zero eigenvalues.
Let us prove here the following lemma.

\begin{lemma}\label{Lemma3.1.}
Let $\Lambda$ be a~regular Hamiltonian family of
\mbox{$m$-phase} solutions of~\eqref{EvInSyst} and $(I^{1},\dots,$ $I^{N})$
be a~complete Hamiltonian set of commuting first integrals of
\eqref{EvInSyst} having the form~\eqref{integ}.
Let for
${\bf U}\in{\cal M}$
\begin{gather*}
{\bf v}^{(l)}_{[{\bf U}]}(\bm{\theta})=
\big(v^{(l)}_{1[{\bf U}]}(\bm{\theta}),\dots,
v^{(l)}_{n[{\bf U}]}(\bm{\theta})\big),\qquad
l=1,\dots,s
\end{gather*}
be the complete set of linearly independent eigenvectors of
the operator ${\hat B}^{ij}_{[0]}({\bf U})$ on the torus
with zero eigenvalues, smoothly depending on
$\bm{\theta}$.
Then
\begin{enumerate}\itemsep=0pt
\item[$1)$] the number of the vectors
${\bf v}^{(l)}_{[{\bf U}]}(\bm{\theta})$
is equal to the number of annihilators of the brac\-ket~\eqref{LocTheorFieldBr} on the submanifold of
\mbox{$m$-phase} solutions of~\eqref{EvInSyst};

\item[$2)$] the functions $A^{i}_{[1][{\bf q}]}(\bm{\theta},X)$
are orthogonal to all the vectors
${\bf v}^{(l)}_{[{\bf U}(X)]}(\bm{\theta})$, i.e.
\begin{gather*}
\int_{0}^{2\pi}\cdots\int_{0}^{2\pi}
v^{(l)}_{i[{\bf U}(X)]}(\bm{\theta})
A^{i}_{[1][{\bf q}]}(\bm{\theta},X)
{{\mathrm d}^{m}\theta\over(2\pi)^{m}}\equiv0.
\end{gather*}
\end{enumerate}
\end{lemma}

\begin{proof}
Consider the values of ${\bf v}^{(l)}_{[{\bf U}]}(\bm{\theta})$
on any of the trajectories of the vector f\/ield
$k^{\alpha}\partial/\partial\theta^{\alpha}$ on the torus
$\mathbb{T}^{m}$.
According to the def\/inition of regular Hamiltonian
family of \mbox{$m$-phase} solutions of~\eqref{EvInSyst} the corresponding
functions $v^{(l)}_{i[{\bf U}]}({\bf k}x+\bm{\theta}_{0})$
should be the variational derivatives of some linear combination
of annihilators of the bracket~\eqref{LocTheorFieldBr}.
We have then
for a~f\/ixed value of $\bm{\theta}_{0}$
\begin{gather*}
v^{(l)}_{i[{\bf U}]}({\bf k}x+\bm{\theta}_{0})=
\sum_{p}\alpha^{l}_{p}({\bf U},\bm{\theta}_{0})\left.
{\delta N^{p}\over\delta\varphi^{i}(x)}
\right|_{\bm{\varphi}=\bm{\Phi}({\bf k}x+\bm{\theta}_{0},{\bf U})}.
\end{gather*}

From relation~\eqref{RazlozhAnn} we have then
\begin{gather}\label{valphankappa}
v^{(l)}_{i[{\bf U}]}({\bf k}({\bf U})x+\bm{\theta}_{0})
=\sum_{p,q}\alpha^{l}_{p}({\bf U},\bm{\theta}_{0})
\, n^{p}_{q}({\bf U})\,
\kappa^{(q)}_{i[{\bf U}]}\left({\bf k}({\bf U})x+\bm{\theta}_{0}\right).
\end{gather}

By def\/inition, for ${\bf U}\in{\cal M}$ the trajectories of the
f\/ield $k^{\alpha}\partial/\partial\theta^{\alpha}$ are
everywhere dense in $\mathbb{T}^{m}$.
Since both the left- and the right-hand parts of
\eqref{valphankappa} are smooth functions on $\mathbb{T}^{m}$,
we get then that they coincide on $\mathbb{T}^{m}$.
We can put then
$\alpha^{l}_{p}({\bf U},\bm{\theta}_{0})=\alpha^{l}_{p}({\bf U})$
and write
\begin{gather}\label{Sootndliavikappa}
v^{(l)}_{i[{\bf U}]}(\bm{\theta})\equiv
\sum_{p,q}\alpha^{l}_{p}({\bf U}) \, n^{p}_{q}({\bf U})
\, \kappa^{(q)}_{i[{\bf U}]}\left(\bm{\theta}\right).
\end{gather}

It is easy to see also that any linear combination of the form~\eqref{Sootndliavikappa} gives a~regular eigenvector of the operator
${\hat B}^{ij}_{[0]}({\bf U}(X))$ with zero eigenvalue.

Statement (2) follows then from relation~\eqref{LeftVecAort}
in view of representation~\eqref{Sootndliavikappa}.
\end{proof}

However, despite the presence of Lemma~\ref{Lemma3.1.}, study of
system~\eqref{BAcondNextChapt} is much more complicated
in the multiphase case if compared with the single-phase case.
Thus, the presence of ``resonances'' for some values of the
parameters ${\bf U}(X)$ may lead to insolubility of
\eqref{BAcondNextChapt} in the space of periodic
(in all $\theta^{\alpha}$) functions.
The set of the ``resonance''
values of ${\bf U}$, as a~rule, has measure zero.
Nevertheless, the possibility of resonant parameters do not allow
to transfer directly the methods of the previous chapter to the
multiphase case.
The following theorem shows, however, that the
procedure of the bracket averaging is in fact insensitive
to the appearance of the resonant values of ${\bf U}$ and can be used
in most multiphase cases, as well as in the single-phase case.

\begin{theorem}\label{Theorem3.1.}
Let the system~\eqref{EvInSyst} be a~local Hamiltonian system
generated by the functional~\eqref{hamfun} in the local
field-theoretic Hamiltonian structure~\eqref{LocTheorFieldBr}.
Let $\Lambda$ be a~regular Hamiltonian family of
\mbox{$m$-phase} solutions of~\eqref{EvInSyst} and $(I^{1},\dots,I^{N})$
be a~complete Hamiltonian set of commuting integrals~\eqref{integ}
for this family.

Let the parameter space ${\bf U}$ of the family $\Lambda$
have a~dense set ${\cal S}\subset{\cal M}$ on which the sys\-tem~\eqref{BAcondNextChapt}, or, equivalently, the multiphase system
\eqref{BbetaZeta}, \eqref{BbetaEta} is solvable in the space
of smooth $2\pi$-periodic in each $\theta^{\alpha}$ functions.
Then the Dubrovin--Novikov bracket, obtained with the aid of
the functionals $(I^{1},\dots,I^{N})$, satisfies the Jacobi
identity.
\end{theorem}

\begin{proof}
As before, for a~smooth compactly supported vector-valued function
 ${\bf q}(X)=(q_{1}(X)$, $\dots,q_{N}(X))$ we def\/ine
the functional
\begin{gather*}
J_{[{\bf q}]}=\int_{-\infty}^{+\infty}
J^{\nu}(X) q_{\nu}(X){\mathrm d}X.
\end{gather*}

Then, for arbitrary smooth, compactly supported in $X$ and
$2\pi$-periodic in each $\theta^{\alpha}$ functions
${\tilde{\bf Q}}(\bm{\theta},X)=
({\tilde Q}_{1}(\bm{\theta},X),\dots,{\tilde Q}_{n}(\bm{\theta},X))$
we def\/ine the functionals
\begin{gather*}
Q_{i}(\bm{\theta},X)={\tilde Q}_{i}(\bm{\theta},X)\\
\hphantom{Q_{i}(\bm{\theta},X)=} {}-
\Phi^{i}_{\theta^{\beta}}(\bm{\theta},{\bf U}(X))
M^{\beta\gamma}({\bf U}(X))
\int_{0}^{2\pi}\cdots\int_{0}^{2\pi}
{\tilde Q}_{j}(\bm{\theta}^{\prime},X)
\Phi^{j}_{\theta^{\prime\gamma}}(\bm{\theta}^{\prime},{\bf U}(X))
{{\mathrm d}^{m}\theta^{\prime}\over(2\pi)^{m}},
\end{gather*}
where the matrix $M^{\beta\gamma}({\bf U})$ is the inverse of the
matrix
\begin{gather*}
\int_{0}^{2\pi}\cdots\int_{0}^{2\pi} \sum_{i=1}^{n}
\Phi^{i}_{\theta^{\beta}}(\bm{\theta},{\bf U})
\Phi^{i}_{\theta^{\gamma}}(\bm{\theta},{\bf U})
{{\mathrm d}^{m}\theta\over(2\pi)^{m}},
\end{gather*}
which is always def\/ined according to the def\/inition of a~complete
regular family of \mbox{$m$-phase} solutions of system~\eqref{EvInSyst}.

By def\/inition, the functions $Q_{i}(\bm{\theta},X)$ are local
functionals of ${\bf U}(X)$
\begin{gather*}
Q_{i}(\bm{\theta},X)\equiv
Q_{i}(\bm{\theta},X,{\bf U}(X))
\end{gather*}
depending also on the arbitrary f\/ixed functions
${\tilde{\bf Q}}(\bm{\theta},X)$.
Everywhere below we will assume
that ${\bf Q}(\bm{\theta},X)$ is a~functional of this type def\/ined
with some function ${\tilde{\bf Q}}(\bm{\theta},X)$.

Easy to see that the values of $Q_{i}(\bm{\theta},X)$ with arbitrary
${\tilde{\bf Q}}(\bm{\theta},X)$ represent for f\/ixed values of the
functionals ${\bf U}(Z)$ all possible smooth, compactly supported
in $X$ and $2\pi$-periodic in each $\theta^{\alpha}$ functions with
the only restriction
\begin{gather}\label{OgranQ}
\int_{0}^{2\pi}\cdots\int_{0}^{2\pi}
Q_{i}(\bm{\theta},X)\Phi^{i}_{\theta^{\alpha}}
(\bm{\theta},{\bf U}(X)){{\mathrm d}^{m}\theta\over(2\pi)^{m}}
=0,\qquad\forall\, X,
\alpha=1,\dots,m.
\end{gather}

For the functionals $Q_{i}(\bm{\theta},X)$ we def\/ine the functionals
\begin{gather*}
g_{[{\bf Q}]}=\int_{-\infty}^{+\infty}
\int_{0}^{2\pi}\cdots\int_{0}^{2\pi} g^{i}(\bm{\theta},X)
Q_{i}\left({{\bf S}(X)\over\epsilon}+\bm{\theta},X\right)
{{\mathrm d}^{m}\theta\over(2\pi)^{m}}{\mathrm d}X.
\end{gather*}

Now, for f\/ixed functions ${\bf q}(X)$, ${\bf p}(X)$, and functional
${\bf Q}(\bm{\theta},X)$ consider the Jacobi identity of the form
\begin{gather}\label{JacobigJJ}
\left\{g_{[{\bf Q}]},\left\{J_{[{\bf q}]},J_{[{\bf p}]}
\right\}\right\}+\left\{J_{[{\bf p}]},\left\{g_{[{\bf Q}]},J_{[{\bf q}]}
\right\}\right\}+\left\{J_{[{\bf q}]},\left\{J_{[{\bf p}]},g_{[{\bf Q}]}
\right\}\right\}\equiv0.
\end{gather}

Expanding the values of the brackets
$\{J^{\nu}(X),J^{\mu}(Y)\}$ in the neighborhood of the submanifold~${\cal K}$, as in the single-phase case, we can write
\begin{gather*}
\left\{J^{\nu}(X),J^{\mu}(Y)\right\}=
\left.
\left\{J^{\nu}(X),J^{\mu}(Y)\right\}\right|_{\cal K}
\\
\qquad{}+\sum_{s\geq0}\left.
\left[
{\delta\over\delta\varphi^{k}(\bm{\theta},W)}
\int_{0}^{2\pi}\cdots\int_{0}^{2\pi}
A^{\nu\mu}_{s}\left(\bm{\varphi}(\bm{\theta}^{\prime},X),
\epsilon\bm{\varphi}_{X}(\bm{\theta}^{\prime},X),\dots
\right){{\mathrm d}^{m}\theta^{\prime}\over(2\pi)^{m}}\right]
\right|_{\cal K}\\
\hphantom{\quad{}+\sum_{s\geq0}}\quad {}\times g^{k}(\bm{\theta},W)
{{\mathrm d}^{m}\theta\over(2\pi)^{m}}{\mathrm d}W \epsilon^{s}
\delta^{(s)}(X - Y)+O\big({\bf g}^{2}\big),
\end{gather*}
where all the values on the submanifold ${\cal K}$ are calculated
at the same values of the functionals
$[{\bf U}(Z)]$ as for the original function
$\bm{\varphi}(\bm{\theta},X)$.

Let us introduce by def\/inition
\begin{gather}
\left.
{\delta\{J_{[{\bf q}]},J_{[{\bf p}]}\}\over
\delta g^{k}(\bm{\theta},W)}\right|_{\cal K}\label{JqJpVarDeriv}\\
{}\equiv
\sum_{s\geq0}\int\!
\Bigg[
{\delta\over\delta\varphi^{k}(\bm{\theta},W)}
\!\int_{0}^{2\pi}\!\!\!\cdots\!\int_{0}^{2\pi}\!\! \!\epsilon^{s}
A^{\nu\mu}_{s}\left(\bm{\varphi}(\bm{\theta}^{\prime},X),
\epsilon\bm{\varphi}_{X}(\bm{\theta}^{\prime},X),\dots\!
\right)\!{{\mathrm d}^{m}\theta^{\prime}\over(2\pi)^{m}}\Bigg]
\Bigg|_{\cal K}\!q_{\nu}(X)p_{\mu,sX}{\mathrm d}X.\nonumber
\end{gather}

Note that the notations $\delta/\delta g^{k}(\bm{\theta},W)$,
generally speaking, are not natural in our situation because of
the dependence of the chosen system of constraints.
Nevertheless,
the preservation of these notations can better clarify the algebraic
structure of the further calculations.

The values def\/ined by~\eqref{JqJpVarDeriv}
can be represented in the form of the graded decompositions at
$\epsilon\rightarrow0$ on the submanifold ${\cal K}$.
By virtue of~\eqref{epsAQRel} it is easy to conclude that the
expansion in $\epsilon$ of the quantities~\eqref{JqJpVarDeriv} begins
with the f\/irst degree in $\epsilon$
\begin{gather}\label{JqJpSkobRazl}
\left.
{\delta\{J_{[{\bf q}]},J_{[{\bf p}]}\}\over
\delta g^{k}(\bm{\theta},W)}\right|_{\cal K}=
\epsilon\left.
{\delta\{J_{[{\bf q}]},J_{[{\bf p}]}\}
\over\delta g^{k}(\bm{\theta},W)}\right|_{{\cal K}[1]}
+\epsilon^{2}
\left.
{\delta\{J_{[{\bf q}]},J_{[{\bf p}]}\}
\over\delta g^{k}(\bm{\theta},W)}\right|_{{\cal K}[2]}
+\cdots.
\end{gather}

The leading term of~\eqref{JqJpSkobRazl} can be divided into two parts,
corresponding to the functions \linebreak
$\epsilon Q^{\nu\mu}_{X}(\bm{\varphi},\epsilon
\bm{\varphi}_{X},\dots)$ and
$A^{\nu\mu}_{1}(\bm{\varphi},\epsilon\bm{\varphi}_{X},\dots)$
and containing the quantities $q_{\nu}(W)p_{\mu}(W)$ and \linebreak
$q_{\nu}(W)p_{\mu,W}(W)$ as local factors, respectively.

The values of $\{J_{[{\bf q}]},J_{[{\bf p}]}\}$, are obviously
invariant under transformations of the form
\begin{gather}\label{DeltaSdvig}
\bm{\varphi}(\bm{\theta},X)\rightarrow
\bm{\varphi}(\bm{\theta}+\Delta\bm{\theta},X).
\end{gather}

From the invariance of the functionals ${\bf U}(X)$ in
such transformations, we can write for the corresponding
increments of the constraints~\eqref{gconstr}
\begin{gather*}
\delta g^{k}(\bm{\theta},X)=
\varphi^{k}(\bm{\theta}+\delta\bm{\theta},X)-
\varphi^{k}(\bm{\theta},X).
\end{gather*}

As a~consequence, we can write on the submanifold ${\cal K}$
\begin{gather}\label{JqJpOrtPhi1}
\iint_{0}^{2\pi}\cdots\int_{0}^{2\pi}
\left.
{\delta\{J_{[{\bf q}]},J_{[{\bf p}]}\}\over
\delta g^{k}(\bm{\theta},W)}\right|_{\cal K}
\Phi^{k}_{\theta^{\alpha}}\left({{\bf S}(W)\over\epsilon}+\bm{\theta},{\bf
U}(W)\right){{\mathrm d}^{m}\theta\over(2\pi)^{m}}
{\mathrm d}W\equiv0,\\
\alpha=1,\dots,m.\nonumber
\end{gather}

The above relation is satisf\/ied to all orders in $\epsilon$.
By the arbitrariness of the functions $q_{\nu}(X)$, the relation
\eqref{JqJpOrtPhi1} in the leading order can be strengthened.
Namely, according to the remark about the form of the leading term of
\eqref{JqJpSkobRazl} we can write for any $W$
\begin{gather}\label{JqJpVarDerOrtRel}
\int_{0}^{2\pi}\!\cdots\int_{0}^{2\pi} \!
\left.
{\delta\{J_{[{\bf q}]},J_{[{\bf p}]}\}\over
\delta g^{k}(\bm{\theta},W)}\right|_{{\cal K}[1]}\!
\Phi^{k}_{\theta^{\alpha}}\!\left({{\bf S}(W)\over\epsilon}+\bm{\theta},{\bf
U}(W)\right){{\mathrm d}^{m}\theta\over(2\pi)^{m}}
\equiv0,\qquad
\alpha=1,\dots,m.\!\!\!
\end{gather}

Quite similarly, we have the relation
\begin{gather*}
\left\{ g_{[{\bf Q}]}, J_{[{\bf q}]} \right\}=
\int Q_{i} \left( {{\bf S}(X) \over \epsilon} +\bm{\theta}, X \right)
\left\{ g^{i} (\bm{\theta}, X), J_{[{\bf q}]} \right\}
{{\mathrm d}^{m} \theta \over (2\pi)^{m}}{\mathrm d}X
\\
\qquad{} + \int g^{i} (\bm{\theta}, X) Q_{i,\theta^{\alpha}}
\left( {{\bf S}(X) \over \epsilon} +\bm{\theta}, X \right)
{1 \over 2\epsilon}   \operatorname{sgn} (X - Y)
\left\{ k^{\alpha}(Y), J_{[{\bf q}]} \right\}
{{\mathrm d}^{m} \theta \over (2\pi)^{m}}{\mathrm d}Y{\mathrm d}X
\\
\qquad\quad{} + \int g^{i} (\bm{\theta}, X) Q_{i, U^{\nu}}
\left( {{\bf S}(X) \over \epsilon} +\bm{\theta}, X \right)
\left\{ U^{\nu}(X), J_{[{\bf q}]} \right\}
{{\mathrm d}^{m} \theta \over (2\pi)^{m}}{\mathrm d}X
\\
\qquad{} =\int
Q_{i} \left( {{\bf S}(X) \over \epsilon} +\bm{\theta}, X \right)
\sum_{l\geq0} \epsilon^{l} C^{i\mu}_{(l)} \left(
\bm{\varphi} (\bm{\theta}, X), \epsilon \bm{\varphi}_{X}
(\bm{\theta}, X), \dots \right) {{\mathrm d}^{m} \theta \over (2\pi)^{m}}
q_{\mu, lX}(X){\mathrm d}X \\
\qquad  {} - \int
Q_{i} \left( {{\bf S}(X) \over \epsilon} +\bm{\theta}, X \right)
\Phi^{i}_{U^{\nu}} \left( {{\bf S}(X) \over \epsilon} +\bm{\theta}, {\bf U}(X) \right)
\left\{ U^{\nu}(X), J_{[{\bf q}]} \right\}
{{\mathrm d}^{m} \theta \over (2\pi)^{m}}  {\mathrm d}X
\\
\qquad{} - \int
Q_{i} \left( {{\bf S}(X) \over \epsilon} +\bm{\theta}, X \right)
\Phi^{i}_{\theta^{\alpha}} \left( {{\bf S}(X) \over \epsilon} +\bm{\theta}, {\bf
U}(X) \right)
\\
\qquad\quad{} \times{1 \over 2\epsilon}  \operatorname{sgn} (X - Y)
\left\{ k^{\alpha}(Y), J_{[{\bf q}]} \right\}
{{\mathrm d}^{m} \theta \over (2\pi)^{m}}  {\mathrm d}Y{\mathrm d}X
\\
\qquad{} +\int g^{i} (\bm{\theta}, X) Q_{i,\theta^{\alpha}}
\left( {{\bf S}(X) \over \epsilon} +\bm{\theta}, X \right)
{1 \over 2\epsilon}   \operatorname{sgn} (X - Y)
\left\{ k^{\alpha}(Y), J_{[{\bf q}]} \right\}
{{\mathrm d}^{m} \theta \over (2\pi)^{m}}  {\mathrm d}Y{\mathrm d}X
\\
\qquad{} + \int g^{i} (\bm{\theta}, X) Q_{i, U^{\nu}}
\left( {{\bf S}(X) \over \epsilon} +\bm{\theta}, X \right)
\left\{ U^{\nu}(X), J_{[{\bf q}]} \right\}
{{\mathrm d}^{m} \theta \over (2\pi)^{m}}{\mathrm d}X.
\end{gather*}

According to relations~\eqref{OgranQ}, we can actually see here
that the third term in the above expression is identically equal to zero.

According to the form of the constraints,
we have again in the neighborhood of ${\cal K}$
\begin{gather*}
\{g_{[{\bf Q}]},J_{[{\bf q}]}\}=
\left.
\{g_{[{\bf Q}]},J_{[{\bf q}]}\}\right|_{\cal K}
\\
\left. {} +
\left[\int Q_{i}\left({{\bf S}(X)\over\epsilon}
+\bm{\theta}^{\prime},X\right)\sum_{l\geq0}\epsilon^{l}
{\delta C^{i\mu}_{(l)}(\bm{\varphi}(\bm{\theta}^{\prime},X),
\epsilon\bm{\varphi}_{X}(\bm{\theta}^{\prime},X),\dots)
\over\delta\varphi^{k}(\bm{\theta},W)}
{{\mathrm d}^{m}\theta^{\prime}\over(2\pi)^{m}}q_{\mu,lX}(X){\mathrm d}X
\right]\right|_{\cal K}
\\
\qquad\quad{}\times g^{k}(\bm{\theta},W)
{{\mathrm d}^{m}\theta\over(2\pi)^{m}}{\mathrm d}W
\\
{}-\int Q_{i}(\bm{\theta}^{\prime},X)
\Phi^{i}_{U^{\nu}}\left(\bm{\theta}^{\prime},{\bf U}(X)\right)
{{\mathrm d}^{m}\theta^{\prime}\over(2\pi)^{m}}\left.
{\delta\{U^{\nu}(X),J_{[{\bf q}]}\}
\over\delta\varphi^{k}(\bm{\theta},W)}\right|_{\cal K}{\mathrm d}X
 \times g^{k}(\bm{\theta},W)
{{\mathrm d}^{m}\theta\over(2\pi)^{m}}{\mathrm d}W
\\
{}+\int
Q_{k,\theta^{\alpha}}\left({{\bf S}(W)\over\epsilon}+\bm{\theta},W\right){1\over2\epsilon}
\operatorname{sgn}(W-Y)\left.
\left\{k^{\alpha}(Y),
J_{[{\bf q}]}\right\}\right|_{\cal K}{\mathrm d}Y
 \times g^{k}(\bm{\theta},W)
{{\mathrm d}^{m}\theta\over(2\pi)^{m}}{\mathrm d}W
\\
{}+\int Q_{i,U^{\nu}}
\left({{\bf S}(W)\over\epsilon}+\bm{\theta},W\right)
\left.
\left\{U^{\nu}(W),J_{[{\bf q}]}\right\}
\right|_{\cal K} \times g^{k}(\bm{\theta},W)
{{\mathrm d}^{m}\theta\over(2\pi)^{m}}{\mathrm d}W
+O\big({\bf g}^{2}\big)
\end{gather*}
provided that all the values on the submanifold ${\cal K}$
are calculated at the same values of the functionals $[{\bf U}(Z)]$.

For the quantities $\{g_{[{\bf Q}]},J_{[{\bf q}]}\}$
we can introduce by def\/inition
\begin{gather}
\left.
{\delta\{g_{[{\bf Q}]},J_{[{\bf q}]}\}\over\delta
g^{k}(\bm{\theta},W)}\right|_{\cal K}\nonumber\\
\left.\equiv
\int Q_{i}\left({{\bf S}(X)\over\epsilon}
+\bm{\theta}^{\prime},X\right)
\left[\sum_{l\geq0}\epsilon^{l}
{\delta C^{i\mu}_{(l)}(\bm{\varphi}(\bm{\theta}^{\prime},X),
\epsilon\bm{\varphi}_{X}(\bm{\theta}^{\prime},X),\dots)
\over\delta\varphi^{k}(\bm{\theta},W)} q_{\mu,lX}(X)
\right]\right|_{\cal K}
{{\mathrm d}^{m}\theta^{\prime}\over(2\pi)^{m}}
{\mathrm d}X\nonumber\\
\qquad{} {} -\int Q_{i}(\bm{\theta}^{\prime},X)
\Phi^{i}_{U^{\nu}}\left(\bm{\theta}^{\prime},{\bf U}(X)\right)
{{\mathrm d}^{m}\theta^{\prime}\over(2\pi)^{m}}\left.
{\delta\{U^{\nu}(X),J_{[{\bf q}]}\}
\over\delta\varphi^{k}(\bm{\theta},W)}\right|_{\cal K}{\mathrm d}X
\nonumber\\
\qquad{}
+Q_{k,\theta^{\alpha}}\left({{\bf
S}(W)\over\epsilon}+\bm{\theta},W\right){1\over2\epsilon}
\int\operatorname{sgn}(W-Y)\left.
\left\{k^{\alpha}(Y),
J_{[{\bf q}]}\right\}\right|_{\cal K}{\mathrm d}Y\nonumber\\
\qquad{} +Q_{i,U^{\nu}}
\left({{\bf S}(W)\over\epsilon}+\bm{\theta},W\right)
\left.
\left\{U^{\nu}(W),J_{[{\bf q}]}\right\}
\right|_{\cal K}.\label{ManyPhasegJvarder}
\end{gather}

The quantities
$\delta\{g_{[{\bf Q}]},J_{[{\bf q}]}\}/\delta
g^{k}(\bm{\theta},W)|_{\cal K}$ have the order $O(1)$ at
$\epsilon\rightarrow0$.
We have also
\begin{gather*}
C^{i\mu}_{(0)}\left(\bm{\varphi}(\bm{\theta}^{\prime},X),
\epsilon\bm{\varphi}_{X}(\bm{\theta}^{\prime},X),
\dots\right)\equiv
S^{i\mu}\left(\bm{\varphi}(\bm{\theta}^{\prime},X),
\epsilon\bm{\varphi}_{X}(\bm{\theta}^{\prime},X),
\dots\right)
\end{gather*}
according to relation~\eqref{C0SmuRel}.

Let us introduce the functions
\begin{gather*}
S^{i\mu}_{k(l)}(\bm{\varphi},\bm{\varphi}_{x},\dots)
\equiv{\partial S^{i\mu}(\bm{\varphi},\bm{\varphi}_{x},\dots)
\over\partial\varphi^{k}_{lx}}.
\end{gather*}

Using now Lemma~\ref{Lemma2.3.}
and relations~\eqref{JqJpSkobRazl} we can write
for the leading term of~\eqref{ManyPhasegJvarder}
\begin{gather}
\left.
{\delta \{ g_{[{\bf Q}]}, J_{[{\bf q}]} \} \over \delta
g^{k} (\bm{\theta}, W)} \right|_{{\cal K}[0]}=
\sum_{l\geq0} (-1)^{l} k^{\alpha_{1}}(W) \cdots k^{\alpha_{l}}(W)
{\partial^{l} \over
\partial \theta^{\alpha_{1}} \cdots \partial \theta^{\alpha_{l}}}
\nonumber\\
\hphantom{\left.
{\delta \{ g_{[{\bf Q}]}, J_{[{\bf q}]} \} \over \delta
g^{k} (\bm{\theta}, W)} \right|_{{\cal K}[0]}=} {}\times \left[ Q_{i} \left( {{\bf S}(W) \over \epsilon} +\bm{\theta}, W \right) \left.
S^{i\mu}_{k(l)} \left(
\bm{\varphi} (\bm{\theta}, W), \epsilon
\bm{\varphi}_{W} (\bm{\theta}, W), \dots \right) \right|_{{\cal K}[0]}
\right] q_{\mu} (W)
\nonumber\\
\hphantom{\left.
{\delta \{ g_{[{\bf Q}]}, J_{[{\bf q}]} \} \over \delta
g^{k} (\bm{\theta}, W)} \right|_{{\cal K}[0]}=}{}+ \omega^{\alpha\mu} (W) q_{\mu} (W)
Q_{k,\theta^{\alpha}} \left( {{\bf S}(W) \over \epsilon} +\bm{\theta}, W \right),\label{gJVarDerNulPor}
\end{gather}
where
\begin{gather*}
\left.
S^{i\mu}_{k(l)} \left(
\bm{\varphi} (\bm{\theta}, W), \epsilon
\bm{\varphi}_{W} (\bm{\theta}, W), \dots \right) \right|_{{\cal K}[0]}
  \\
\qquad{}\equiv   S^{i\mu}_{k(l)} \left(
\bm{\Phi} \left( {{\bf S}(W) \over \epsilon} +\bm{\theta}, {\bf U}(W) \right),
k^{\gamma} \bm{\Phi}_{\theta^{\gamma}}
\left( {{\bf S}(W) \over \epsilon} +\bm{\theta}, {\bf U}(W)
\right), \dots \right).
\end{gather*}

Note one more property of the values
$\delta\{g_{[{\bf Q}]},J_{[{\bf q}]}\}/\delta
g^{k}(\bm{\theta},W)|_{{\cal K}[0]}$.
As we saw earlier,
the values $\{g_{[{\bf Q}]},J_{[{\bf q}]}\}$ are of the order
$O(\epsilon)$ at $\epsilon\rightarrow0$ on the submanifold~${\cal K}$.
This property is preserved also under the overall shift
of the initial phase~\eqref{DeltaSdvig}.

Indeed, for
\begin{gather}\label{PhiDeltaTheta}
\varphi^{i}(\bm{\theta},X)=
\Phi^{i}\left({{\bf S}(X)\over\epsilon}+\bm{\theta}+\Delta\bm{\theta},{\bf U}(X)\right)
\end{gather}
we can write
\begin{gather*}
\left\{g_{[{\bf Q}]}, J_{[{\bf q}]} \right\}=
\int Q_{i, \theta^{\alpha}} \left( {{\bf S}(Z) \over \epsilon}
+\bm{\theta}, Z \right) {1 \over 2 \epsilon} \operatorname{sgn}
(Z - W) \\
\hphantom{\left\{g_{[{\bf Q}]}, J_{[{\bf q}]} \right\}=}{} \times \left\{ k^{\alpha}(W), J_{[{\bf q}]} \right\}
\Phi^{i} \left( {{\bf S}(Z) \over \epsilon} +\bm{\theta} +\Delta \bm{\theta}, {\bf U}(Z) \right)
{{\mathrm d}^{m} \theta \over (2\pi)^{m}}{\mathrm d}Z{\mathrm d}W
\\
+\!\int\!\! Q_{i}\!  \left(\! {{\bf S}(Z) \over \epsilon}
+\bm{\theta}, Z \!\right)\!
C^{i\mu}_{(0)} \! \left(\!
\bm{\Phi} \!\left(\! {{\bf S}(Z) \over \epsilon}\! +\bm{\theta} +\Delta \bm{\theta}, {\bf
U} (Z) \!\right)\!, k^{\gamma}
\bm{\Phi}_{\theta^{\gamma}}\!\! \left(\! {{\bf
S}(Z) \over \epsilon}\! +\bm{\theta} +\Delta \bm{\theta}, {\bf U} (Z) \!\right)\!,
\dots \!\right)\!\!
\\
\hphantom{\left\{g_{[{\bf Q}]}, J_{[{\bf q}]} \right\}=}{}\times q_{\mu} (Z) {{\mathrm d}^{m} \theta \over (2\pi)^{m}}{\mathrm d}Z
+ O (\epsilon).
\end{gather*}

Because of the invariance under translations~\eqref{DeltaSdvig}
the value of the bracket $\{k^{\alpha}(W),J_{[{\bf q}]}\}$
on the functions~\eqref{PhiDeltaTheta} is equal to its value on
${\cal K}$
\begin{gather*}
\left\{k^{\alpha}(W),J_{[{\bf q}]}\right\}=
\epsilon\left(\omega^{\alpha\nu}(W)q_{\nu}(W)\right)_{W}
+O\big(\epsilon^{2}\big).
\end{gather*}

Similarly, we have on the functions~\eqref{PhiDeltaTheta}
\begin{gather*}
C^{i\mu}_{(0)} \left(
\bm{\Phi} \left( {{\bf S}(Z) \over \epsilon} +\bm{\theta} +\Delta \bm{\theta}, {\bf U} (Z) \right),
k^{\gamma}
\bm{\Phi}_{\theta^{\gamma}} \left( {{\bf
S}(Z) \over \epsilon} +\bm{\theta} +\Delta \bm{\theta}, {\bf U} (Z) \right) \dots
\right)
\\
\qquad{}=\omega^{\alpha\mu}(Z)
\Phi^{i}_{\theta^{\alpha}} \left( {{\bf
S}(Z) \over \epsilon} +\bm{\theta} +\Delta \bm{\theta}, {\bf U} (Z) \right).
\end{gather*}

We then obtain
\begin{gather*}
\left\{g_{[{\bf Q}]},J_{[{\bf q}]}\right\}=\int
q_{\mu}(Z)\omega^{\alpha\mu}(Z){\partial\over\partial
\theta^{\alpha}}\left[Q_{i}(\bm{\theta},Z)\Phi^{i}
(\bm{\theta}+\Delta\bm{\theta},{\bf U}(Z))\right]
{{\mathrm d}^{m}\theta\over(2\pi)^{m}}{\mathrm d}Z+O(\epsilon)=O(\epsilon).
\end{gather*}

As a~consequence, we can write
\begin{gather}\label{SlabVarDerRel}
\iint_{0}^{2\pi}\cdots\int_{0}^{2\pi}
\left.
{\delta\{g_{[{\bf Q}]},J_{[{\bf q}]}\}\over
\delta g^{k}(\bm{\theta},W)}\right|_{{\cal K}[0]}
\Phi^{k}_{\theta^{\alpha}}\left({{\bf S}(W)\over\epsilon}+\bm{\theta},{\bf
U}(W)\right){{\mathrm d}^{m}\theta\over(2\pi)^{m}}
{\mathrm d}W\equiv0,\\
\alpha=1,\dots,m\nonumber
\end{gather}
for the main part of $\delta\{g_{[{\bf Q}]},J_{[{\bf q}]}\}/
\delta g^{k}(\bm{\theta},W)$ on ${\cal K}$.

Using again the fact that the relation~\eqref{SlabVarDerRel} contains
arbitrary functions $q_{\mu}(W)$, appearing in the integrand
expression in the form of local factors, we can rewrite
\eqref{SlabVarDerRel} in a~stronger form
\begin{gather}\label{SilnVarDerRel}
\int_{0}^{2\pi}\cdots\int_{0}^{2\pi}
\left.
{\delta\{g_{[{\bf Q}]},J_{[{\bf q}]}\}\over
\delta g^{k}(\bm{\theta},W)}\right|_{{\cal K}[0]}
\Phi^{k}_{\theta^{\alpha}}\left({{\bf S}(W)\over\epsilon}+\bm{\theta},{\bf
U}(W)\right){{\mathrm d}^{m}\theta\over(2\pi)^{m}}
\equiv0,\\
\forall\, W,\quad
\alpha=1,\dots,m.\nonumber
\end{gather}

We now turn back to the Jacobi identity~\eqref{JacobigJJ} for
the functionals $g_{[{\bf Q}]}$, $J_{[{\bf q}]}$, and $J_{[{\bf p}]}$.
It is not dif\/f\/icult to see that after the restriction on ${\cal K}$
the leading term (${\sim}\epsilon$) of the relation~\eqref{JacobigJJ}
can be written as
\begin{gather*}
\int
\big\{g_{[{\bf Q}]},g^{k}(\bm{\theta},W)
\big\}\big|_{{\cal K}[0]}\left.
{\delta\{J_{[{\bf q}]},J_{[{\bf p}]}\}\over
\delta g^{k}(\bm{\theta},W)}\right|_{{\cal K}[1]}
{{\mathrm d}^{m}\theta\over(2\pi)^{m}}{\mathrm d}W
\\
\qquad
{}+\int
\big\{J_{[{\bf p}]},
g^{k}(\bm{\theta},W)\big\}\big|_{{\cal K}[1]}\left.
{\delta\{g_{[{\bf Q}]},J_{[{\bf q}]}\}\over
\delta g^{k}(\bm{\theta},W)}\right|_{{\cal K}[0]}
{{\mathrm d}^{m}\theta\over(2\pi)^{m}}{\mathrm d}W
\\
\qquad
{}-\int
\big\{J_{[{\bf q}]},
g^{k}(\bm{\theta},W)\big\}\big|_{{\cal K}[1]}\left.
{\delta\{g_{[{\bf Q}]},J_{[{\bf p}]}\}\over
\delta g^{k}(\bm{\theta},W)}\right|_{{\cal K}[0]}
{{\mathrm d}^{m}\theta\over(2\pi)^{m}}{\mathrm d}W\equiv0.
\end{gather*}

The above identity can again be written in a~stronger form.
Namely, making the change
${\tilde Q}_{i}(\bm{\theta},W)\rightarrow
{\tilde Q}_{i}(\bm{\theta},W)\mu_{i}(W)$,
we get the corresponding change,
$Q_{i}(\bm{\theta},W)\rightarrow Q_{i}(\bm{\theta},W)\mu_{i}(W)$,
where $\mu_{i}(W)$ are arbitrary smooth functions of $W$.
By virtue of~\eqref{OgranQ},~\eqref{JqJpVarDerOrtRel}, and
\eqref{gJVarDerNulPor} it is easy to see then that the integrands
are smooth functions of $\bm{\theta}$ and $W$, containing $\mu_{i}(W)$
in the form of local factors.
By the arbitrariness of $\mu_{i}(W)$,
we can omit the integration over $W$ in the above integrals and
write for every $W$
\begin{gather*}
\int_{0}^{2\pi}\cdots\int_{0}^{2\pi}
\big\{g_{[{\bf Q}]},g^{k}(\bm{\theta},W)
\big\} \big|_{{\cal K}[0]}\left.
{\delta\{J_{[{\bf q}]},J_{[{\bf p}]}\}\over
\delta g^{k}(\bm{\theta},W)}\right|_{{\cal K}[1]}
{{\mathrm d}^{m}\theta\over(2\pi)^{m}}
\\
\qquad
{}+\int_{0}^{2\pi}\cdots\int_{0}^{2\pi}
\big\{J_{[{\bf p}]},
g^{k}(\bm{\theta},W)\big\}\big|_{{\cal K}[1]}\left.
{\delta\{g_{[{\bf Q}]},J_{[{\bf q}]}\}\over
\delta g^{k}(\bm{\theta},W)}\right|_{{\cal K}[0]}
{{\mathrm d}^{m}\theta\over(2\pi)^{m}}
\\
\qquad
{}-\int_{0}^{2\pi}\cdots\int_{0}^{2\pi}
\big\{J_{[{\bf q}]},
g^{k}(\bm{\theta},W)\big\}\big|_{{\cal K}[1]}\left.
{\delta\{g_{[{\bf Q}]},J_{[{\bf p}]}\}\over
\delta g^{k}(\bm{\theta},W)}\right|_{{\cal K}[0]}
{{\mathrm d}^{m}\theta\over(2\pi)^{m}}\equiv0.
\end{gather*}

Finally, using the relations~\eqref{giJmu1skob} and
\eqref{SilnVarDerRel}, we can write the above identity as
\begin{gather*}
\int_{0}^{2\pi}\cdots\int_{0}^{2\pi}
\big\{g_{[{\bf Q}]},g^{k}(\bm{\theta},W)
\big\}\big|_{{\cal K}[0]}\left.
{\delta\{J_{[{\bf q}]},J_{[{\bf p}]}\}\over
\delta g^{k}(\bm{\theta},W)}\right|_{{\cal K}[1]}
{{\mathrm d}^{m}\theta\over(2\pi)^{m}}
\\
\qquad
{}-\int_{0}^{2\pi}\cdots\int_{0}^{2\pi}
A^{k}_{[1][{\bf p}]}\left({{\bf S}(W)\over\epsilon}+\bm{\theta},W\right)\left.
{\delta\{g_{[{\bf Q}]},J_{[{\bf q}]}\}\over
\delta g^{k}(\bm{\theta},W)}\right|_{{\cal K}[0]}
{{\mathrm d}^{m}\theta\over(2\pi)^{m}}
\\
\qquad
{}+\int_{0}^{2\pi}\cdots\int_{0}^{2\pi}
A^{k}_{[1][{\bf q}]}\left({{\bf S}(W)\over\epsilon}+\bm{\theta},W\right)\left.
{\delta\{g_{[{\bf Q}]},J_{[{\bf p}]}\}\over
\delta g^{k}(\bm{\theta},W)}\right|_{{\cal K}[0]}
{{\mathrm d}^{m}\theta\over(2\pi)^{m}}\equiv0,
\end{gather*}
where the functions $A^{k}_{[1][{\bf q}]}(\bm{\theta},W)$ are
introduced by formula~\eqref{FunktsiiA}.

Now assume that the values
$A^{k}_{[1][{\bf q}]}(\bm{\theta},W)$ for
${\bf U}(W)\in{\cal S}$, according to~\eqref{BAcondNextChapt},
can be represented in the form
\begin{gather}\label{ABBPredstav}
A^{k}_{[1][{\bf q}]}(\bm{\theta},W)=-
{\hat B}^{kj}_{[0]}(W) B_{j[{\bf q}](1)}(\bm{\theta},W)
\end{gather}
with some smooth in $\bm{\theta}$, $2\pi$-periodic in each
$\theta^{\alpha}$ functions $B_{j[{\bf q}](1)}(\bm{\theta},W)$.

We can then write for ${\bf U}(W)\in{\cal S}$
\begin{gather}\label{TozhdgQJqJp}
\int_{0}^{2\pi}\cdots\int_{0}^{2\pi}
\big\{g_{[{\bf Q}]},g^{k}(\bm{\theta},W)
\big\}\big|_{{\cal K}[0]}\left.
{\delta\{J_{[{\bf q}]},J_{[{\bf p}]}\}\over
\delta g^{k}(\bm{\theta},W)}\right|_{{\cal K}[1]}
{{\mathrm d}^{m}\theta\over(2\pi)^{m}} \\
\qquad{} +\int_{0}^{2\pi}\cdots\int_{0}^{2\pi}
\left[{\hat B}^{kj}_{[0][{\bf S}]}(W)
B_{j[{\bf p}](1)}\left({{\bf S}(W)\over\epsilon}+\bm{\theta},
W\right)\right] \left.
{\delta\{g_{[{\bf Q}]},J_{[{\bf q}]}\}\over
\delta g^{k}(\bm{\theta},W)}\right|_{{\cal K}[0]}
{{\mathrm d}^{m}\theta\over(2\pi)^{m}}\nonumber\\
\qquad{} -\int_{0}^{2\pi}\cdots\int_{0}^{2\pi}
\left[{\hat B}^{kj}_{[0][{\bf S}]}(W)
B_{j[{\bf q}](1)}\left({{\bf S}(W)\over\epsilon}+\bm{\theta},
W\right)\right] \left.
{\delta\{g_{[{\bf Q}]},J_{[{\bf p}]}\}\over
\delta g^{k}(\bm{\theta},W)}\right|_{{\cal K}[0]}
{{\mathrm d}^{m}\theta\over(2\pi)^{m}}\equiv0,
\nonumber
\end{gather}
where
\begin{gather*}
{\hat B}^{ij}_{[0][{\bf S}]} (W) \equiv  \sum_{s\geq0}
B^{ij}_{(s)} \left( \bm{\Phi} \left( {{\bf S}(W) \over \epsilon} +\bm{\theta}, {\bf U}(W)\right),
k^{\gamma}(W) \bm{\Phi}_{\theta^{\gamma}} \left(
{{\bf S}(W) \over \epsilon} +\bm{\theta}, {\bf U}(W)\right),
\dots \right)
\\
\hphantom{{\hat B}^{ij}_{[0][{\bf S}]} (W) \equiv}{}
\times k^{{\alpha}_{1}}(W) \cdots k^{{\alpha}_{s}}(W)
{\partial^{s} \over \partial \theta^{{\alpha}_{1}} \cdots
\partial \theta^{{\alpha}_{s}}}.
\end{gather*}
Using the expression~\eqref{SkobSv} for the bracket of constraints
on ${\cal K}$, as well as the relations~\eqref{QphiUbr}, \eqref{OgranQ},
\eqref{JqJpVarDerOrtRel}, we can obtain also the following relation
\begin{gather*}
\int_{0}^{2\pi}\cdots\int_{0}^{2\pi}
\big\{ g_{[{\bf Q}]}, g^{k} (\bm{\theta}, W)
\big\} \big|_{{\cal K}[0]} \left.
{\delta \{ J_{[{\bf q}]}, J_{[{\bf p}]} \} \over
\delta g^{k} (\bm{\theta},W)} \right|_{{\cal K}[1]}
{{\mathrm d}^{m} \theta \over (2\pi)^{m}}
\\
\qquad{} =- \int_{0}^{2\pi}\cdots\int_{0}^{2\pi}
\left[ {\hat B}^{kj}_{[0][{\bf S}]} (W)
Q_{j} \left({{\bf S}(W) \over \epsilon} +\bm{\theta},
W \right) \right] \left.
{\delta \{ J_{[{\bf q}]}, J_{[{\bf p}]} \} \over
\delta g^{k} (\bm{\theta},W)} \right|_{{\cal K}[1]}
{{\mathrm d}^{m} \theta \over (2\pi)^{m}}.
\end{gather*}

Expanding also the remaining terms of the identity
\eqref{TozhdgQJqJp} according to~\eqref{gJVarDerNulPor},
we get for ${\bf U}(W)\in{\cal S}$
\begin{gather}
\int_{0}^{2\pi}\cdots\int_{0}^{2\pi}
\left[{\hat B}^{kj}_{[0][{\bf S}]}(W)
Q_{j}\left({{\bf S}(W)\over\epsilon}+\bm{\theta},
W\right)\right]\left.
{\delta\{J_{[{\bf q}]},J_{[{\bf p}]}\}\over
\delta g^{k}(\bm{\theta},W)}\right|_{{\cal K}[1]}
{{\mathrm d}^{m}\theta\over(2\pi)^{m}}\nonumber\\
\qquad{} -\int_{0}^{2\pi}\cdots\int_{0}^{2\pi}
{{\mathrm d}^{m}\theta\over(2\pi)^{m}}
\left[{\hat B}^{kj}_{[0][{\bf S}]}(W)
B_{j[{\bf p}](1)}\left({{\bf S}(W)\over\epsilon}+\bm{\theta},
W\right)\right] \nonumber\\
\qquad{} \times\Bigg[
\sum_{l\geq0}(-1)^{l}k^{\alpha_{1}}(W)\cdots k^{\alpha_{l}}(W)
{\partial^{l}\over
\partial\theta^{\alpha_{1}}\cdots\partial\theta^{\alpha_{l}}}
\nonumber\\
\qquad\qquad{} \times\left[Q_{i}\left({{\bf S}(W)\over\epsilon}+\bm{\theta},W\right)\left.
S^{i\mu}_{k(l)}\left(
\bm{\varphi}(\bm{\theta},W),\dots\right)\right|_{{\cal K}[0]}
\right]q_{\mu}(W)\nonumber\\
 \qquad\qquad\qquad
{}+\omega^{\alpha\mu}(W)
q_{\mu}(W)Q_{k,\theta^{\alpha}}
\left({{\bf S}(W)\over\epsilon}+\bm{\theta},
W\right)\Bigg]\nonumber\\
\qquad{} +\int_{0}^{2\pi}\cdots\int_{0}^{2\pi}
{{\mathrm d}^{m}\theta\over(2\pi)^{m}}
\left[{\hat B}^{kj}_{[0][{\bf S}]}(W)
B_{j[{\bf q}](1)}\left({{\bf S}(W)\over\epsilon}+\bm{\theta},
W\right)\right]\nonumber\\
\qquad{} \times\Bigg[
\sum_{l\geq0}(-1)^{l}k^{\alpha_{1}}(W)\cdots k^{\alpha_{l}}(W)
{\partial^{l}\over
\partial\theta^{\alpha_{1}}\cdots\partial\theta^{\alpha_{l}}}
\nonumber\\
\qquad\qquad{}\times\left[Q_{i}\left({{\bf S}(W)\over\epsilon}+\bm{\theta},W\right)\left.
S^{i\mu}_{k(l)}\left(
\bm{\varphi}(\bm{\theta},W),\dots\right)\right|_{{\cal K}[0]}
\right]p_{\mu}(W)\nonumber\\
\qquad\qquad\qquad
{}
+\omega^{\alpha\mu}(W)
p_{\mu}(W)Q_{k,\theta^{\alpha}}
\left({{\bf S}(W)\over\epsilon}+\bm{\theta},
W\right)\Bigg]\equiv0.\label{JacobigJJmain}
\end{gather}

Consider now the Jacobi identity of the form
\begin{gather}\label{JacobiggJ}
\left\{g_{[{\bf P}]},\left\{g_{[{\bf Q}]},
J_{[{\bf q}]}\right\}\right\}+\left\{g_{[{\bf Q}]},\left\{J_{[{\bf q}]},
g_{[{\bf P}]}\right\}\right\}+\left\{J_{[{\bf q}]},\left\{g_{[{\bf P}]},
g_{[{\bf Q}]}\right\}\right\}\equiv0
\end{gather}
for arbitrary f\/ixed functions ${\bf q}(X)$, and the functionals
${\bf P}(\bm{\theta},X)$ and ${\bf Q}(\bm{\theta},X)$
def\/ined as before with the aid of arbitrary functions
${\tilde{\bf P}}(\bm{\theta},X)$, ${\tilde{\bf Q}}(\bm{\theta},X)$.

According to the relations
\begin{gather*}
\big\{g_{[{\bf Q},{\bf P}]},
U^{\mu}(W)\big\}\big|_{\cal K}=O(\epsilon)
,\qquad
\big\{J_{[{\bf q}]},g^{k}(\bm{\theta},W)
\big\} \big|_{\cal K}=O(\epsilon)
,\qquad
\big\{J_{[{\bf q}]},
U^{\mu}(W)\big\}\big|_{\cal K}=O(\epsilon),
\end{gather*}
it's not dif\/f\/icult to see that after the restriction on ${\cal K}$
the major term (in $\epsilon$) of~\eqref{JacobiggJ} will be written
as
\begin{gather*}
\int
\big\{g_{[{\bf P}]},g^{k}(\bm{\theta},W)
\big\}\big|_{{\cal K}[0]}\left.
{\delta\{g_{[{\bf Q}]},J_{[{\bf q}]}\}\over
\delta g^{k}(\bm{\theta},W)}\right|_{{\cal K}[0]}
{{\mathrm d}^{m}\theta\over(2\pi)^{m}}{\mathrm d}W
\\
\qquad
{}-\int
\big\{g_{[{\bf Q}]},
g^{k}(\bm{\theta},W)\big\}\big|_{{\cal K}[0]}\left.
{\delta\{g_{[{\bf P}]},J_{[{\bf q}]}\}\over
\delta g^{k}(\bm{\theta},W)}\right|_{{\cal K}[0]}
{{\mathrm d}^{m}\theta\over(2\pi)^{m}}{\mathrm d}W\equiv0.
\end{gather*}

Again recalling that $q_{\nu}(W)$ are arbitrary functions of~$W$
appearing in the integrand in the form of local factors,
we can write the above relation in a~stronger form.
That is, for every $W$
\begin{gather}
\int_{0}^{2\pi}\cdots\int_{0}^{2\pi}
\big\{g_{[{\bf P}]},g^{k}(\bm{\theta},W)
\big\}\big|_{{\cal K}[0]}\left.
{\delta\{g_{[{\bf Q}]},J_{[{\bf q}]}\}\over
\delta g^{k}(\bm{\theta},W)}\right|_{{\cal K}[0]}
{{\mathrm d}^{m}\theta\over(2\pi)^{m}} \nonumber\\
\qquad
{}-\int_{0}^{2\pi}\cdots\int_{0}^{2\pi}
\big\{g_{[{\bf Q}]},
g^{k}(\bm{\theta},W)\big\}\big|_{{\cal K}[0]}\left.
{\delta\{g_{[{\bf P}]},J_{[{\bf q}]}\}\over
\delta g^{k}(\bm{\theta},W)}\right|_{{\cal K}[0]}
{{\mathrm d}^{m}\theta\over(2\pi)^{m}}\equiv0.\label{JacobiggJmain}
\end{gather}

As well as in the case of the identity~\eqref{JacobigJJmain},
using the relations~\eqref{SkobSv},~\eqref{QphiUbr},~\eqref{OgranQ},
\eqref{SilnVarDerRel}, we can write the identity
\eqref{JacobiggJmain} in the form
\begin{gather}
\int_{0}^{2\pi}\cdots\int_{0}^{2\pi}
{{\mathrm d}^{m}\theta\over(2\pi)^{m}}
\left[{\hat B}^{kj}_{[0][{\bf S}]}(W)
P_{j}\left({{\bf S}(W)\over\epsilon}+\bm{\theta},
W\right)\right]\nonumber\\
\qquad{}\times\Bigg[
\sum_{l\geq0}(-1)^{l}k^{\alpha_{1}}(W)\cdots k^{\alpha_{l}}(W)
{\partial^{l}\over
\partial\theta^{\alpha_{1}}\cdots\partial\theta^{\alpha_{l}}}
\nonumber\\
\qquad\qquad{}\times
\left[Q_{i}\left({{\bf S}(W)\over\epsilon}+\bm{\theta},W\right)\left.
S^{i\mu}_{k(l)}\left(
\bm{\varphi}(\bm{\theta},W),\dots\right)\right|_{{\cal K}[0]}
\right]q_{\mu}(W)\nonumber\\
\qquad\qquad\qquad
{}+\omega^{\alpha\mu}(W)
q_{\mu}(W)Q_{k,\theta^{\alpha}}
\left({{\bf S}(W)\over\epsilon}+\bm{\theta},
W\right)\Bigg]\nonumber\\
\qquad{} -\int_{0}^{2\pi}\cdots\int_{0}^{2\pi}
{{\mathrm d}^{m}\theta\over(2\pi)^{m}}
\left[{\hat B}^{kj}_{[0][{\bf S}]}(W)
Q_{j}\left({{\bf S}(W)\over\epsilon}+\bm{\theta},
W\right)\right] \nonumber\\
\qquad{} \times\Bigg[
\sum_{l\geq0}(-1)^{l}k^{\alpha_{1}}(W)\cdots k^{\alpha_{l}}(W)
{\partial^{l}\over
\partial\theta^{\alpha_{1}}\cdots\partial\theta^{\alpha_{l}}}
\nonumber\\
\qquad\qquad{} \times\left[P_{i}\left({{\bf S}(W)\over\epsilon}+\bm{\theta},W\right)\left.
S^{i\mu}_{k(l)}\left(
\bm{\varphi}(\bm{\theta},W),\dots\right)\right|_{{\cal K}[0]}
\right]q_{\mu}(W)\nonumber\\
\qquad\qquad\qquad
{}
+\omega^{\alpha\mu}(W)
q_{\mu}(W)P_{k,\theta^{\alpha}}
\left({{\bf S}(W)\over\epsilon}+\bm{\theta},
W\right)\Bigg]\equiv0.\label{ggJRasp}
\end{gather}

Note now that the values of
${\bf Q}(\bm{\theta},X)$ and ${\bf P}(\bm{\theta},X)$
are arbitrary $2\pi$-periodic functions of $\bm{\theta}$,
satisfying the conditions~\eqref{OgranQ}.
In particular, we can put in~\eqref{ggJRasp} at
${\bf U}(W)\in{\cal S}$
\begin{gather}\label{PBPB}
{\bf P}(\bm{\theta},W)=
{\bf B}_{[{\bf p}](1)}(\bm{\theta},W)
\qquad\text{or}\qquad
{\bf P}(\bm{\theta},W)=
{\bf B}_{[{\bf q}](1)}(\bm{\theta},W).
\end{gather}

By analogy with~\eqref{gJVarDerNulPor} we introduce for convenience
the notation for ${\bf U}(W)\in{\cal S}$
\begin{gather}
\left.
{\delta\{g_{[\epsilon{\bf B}_{[{\bf p}](1)}]},
J_{[{\bf q}]}\}\over\delta g^{k}(\bm{\theta},W)}
\right|_{{\cal K}[1]}
\equiv\Bigg[
\sum_{l\geq0}(-1)^{l}k^{\alpha_{1}}(W)\cdots k^{\alpha_{l}}(W)
{\partial^{l}\over
\partial\theta^{\alpha_{1}}\cdots\partial\theta^{\alpha_{l}}}
\nonumber\\
\hphantom{\left.
{\delta\{g_{[\epsilon{\bf B}_{[{\bf p}](1)}]},
J_{[{\bf q}]}\}\over\delta g^{k}(\bm{\theta},W)}
\right|_{{\cal K}[1]}
\equiv}{}
\times
\left[B_{i[{\bf p}](1)}\left({{\bf S}(W)\over\epsilon}
+\bm{\theta},W\right)\left.
S^{i\mu}_{k(l)}\left(
\bm{\varphi}(\bm{\theta},W),\dots\right)\right|_{{\cal K}[0]}
\right]q_{\mu}(W)
\nonumber\\
\hphantom{\left.
{\delta\{g_{[\epsilon{\bf B}_{[{\bf p}](1)}]},
J_{[{\bf q}]}\}\over\delta g^{k}(\bm{\theta},W)}
\right|_{{\cal K}[1]}
\equiv}{}
 +\omega^{\alpha\mu}(W)
q_{\mu}(W)B_{k[{\bf p}](1),\theta^{\alpha}}
\left({{\bf S}(W)\over\epsilon}+\bm{\theta},
W\right)\Bigg]\label{gBJqVarDerDefinition}
\end{gather}
for arbitrary smooth functions ${\bf q}(X)$, ${\bf p}(X)$.

Note that the functional $g_{[\epsilon{\bf B}_{[{\bf p}](1)}]}$ is
not def\/ined on the whole functional space, so the rela\-tion~\eqref{gBJqVarDerDefinition} plays just a~role of a~formal notation
for ${\bf U}(W)\in{\cal S}$.

Using now~\eqref{ggJRasp} for the functions~\eqref{PBPB},
we can rewrite~\eqref{JacobigJJmain} in the form
\begin{gather*}
\int_{0}^{2\pi}\cdots\int_{0}^{2\pi}
{{\mathrm d}^{m} \theta \over (2\pi)^{m}}
\left[ {\hat B}^{kj}_{[0][{\bf S}]} (W)
Q_{j} \left({{\bf S}(W) \over \epsilon} +\bm{\theta},
W \right) \right]   \\
\qquad{} \times\left( \left.
{\delta \{ J_{[{\bf q}]}, J_{[{\bf p}]} \} \over
\delta g^{k} (\bm{\theta},W)} \right|_{{\cal K}[1]} -
\left.
{\delta \{ g_{[\epsilon {\bf B}_{[{\bf p}](1)}]},
J_{[{\bf q}]} \} \over \delta g^{k} (\bm{\theta},W)}
\right|_{{\cal K}[1]} +\left.
{\delta \{ g_{[\epsilon {\bf B}_{[{\bf q}](1)}]},
J_{[{\bf p}]} \} \over \delta g^{k} (\bm{\theta},W)}
\right|_{{\cal K}[1]} \right) \equiv   0
\end{gather*}
provided that ${\bf U}(W)\in{\cal S}$.

Using the skew-symmetry of the Hamiltonian operator
${\hat B}^{kj}_{[0][{\bf S}]}(W)$
on $\Lambda$ we can then write
\begin{gather*}
\int_{0}^{2\pi}\cdots\int_{0}^{2\pi}
{{\mathrm d}^{m} \theta \over (2\pi)^{m}}
Q_{j} \left( {{\bf S}(W) \over \epsilon} +\bm{\theta}, W \right)
\\
\times\Bigg[ {\hat B}^{jk}_{[0][{\bf S}]} (W)
\Bigg( \left.
{\delta \{ J_{[{\bf q}]}, J_{[{\bf p}]} \} \over
\delta g^{k} (\bm{\theta},W)} \right|_{{\cal K}[1]}
 \left.  -
{\delta \{ g_{[\epsilon {\bf B}_{[{\bf p}](1)}]},
J_{[{\bf q}]} \} \over \delta g^{k} (\bm{\theta},W)}
\right|_{{\cal K}[1]} +\left.
{\delta \{ g_{[\epsilon {\bf B}_{[{\bf q}](1)}]},
J_{[{\bf p}]} \} \over \delta g^{k} (\bm{\theta},W)}
\right|_{{\cal K}[1]} \Bigg) \Bigg] \equiv  0.
\end{gather*}

The values $Q_{j}(\bm{\theta},W)$ are arbitrary smooth
$2\pi$-periodic functions of $\bm{\theta}$ with the only
restriction~\eqref{OgranQ}.
We know also that the values in the
brackets are smooth $2\pi$-periodic in each $\theta^{\alpha}$
functions of $\bm{\theta}$ for ${\bf U}(W)\in{\cal S}$.
As a~consequence, we can write for ${\bf U}(W)\in{\cal S}$
\begin{gather*}
{\hat B}^{jk}_{[0][{\bf S}]} (W) \left( \left.
{\delta \{ J_{[{\bf q}]}, J_{[{\bf p}]} \} \over
\delta g^{k} (\bm{\theta},W)} \right|_{{\cal K}[1]} -
\left.
{\delta \{ g_{[\epsilon {\bf B}_{[{\bf p}](1)}]},
J_{[{\bf q}]} \} \over \delta g^{k} (\bm{\theta},W)}
\right|_{{\cal K}[1]} +\left.
{\delta \{ g_{[\epsilon {\bf B}_{[{\bf q}](1)}]},
J_{[{\bf p}]} \} \over \delta g^{k} (\bm{\theta},W)}
\right|_{{\cal K}[1]} \right)
\\
\qquad{} \equiv \sum_{\alpha=1}^{m} a^{\alpha} ({\bf U}(W), {\bf U}_{W}(W))
\Phi^{j}_{\theta^{\alpha}} \left(
{{\bf S}(W) \over \epsilon} +\bm{\theta}, {\bf U}(W) \right)
\end{gather*}
with some coef\/f\/icients $a^{\alpha}({\bf U}(W),{\bf U}_{W}(W))$.

The values in parentheses are smooth $2\pi$-periodic
functions of $\bm{\theta}$ at ${\bf U}(W)\in{\cal S}$.
At the same time the trajectories of the vector f\/ield
$(k^{1}(W),\dots,k^{m}(W))$ are completely irrational windings
of the torus $\mathbb{T}^{m}$.
We can therefore say in the case of a
regular Hamiltonian family $\Lambda$, that up to a~linear combination of
the regular eigenvectors
${\bf v}^{(l)}_{[{\bf U}(W)]}({\bf S}(W)/\epsilon+\bm{\theta})$
of ${\hat B}^{jk}_{[0][{\bf S}]}(W)
={\hat B}^{jk}_{[0][{\bf S}]}({\bf U}(W))$,
corresponding to zero eigenvalues,
the value in parentheses is a~linear combination of the
variational derivatives~\eqref{VarDer}, generating linear shifts
of the phases on $\Lambda$.
For complete Hamiltonian set of the
integrals $(I^{1},\dots,I^{N})$ we can then write according to
\eqref{SviazZetaKappa} and~\eqref{Sootndliavikappa}
\begin{gather}
\left.
{\delta \{ J_{[{\bf q}]}, J_{[{\bf p}]} \} \over
\delta g^{k} (\bm{\theta},W)} \right|_{{\cal K}[1]} -
\left.
{\delta \{ g_{[\epsilon {\bf B}_{[{\bf p}](1)}]},
J_{[{\bf q}]} \} \over \delta g^{k} (\bm{\theta},W)}
\right|_{{\cal K}[1]} +\left.
{\delta \{ g_{[\epsilon {\bf B}_{[{\bf q}](1)}]},
J_{[{\bf p}]} \} \over \delta g^{k} (\bm{\theta},W)}
\right|_{{\cal K}[1]}
\nonumber\\
\qquad{} \equiv \sum_{q} b_{q} ({\bf U}(W), {\bf U}_{W}(W))
\kappa^{(q)}_{k[{\bf U}(W)]}
\left( {{\bf S}(W) \over \epsilon} +\bm{\theta}\right)\label{PredstavVarDer}
\end{gather}
with some coef\/f\/icients $b_{q}({\bf U}(W),{\bf U}_{W}(W))$ at
${\bf U}(W)\in{\cal S}$.

Consider the Jacobi identity of the form
\begin{gather*}
\left\{\left\{J_{[{\bf q}]},J_{[{\bf p}]}\right\},
J_{[{\bf r}]}\right\}+\left\{\left\{J_{[{\bf p}]},J_{[{\bf r}]}\right\},
J_{[{\bf q}]}\right\}+\left\{\left\{J_{[{\bf r}]},J_{[{\bf q}]}\right\},
J_{[{\bf p}]}\right\}\equiv0
\end{gather*}
with arbitrary smooth functions
${\bf q}(X)$, ${\bf p}(X)$, ${\bf r}(X)$.

In the main (${\sim}\epsilon^{2}$) order on ${\cal K}$ given
identity leads to the relations
\begin{gather}
\int\left.
{\delta\{J_{[{\bf q}]},J_{[{\bf p}]}\}\over
\delta U^{\gamma}(W)}\right|_{{\cal K}[1]}
\big\{U^{\gamma}(W),J_{[{\bf r}]}\big\}
\big|_{{\cal K}[1]}{\mathrm d}W+\text{c.p.}\nonumber
\\
\qquad
{}+\iint_{0}^{2\pi}\cdots\int_{0}^{2\pi}
\left.
{\delta\{J_{[{\bf q}]},J_{[{\bf p}]}\}\over
\delta g^{k}(\bm{\theta},W)}\right|_{{\cal K}[1]}
\big\{g^{k}(\bm{\theta},W),J_{[{\bf r}]}\big\}
\big|_{{\cal K}[1]}{{\mathrm d}^{m}\theta\over(2\pi)^{m}}{\mathrm d}W
+\text{c.p.}
\equiv0.\label{JJJskobrazl}
\end{gather}

Again, using the relations~\eqref{giJmu1skob},~\eqref{FunktsiiA} and
\eqref{JqJpVarDerOrtRel}, we can replace the identity~\eqref{JJJskobrazl}
by the following relation
\begin{gather*}
\int\left.
{\delta\{J_{[{\bf q}]},J_{[{\bf p}]}\}\over
\delta U^{\gamma}(W)}\right|_{{\cal K}[1]}
\big\{U^{\gamma}(W),J_{[{\bf r}]}\big\}
\big|_{{\cal K}[1]}{\mathrm d}W+\text{c.p.}
\\
\qquad
{}+\iint_{0}^{2\pi}\cdots\int_{0}^{2\pi}
\left.
{\delta\{J_{[{\bf q}]},J_{[{\bf p}]}\}\over
\delta g^{k}(\bm{\theta},W)}\right|_{{\cal K}[1]}
A^{k}_{[1][{\bf r}]}\left(
{{\bf S}(W)\over\epsilon}+\bm{\theta},W\right)
{{\mathrm d}^{m}\theta\over(2\pi)^{m}}{\mathrm d}W
+\text{c.p.}
\equiv0.
\end{gather*}

Using relations~\eqref{LeftVecAort} and representations
\eqref{ABBPredstav} and~\eqref{PredstavVarDer}, we can write for
${\bf U}(W)\in{\cal S}$
\begin{gather}
\int_{0}^{2\pi}\cdots\int_{0}^{2\pi}
\left.
{\delta\{J_{[{\bf q}]},J_{[{\bf p}]}\}\over
\delta g^{k}(\bm{\theta},W)}\right|_{{\cal K}[1]}
A^{k}_{[1][{\bf r}]}\left(
{{\bf S}(W)\over\epsilon}+\bm{\theta},W\right)
{{\mathrm d}^{m}\theta\over(2\pi)^{m}}
+\text{c.p.}\nonumber
 \\
\qquad{} \equiv\int_{0}^{2\pi}\cdots\int_{0}^{2\pi}
\left[{\hat B}^{kj}_{[0][{\bf S}]}(W)B_{j[{\bf r}](1)}
\left({{\bf S}(W)\over\epsilon}+\bm{\theta},
W\right)\right] \nonumber\\
\qquad\qquad\qquad{} \times\left[
\left.
{\delta\{g_{[\epsilon{\bf B}_{[{\bf q}](1)}]},
J_{[{\bf p}]}\}\over\delta g^{k}(\bm{\theta},W)}
\right|_{{\cal K}[1]}-
\left.
{\delta\{g_{[\epsilon{\bf B}_{[{\bf p}](1)}]},
J_{[{\bf q}]}\}\over\delta g^{k}(\bm{\theta},W)}
\right|_{{\cal K}[1]}\right]{{\mathrm d}^{m}\theta\over(2\pi)^{m}}
\nonumber\\
\quad\qquad{} +\int_{0}^{2\pi}\cdots\int_{0}^{2\pi}
\left[{\hat B}^{kj}_{[0][{\bf S}]}(W)B_{j[{\bf q}](1)}
\left({{\bf S}(W)\over\epsilon}+\bm{\theta},
W\right)\right] \nonumber\\
\qquad \qquad\qquad{}\times\left[
\left.
{\delta\{g_{[\epsilon{\bf B}_{[{\bf p}](1)}]},
J_{[{\bf r}]}\}\over\delta g^{k}(\bm{\theta},W)}
\right|_{{\cal K}[1]}-
\left.
{\delta\{g_{[\epsilon{\bf B}_{[{\bf r}](1)}]},
J_{[{\bf p}]}\}\over\delta g^{k}(\bm{\theta},W)}
\right|_{{\cal K}[1]}\right]{{\mathrm d}^{m}\theta\over(2\pi)^{m}}
\nonumber\\
\quad \qquad{} +\int_{0}^{2\pi}\cdots\int_{0}^{2\pi} \left[
{\hat B}^{kj}_{[0][{\bf S}]}(W)B_{j[{\bf p}](1)}
\left({{\bf S}(W)\over\epsilon}+\bm{\theta},
W\right)\right] \nonumber\\
\qquad \qquad\qquad{} \times\left[
\left.
{\delta\{g_{[\epsilon{\bf B}_{[{\bf r}](1)}]},
J_{[{\bf q}]}\}\over\delta g^{k}(\bm{\theta},W)}
\right|_{{\cal K}[1]}-
\left.
{\delta\{g_{[\epsilon{\bf B}_{[{\bf q}](1)}]},
J_{[{\bf r}]}\}\over\delta g^{k}(\bm{\theta},W)}
\right|_{{\cal K}[1]}\right]{{\mathrm d}^{m}\theta\over(2\pi)^{m}}.
\label{gchastTozhdJnak}
\end{gather}

Similarly to earlier arguments, substituting now in the identity \eqref{ggJRasp}
\begin{gather*}
{\bf Q}(\bm{\theta},W)=
{\bf B}_{[{\bf r}](1)}(\bm{\theta},W),\qquad
{\bf P}(\bm{\theta},W)=
{\bf B}_{[{\bf p}](1)}(\bm{\theta},W)
\end{gather*}
for ${\bf U}(W)\in{\cal S}$ we obtain the identity
\begin{gather}
\int_{0}^{2\pi}\cdots\int_{0}^{2\pi}
\left[{\hat B}^{kj}_{[0][{\bf S}]}(W) B_{j[{\bf p}](1)}
\left({{\bf S}(W)\over\epsilon}+\bm{\theta},
W\right)\right] \left.
{\delta\{g_{[\epsilon{\bf B}_{[{\bf r}](1)}]},
J_{[{\bf q}]}\}\over\delta g^{k}(\bm{\theta},W)}
\right|_{{\cal K}[1]}{{\mathrm d}^{m}\theta\over(2\pi)^{m}} \label{ParnTozhd}\\
 {} -\int_{0}^{2\pi}\cdots\int_{0}^{2\pi}
\left[{\hat B}^{kj}_{[0][{\bf S}]}(W) B_{j[{\bf r}](1)}
\left({{\bf S}(W)\over\epsilon}+\bm{\theta},
W\right)\right] \left.
{\delta\{g_{[\epsilon{\bf B}_{[{\bf p}](1)}]},
J_{[{\bf q}]}\}\over\delta g^{k}(\bm{\theta},W)}
\right|_{{\cal K}[1]}{{\mathrm d}^{m}\theta\over(2\pi)^{m}}
=0.\nonumber
\end{gather}

Using the cyclic permutations of the functions ${\bf q}(X)$,
${\bf p}(X)$, and ${\bf r}(X)$ in the identity~\eqref{ParnTozhd},
it's not dif\/f\/icult to see that the right-hand part of the relation~\eqref{gchastTozhdJnak} is identically equal to zero at
${\bf U}(W)\in{\cal S}$.
It's not dif\/f\/icult to see also that
the left-hand side of the expression~\eqref{gchastTozhdJnak}
is a~smooth regular function of the parameters~${\bf U}(W)$.
Using the fact that the set
${\cal S}$ is everywhere dense in the parameter space
${\bf U}$, we can conclude that the left-hand side of equation~\eqref{gchastTozhdJnak} is identically equal to zero under the
conditions of the theorem.

We have, therefore, that under the conditions of the theorem,
the identity~\eqref{JJJskobrazl} implies the relation
\begin{gather*}
\int\left.
{\delta\{J_{[{\bf q}]},J_{[{\bf p}]}\}
\over\delta U^{\gamma}(W)}\right|_{{\cal K}[1]}
\big\{U^{\gamma}(W),J_{[{\bf r}]}\big\}
\big|_{{\cal K}[1]}{\mathrm d}W+\text{c.p.}
\equiv0.
\end{gather*}

Using the relations
\begin{gather*}
\left.
{\delta\{J_{[{\bf q}]},J_{[{\bf p}]}\}
\over\delta U^{\gamma}(W)}\right|_{{\cal K}[1]}=
{\delta\{U_{[{\bf q}]},U_{[{\bf p}]}\}_{\rm DN}
\over\delta U^{\gamma}(W)},\qquad
\big\{U^{\gamma}(W),J_{[{\bf r}]}\big\}
\big|_{{\cal K}[1]}=
\big\{U^{\gamma}(W),U_{[{\bf r}]}\big\}_{\rm DN},
\end{gather*}
we obtain the Jacobi identity for the Dubrovin--Novikov bracket on
the space of the functions~${\bf U}(X)$.
\end{proof}

\begin{remark}
According to relations~\eqref{kUmuskob} the functionals
\begin{gather*}
K^{\alpha}=\int_{-\infty}^{+\infty}
k^{\alpha}\left({\bf U}(X)\right){\mathrm d}X
\end{gather*}
are annihilators of the Dubrovin--Novikov bracket~\eqref{dubrnovb}
since we have $\{K^{\alpha},U_{[{\bf q}]}\}=0$ for any
functional $U_{[{\bf q}]}$.
As a~consequence, we can claim that the
functions $k^{\alpha}({\bf U})$ represent a~part of the f\/lat
coordinates for the metric
$g^{\nu\mu}({\bf U})=\langle A^{\nu\mu}_{1}\rangle$
connected with the Dubrovin--Novikov bracket for the Whitham system.
Besides that, we have
\begin{gather*}
\big\{k^{\alpha}(X),k^{\beta}(Y)\big\}_{\rm DN}
=0
\end{gather*}
according to relations~\eqref{kknulrel}.

The remaining part of the f\/lat coordinates of $g^{\nu\mu}({\bf U})$
and the corresponding annihilators of the bracket~\eqref{dubrnovb}
are def\/ined by concrete form of the initial bracket
\eqref{LocTheorFieldBr} and the family of \mbox{$m$-phase} solutions of
system~\eqref{EvInSyst}.
\end{remark}

Let us prove now the second theorem justifying the invariance of
the Dubrovin--Novikov procedure.

\begin{theorem}\label{Theorem3.2.}
Let $\Lambda$ be a~regular Hamiltonian family of \mbox{$m$-phase}
solutions of~\eqref{EvInSyst}.
Let $(I^{1},\dots,I^{N})$
and $(I^{\prime1},\dots,I^{\prime N})$ be two different complete
Hamiltonian sets of commuting first integrals of~\eqref{EvInSyst}
having the form~\eqref{integ}.
Then the Dubrovin--Novikov brackets
obtained using the sets $(I^{1},\dots,I^{N})$ and
$(I^{\prime1},\dots,I^{\prime N})$ coincide with each other.
\end{theorem}

\begin{proof}
Note that the sets $(I^{1},\dots,I^{N})$,
$(I^{\prime1},\dots,I^{\prime N})$ correspond to the two
systems of coordinates $(U^{1},\dots,U^{N})$,
$(U^{\prime1},\dots,U^{\prime N})$ on the family~$\Lambda$,
given by the averages of the functionals~${\bf I}$ and~${\bf I}^{\prime}$.
We have to show then that the Dubrovin--Novikov
brackets obtained using the set
$(I^{\prime1},\dots,I^{\prime N})$ coincides with the bracket,
obtained using the set $(I^{1},\dots,I^{N})$, after the
corresponding change of coordinates,
\begin{gather*}
U^{\prime\nu}=U^{\prime\nu}({\bf U}).
\end{gather*}

Consider the values of the functionals
\begin{gather*}
J^{\nu}(X)=\int_{0}^{2\pi}\cdots\int_{0}^{2\pi}
P^{\nu}(\bm{\varphi},\epsilon\bm{\varphi}_{X},\dots)
{{\mathrm d}^{m}\theta\over(2\pi)^{m}},
\end{gather*}
and the values of constraints $g^{i}(\bm{\theta},X)$, introduced
with the aid of the functionals ${\bf J}(X)$ by formula~\eqref{gconstr}, as a~coordinate system in the neighborhood of the
submani\-fold~${\cal K}$.

For values of the functionals $J^{\prime\nu}(X)$ on the submanifold
${\cal K}$ we then have the relations
\begin{gather*}
\left.
J^{\prime\nu}(X)\right|_{\cal K}=
U^{\prime\nu}\left({\bf J}(X)\right)+\sum_{l\geq1}\epsilon^{l}j^{\prime\nu}_{(l)}
\left({\bf J},{\bf J}_{X},\dots\right),
\end{gather*}
where $j^{\prime\nu}_{(l)}$ are smooth functions of
$({\bf J},{\bf J}_{X},\dots)$, polynomial in the derivatives and
having degree~$l$.

Expanding the values of $J^{\prime\nu}(X)$ in the neighborhood of
the submanifold ${\cal K}$, we can write
\begin{gather*}
J^{\prime\nu}(X)=
U^{\prime\nu}\left({\bf J}(X)\right)+\sum_{l\geq1}\epsilon^{l}j^{\prime\nu}_{(l)}
\left({\bf J},{\bf J}_{X},\dots\right)
\\
\hphantom{J^{\prime\nu}(X)=}{}
+\int_{-\infty}^{+\infty}
\int_{0}^{2\pi}\cdots\int_{0}^{2\pi}
T^{\prime\nu}_{i}(X,\bm{\theta},Y,\epsilon)\,
g^{i}(\bm{\theta},Y){{\mathrm d}^{m}\theta\over(2\pi)^{m}}{\mathrm d}Y
+O\big({\bf g}^{2}\big),
\end{gather*}
where, according to the form of the constraints~\eqref{gconstr},
we can put
\begin{gather*}
T^{\prime\nu}_{i}(X,\bm{\theta},Y,\epsilon)=
\sum_{l\geq0}\left.
\Pi^{\prime\nu}_{i(l)}\left(
\bm{\varphi}(\bm{\theta},X),
\epsilon\bm{\varphi}_{X}(\bm{\theta},X),\dots\right)
\right|_{\cal K}\epsilon^{l}\delta^{(l)}(X - Y)
\\
\hphantom{T^{\prime\nu}_{i}(X,\bm{\theta},Y,\epsilon)}{}
=\sum_{l\geq0}\left.
{\partial P^{\prime\nu}\over\partial\varphi^{i}_{lx}}
\left(\bm{\varphi}(\bm{\theta},X),
\epsilon\bm{\varphi}_{X}(\bm{\theta},X),\dots\right)
\right|_{\cal K}\epsilon^{l}\delta^{(l)}(X - Y).
\end{gather*}

Considering the functionals
$J^{\prime}_{[{\bf q}]}=\int q_{\nu}(X)J^{\prime\nu}(X){\mathrm d}X$
with arbitrary smooth (compactly supported) functions $q_{\nu}(X)$,
we can write in the vicinity of ${\cal K}$
\begin{gather}\label{qnuJprimenu}
J^{\prime}_{[{\bf q}]}=\int q_{\nu}(X)\bigg[
U^{\prime\nu}\left({\bf J}(X)\right)+\sum_{l\geq1}\epsilon^{l}j^{\prime\nu}_{(l)}
\left({\bf J},{\bf J}_{X},\dots\right)\bigg]{\mathrm d}X \\
{}
+\int\sum_{l\geq0}(-1)^{l}
\epsilon^{l}\left[{{\mathrm d}^{l}\over{\mathrm d}Y^{l}}q_{\nu}(Y)\left.
\Pi^{\prime\nu}_{i(l)}\left(\bm{\varphi}(\bm{\theta},Y),
\epsilon\bm{\varphi}_{Y}(\bm{\theta},Y),\dots\right)
\right|_{\cal K}\right]g^{i}(\bm{\theta},Y)
{{\mathrm d}^{m}\theta\over(2\pi)^{m}}{\mathrm d}Y
+O\big({\bf g}^{2}\big).\nonumber
\end{gather}

The leading term (in $\epsilon$) in the second part of the
expression~\eqref{qnuJprimenu} is given by the expression
\begin{gather*}
\int \! q_{\nu}(Y)\sum_{l\geq0}(-1)^{l}
k^{\alpha_{1}}(Y)\cdots k^{\alpha_{l}}(Y)
\Pi^{\prime\nu}_{i(l)\theta^{\alpha_{1}}\cdots\theta^{\alpha_{l}}}
\! \left(\bm{\Phi}\left(
{{\bf S}(Y)\over\epsilon}+\bm{\theta},Y\right),
\dots\right)g^{i}(\bm{\theta},Y)
{{\mathrm d}^{m}\theta\over(2\pi)^{m}}{\mathrm d}Y
\end{gather*}
and coincides with the value
\begin{gather*}
\iint_{0}^{2\pi}\cdots\int_{0}^{2\pi}
q_{\nu}(Y)\zeta^{\prime(\nu)}_{i[{\bf U}(Y)]}
\left({{\bf S}(Y)\over\epsilon}+\bm{\theta}\right)
g^{i}(\bm{\theta},Y)
{{\mathrm d}^{m}\theta\over(2\pi)^{m}}{\mathrm d}Y,
\end{gather*}
where
\begin{gather*}
\zeta^{\prime(\nu)}_{i[{\bf U}]}(\bm{\theta})=
\left.
\left[{\delta\over\delta\varphi^{i}(\bm{\theta})}
\int_{0}^{2\pi}\cdots\int_{0}^{2\pi} P^{\prime\nu}
\big(\bm{\varphi},k^{\beta}\bm{\varphi}_{\theta^{\beta}},\dots\big)
{{\mathrm d}^{m}\theta\over(2\pi)^{m}}\right]
\right|_{\bm{\varphi}(\bm{\theta})=\bm{\Phi}(\bm{\theta},{\bf U})}.
\end{gather*}

As the values $\zeta^{(\nu)}_{i[{\bf U}]}(\bm{\theta})$, the values
$\zeta^{\prime(\nu)}_{i[{\bf U}]}(\bm{\theta})$ represent regular left
eigenvectors of the opera\-tor~${\hat L}^{i}_{j[{\bf U}]}$ corresponding
to the zero eigenvalues.
In the case of a~complete regular
family of \mbox{$m$-phase} solutions, we have therefore
\begin{gather*}
\zeta^{\prime(\nu)}_{i[{\bf U}]}(\bm{\theta})=
\sum_{q}\Gamma^{\nu}_{q}({\bf U})
\kappa^{(q)}_{i[{\bf U}]}(\bm{\theta})
\end{gather*}
for some functions $\Gamma^{\nu}_{q}({\bf U})$.

We can write, therefore, up to quadratic
terms in ${\bf g}(\bm{\theta},X)$
\begin{gather*}
\int q_{\nu}(X)J^{\prime\nu}(X){\mathrm d}X
=\int q_{\nu}(X)\bigg[
U^{\prime\nu}\left({\bf J}(X)\right)+\sum_{l\geq1}\epsilon^{l}j^{\prime\nu}_{(l)}
\left({\bf J},{\bf J}_{X},\dots\right)\bigg]{\mathrm d}X
\\
\qquad{} +\int q_{\nu}(X)\left[\sum_{q}
\Gamma^{\nu}_{q}({\bf U})\kappa^{(q)}_{i[{\bf U}(X)]}
\left({{\bf S}(X)\over\epsilon}+\bm{\theta}\right)+
O(\epsilon)\right]g^{i}(\bm{\theta},X)
{{\mathrm d}^{m}\theta\over(2\pi)^{m}}{\mathrm d}X
+O\big({\bf g}^{2}\big).
\end{gather*}

Consider the Poisson brackets
\begin{gather*}
\big\{J^{\prime}_{[{\bf q}]},J^{\prime}_{[{\bf p}]}
\big\}\big|_{\cal K}=
\left.
\left\{\int q_{\nu}(X)J^{\prime\nu}(X){\mathrm d}X
, \int p_{\mu}(Y)J^{\prime\mu}(Y){\mathrm d}Y\right\}
\right|_{\cal K}
\\
\qquad{}
=\int q_{\nu}(X)
{\partial U^{\prime\nu}\over\partial U^{\lambda}}(X)\left[
\left.
\left\{J^{\lambda}(X),
\int p_{\mu}(Y)J^{\prime\mu}(Y){\mathrm d}Y\right\}
\right|_{\cal K}+O\big(\epsilon^{2}\big)\right]{\mathrm d}X
\\
{}
+\int
q_{\nu}(X)\sum_{q}\Gamma^{\nu}_{q}({\bf U}(X))\left[
\kappa^{(q)}_{i[{\bf U}(X)]}
\left({{\bf S}(X)\over\epsilon}+\bm{\theta}\right)+
O(\epsilon)\right]
\\
\left.
\qquad{}
 \times
\left\{g^{i}(\bm{\theta},X),
J^{\prime\mu}(Y)\right\}\right|_{\cal K}p_{\mu}(Y)
{{\mathrm d}^{m}\theta\over(2\pi)^{m}}{\mathrm d}X{\mathrm d}Y.
\end{gather*}

By Lemma 2.4$^{\prime}$ we have the relation
$\{g^{i}(\bm{\theta},X),J^{\prime}_{[{\bf p}]}\}|_{\cal K}
=O(\epsilon)$ on the submanifold ${\cal K}$.
In addition,
completely analogous to the relation~\eqref{LeftVecAort} holds
the relation
\begin{gather*}
\int_{0}^{2\pi}\cdots\int_{0}^{2\pi}
\kappa^{(q)}_{i[{\bf U}(X)]}
\left({{\bf S}(X)\over\epsilon}+\bm{\theta}\right)
\big\{g^{i}(\bm{\theta},X),
J^{\prime}_{[{\bf p}]}\big\}|_{{\cal K}[1]}
{{\mathrm d}^{m}\theta\over(2\pi)^{m}}\equiv 0
\end{gather*}
by virtue of the original dependence of the constraints
$g^{i}(\bm{\theta},X)$.
We thus obtain
\begin{gather*}
\iint q_{\nu}(X)
\big\{J^{\prime\nu}(X)
,J^{\prime\mu}(Y)\big\}\big|_{\cal K}p_{\mu}(Y)
{\mathrm d}X{\mathrm d}Y
\\
\qquad{} =\int q_{\nu}(X)
{\partial U^{\prime\nu}\over\partial U^{\lambda}}(X)
\left.
\left\{J^{\lambda}(X),
\int p_{\mu}(Y)J^{\prime\mu}(Y){\mathrm d}Y\right\}\right|_{\cal K}
{\mathrm d}X+O\big(\epsilon^{2}\big).
\end{gather*}

Repeating the arguments for the functional
$\int p_{\mu}(Y)J^{\prime\mu}(Y){\mathrm d}Y$
we f\/inally obtain
\begin{gather}
\iint q_{\nu}(X)
\big\{J^{\prime\nu}(X)
,J^{\prime\mu}(Y)\big\}\big|_{\cal K}p_{\mu}(Y)
{\mathrm d}X{\mathrm d}Y \nonumber\\
\qquad
=\iint q_{\nu}(X)
{\partial U^{\prime\nu}\over\partial U^{\lambda}}(X)
\big\{J^{\lambda}(X),
J^{\gamma}(Y)\big\}\big|_{\cal K}
{\partial U^{\prime\mu}\over\partial U^{\gamma}}(Y)
p_{\mu}(Y){\mathrm d}X{\mathrm d}Y+O\big(\epsilon^{2}\big).\label{JprimeJrel}
\end{gather}

Given that the principal (in $\epsilon$) terms in the expressions
$\{J^{\lambda}(X),J^{\gamma}(Y)\}|_{\cal K}$ and \linebreak
$\{J^{\prime\nu}(X),J^{\prime\mu}(Y)\}|_{\cal K}$
coincide with the Dubrovin--Novikov brackets, obtained with the aid of
the sets $(I^{1},\dots,I^{N})$ and $(I^{\prime1},\dots,I^{\prime N})$
respectively, we conclude from~\eqref{JprimeJrel} that
\begin{gather*}
\big\{U^{\prime\nu}(X),
U^{\prime\mu}(Y)\big\}^{\prime}_{\rm DN}
={\partial U^{\prime\nu}\over\partial U^{\lambda}}(X)
\big\{U^{\lambda}(X),U^{\gamma}(Y)\big\}_{\rm DN}
{\partial U^{\prime\mu}\over\partial U^{\gamma}}(Y),
\end{gather*}
which means the coinciding of the brackets $\{\cdot,\cdot\}_{\rm DN}$
and $\{\cdot,\cdot\}^{\prime}_{\rm DN}$.
\end{proof}

Finally, we prove the theorem about the Hamiltonian properties of the
Whitham system~\eqref{ConsWhitham} under the same conditions as before.

\begin{theorem}\label{Theorem3.3.}
Let $\Lambda$ be a~regular Hamiltonian family of \mbox{$m$-phase}
solutions of~\eqref{EvInSyst}.
Let $(I^{1},\dots,I^{N})$ be
a complete Hamiltonian set of commuting first integrals of
\eqref{EvInSyst} having the form~\eqref{integ} and $H$ be the
Hamiltonian function of the system~\eqref{EvInSyst} having the form~\eqref{hamfun}.
Then the Whitham system~\eqref{ConsWhitham}
is Hamiltonian with respect to the corresponding Dubrovin--Novikov
bracket~\eqref{dubrnovb} with the Hamiltonian function~\eqref{UsrHamFunc}
\begin{gather*}
H^{av}=\int_{-\infty}^{+\infty}\langle P_{H}\rangle
\left({\bf U}(X)\right){\mathrm d}X.
\end{gather*}
\end{theorem}

\begin{proof}
By Theorem~\ref{Theorem3.2.},
without loss of generality we can assume that
the Hamiltonian functional $H$ belongs to the set $(I^{1},\dots,I^{N})$,
$H=I^{\mu_{0}}$.
It is easy to verify that the corresponding
Hamiltonian $H^{av}$ generates in this case the system
\begin{gather*}
U^{\nu}_{T}={d\over{\mathrm d}X}\left[
\langle Q^{\nu\mu_{0}}\rangle\left({\bf U}(X)\right)\right],
\end{gather*}
i.e.\
exactly system~\eqref{ConsWhitham}.
\end{proof}

As the simplest example of Theorems~\ref{Theorem3.1.}--\ref{Theorem3.3.}
consider the procedure of averaging of the Gardner--Zakharov--Faddeev
bracket for the KdV equation.

Consider the Gardner--Zakharov--Faddeev bracket
\begin{gather}\label{GardZakhFadd}
\{\varphi(x),\varphi(y)\}=
\delta^{\prime}(x-y)
\end{gather}
for the KdV equation
\begin{gather*}
\varphi_{t}=\varphi\varphi_{x}-
\varphi_{xxx}
\end{gather*}
with the Hamiltonian functional
\begin{gather*}
H=\int\left({\varphi^{3}\over6}+
{\varphi_{x}^{2}\over2}\right){\rm d} x.
\end{gather*}

As is well known, the KdV equation has a~family of \mbox{$m$-phase} solutions
for any $m$, given by the Novikov potentials, which represent the stationary
points for the higher KdV f\/lows.
According to the Hamiltonian structure~\eqref{GardZakhFadd} it is also equivalent to the extremality of a~linear
combination of higher integrals of the KdV
\begin{gather}\label{NovPot}
c_{1}\delta I^{1}+c_{2}\delta I^{2}+\cdots
+c_{m+2}\delta I^{m+2}=0,
\end{gather}
where
\begin{gather*}
I^{1}=N=
\int\varphi {\mathrm d}x
\end{gather*}
is the annihilator of the bracket~\eqref{GardZakhFadd},
\begin{gather*}
I^{2}=P=
\int{\varphi^{2}\over2}{\mathrm d}x
\end{gather*}
is the momentum functional of the bracket~\eqref{GardZakhFadd},
$I^{3}=H$, and $I^{k}$, $k\geq4$ are higher integrals of the KdV
equation.

As was shown in~\cite{NovikovFuncAn}, all the systems~\eqref{NovPot}
are completely integrable f\/inite-dimensional systems with
quasiperiodic solutions in the general case.
As is well known, the
corresponding \mbox{$m$-phase} solutions of KdV can be represented in the form~\eqref{phasesol} with the theta-functional expressions for the functions
$\Phi(\bm{\theta},{\bf U})$,
satisfying the system~\cite{DubrNovZhETP, DubrNovDokl, ItsMatveev1, ItsMatveev2,
Dubrovin1, DubrMatvNov, Dubrovin2}
\begin{gather}\label{mphaseKdV}
\omega^{\alpha}({\bf U})\Phi_{\theta^{\alpha}}=
k^{\alpha}({\bf U})\Phi\Phi_{\theta^{\alpha}}-
k^{\alpha}({\bf U})k^{\beta}({\bf U})
k^{\gamma}({\bf U})
\Phi_{\theta^{\alpha}\theta^{\beta}\theta^{\gamma}}.
\end{gather}

The parameters of the solution $(E_{1},\dots,E_{2m+1})$,
$E_{1}<E_{2}<\dots<E_{2m+1}$ represent the branching points
of the Riemann surface of genus $m$, and,
together with the initial phases
$(\theta_{0}^{1},\dots,\theta_{0}^{m})$, completely determine the
corresponding solution $\varphi(x,t)$ \cite{NovikovFuncAn}.

The eigenmodes of the linearized operator~\eqref{mphaseKdV},
as well as the adjoint operator, were studied in detail
 \cite{KricheverDAN,dobr1,dobr2,krichev1,krichev2}.
In particular, we
can state that the families of \mbox{$m$-phase} solutions of KdV are complete
regular families in the sense of Def\/inition~\ref{Definition1.1.}
and the corresponding
systems~\eqref{ConsWhitham} for any independent set of $2m+1$
higher integrals of KdV $(I^{k_{1}},\dots,I^{k_{2m+1}})$
represent the regular Whitham systems for the KdV equation
in the sense def\/ined above.
It is well known~\cite{whith3, ffm}, that the Whitham
system for KdV in \mbox{$m$-phase} case can be written in the diagonal
form where the parameters $E_{1},\dots,E_{2m+1}$ are its
Riemann invariants.

Regular zero modes of the linearized operator~\eqref{mphaseKdV}
on the torus $\mathbb{T}^{m}$ are given by the functions
$\Phi_{\theta^{\alpha}}(\bm{\theta},{\bf U})$,
$\theta^{\alpha}=1,\dots,m$ and by the function
$\partial\Phi/\partial n$ (at constant ${\bf k}$ and
$\bm{\omega}$), where $n$ is the value of the functional $N$
on the corresponding solution.
Similarly, regular zero modes of the
adjoint linear operator on the torus $\mathbb{T}^{m}$ are given by
any $m+1$ linearly independent variational derivatives of the
higher integrals of KdV on the family of the functions
$\Phi(\bm{\theta},{\bf U})$.

The families of Novikov potentials are obviously regular Hamiltonian
families with respect to the bracket~\eqref{GardZakhFadd}, while any
set of $2m+1$ independent higher integrals of KdV is a~complete set
of commuting functionals~\eqref{integ} on such a~family.
We also note
that the variational derivative of the annihilator of the bracket
\eqref{GardZakhFadd} always appears on the family of \mbox{$m$-phase} solutions
in the form of a~linear combination of $m+1$ variational derivatives
of higher integrals of KdV according to the original construction of
the Novikov potentials.

Investigation of the system~\eqref{BAcondNextChapt} can be carried out
in this case as follows.
We f\/irst note that the operator
${\hat B}_{[0]}(X)$ is given in this case as
\begin{gather*}
{\hat B}_{[0]}(X)=
k^{1}(X){\partial\over\partial\theta^{1}}+\cdots+
k^{m}(X){\partial\over\partial\theta^{m}}
\end{gather*}
and, thus, is the derivative along the vector f\/ield
$(k^{1}(X),\dots,k^{m}(X))$ on the torus $\mathbb{T}^{m}$.

The regular zero mode of the operator ${\hat B}_{[0]}(X)$ is
a constant function on the torus and is orthogonal to the functions
$A_{[1][{\bf q}]}(\bm{\theta},X)$ by Lemma~\ref{Lemma3.1.}.
This
fact is easily verif\/ied also by direct computation.
Indeed, the functions
$A_{[1][{\bf q}]}(\bm{\theta},X)$ have in this case the form
\begin{gather*}
A_{[1][{\bf q}]}\!\left({{\bf S}(X)\over\epsilon}+\bm{\theta},
X\right)\!
=
\big\{\varphi(\bm{\theta},X),
J_{[{\bf q}]}\big\}\big|_{{\cal K}[1]}\!-
\Phi_{U^{\nu}}\!\left({{\bf S}(X)\over\epsilon}+\bm{\theta},
{\bf U}(X)\right) \!
\big\{U^{\nu}(X),J_{[{\bf q}]}\big\}
\big|_{{\cal K}[1]}.
\end{gather*}

Let us assume for simplicity that the functional
\begin{gather*}
N(X)=
\int_{0}^{2\pi}\cdots\int_{0}^{2\pi}
\varphi(\bm{\theta},X){{\mathrm d}^{m}\theta\over(2\pi)^{m}}
\end{gather*}
is included in the coordinate system ${\bf U}(X)$, $N(X)=U^{1}(X)$,
which, by Theorem~\ref{Theorem3.2.},
does not af\/fect the Dubrovin--Novikov bracket.
We have then
\begin{gather*}
\int_{0}^{2\pi}\!\cdots\int_{0}^{2\pi} \!
\Phi_{U^{1}}(\bm{\theta},{\bf U}(X))
{{\mathrm d}^{m}\theta\over(2\pi)^{m}}\equiv1
,\!\qquad\!
\int_{0}^{2\pi}\!\cdots\int_{0}^{2\pi} \!
\Phi_{U^{\nu}}(\bm{\theta},{\bf U}(X))
{{\mathrm d}^{m}\theta\over(2\pi)^{m}}\equiv0
,\!\qquad\nu\neq1.
\end{gather*}

It is easy to see then that
\begin{gather*}
\int_{0}^{2\pi}\cdots\int_{0}^{2\pi}
A_{[1][{\bf q}]}(\bm{\theta},X)
{{\mathrm d}^{m}\theta\over(2\pi)^{m}}=
\big\{N(X),
J_{[{\bf q}]}\big\}\big|_{{\cal K}[1]}
-
\big\{N(X),
J_{[{\bf q}]}\big\}\big|_{{\cal K}[1]}
\equiv0.
\end{gather*}

The operator ${\hat B}_{[0]}(X)$, however, has also irregular zero
modes which arise in the cases when the closure of trajectories of the
vector f\/ield $(k^{1}(X),\dots,k^{m}(X))$ on $\mathbb{T}^{m}$
represent the lower-dimensional tori
$\mathbb{T}^{k}\subset\mathbb{T}^{m}$.
Modes of this type arise
obviously on a~set of measure zero in the parameter space
${\bf U}$, i.e.~$\{{\bf U}\}/{\cal M}$.

We assume here, without proof, following easily verif\/iable from
the theta-functional representation fact.
The Fourier harmonics $A^{n_{1}\dots n_{m}}_{[1][{\bf q}]}(X)$
of the functions $A_{[1][{\bf q}]}(\bm{\theta},X)$
in the variables~$\bm{\theta}$
\begin{gather*}
A_{[1][{\bf q}]}(\bm{\theta},X)\equiv
\sum_{n_{1},\dots,n_{m}}
A^{n_{1}\dots n_{m}}_{[1][{\bf q}]}(X)
\exp\big(i n_{1}\theta^{1}+\cdots+
i n_{m}\theta^{m}\big)
\end{gather*}
decay faster than any power of $|{\bf n}|$ at
$|{\bf n}|\rightarrow\infty$, where
\begin{gather*}
|{\bf n}|\equiv
\sqrt{n_{1}^{2}+\cdots+n_{m}^{2}}.
\end{gather*}

Let us def\/ine the Diophantine conditions for an arbitrary set of
values $(k^{1}({\bf U}),\dots,k^{m}({\bf U}))$.
The vector
$(k^{1}({\bf U}),\dots,k^{m}({\bf U}))$ is a~Diophantine vector
with the index $\nu>0$ and the coef\/f\/icient $A>0$, if
\begin{gather*}
\left|n_{1}k^{1}({\bf U})+\cdots+
n_{m}k^{m}({\bf U})\right|\geq
A|{\bf n}|^{-\nu}
\end{gather*}
for all $(n_{1},\dots,n_{m})\in\mathbb{Z}^{m}$
$((n_{1},\dots,n_{m})\neq(0,\dots,0))$.

Let us use now the following well known theorem (see, e.g.,~\cite{Arnold, Shmidt}):

{\it For $\nu>m-1$ the measure of the set of non-Diophantine vectors
$(k^{1},\dots,k^{m})$ in $\mathbb{R}^{m}$ is equal to zero.}

Using the condition
$\operatorname{rank}||\partial k^{\alpha}/\partial U^{\nu}||=m$
we can also state that the measure of the corresponding parameters
${\bf U}$, such that ${\bf k}({\bf U})$ are non-Diophantine vectors
with index $\nu>m-1$, is also zero in the parameter space ${\bf U}$.

We can now def\/ine the sets ${\cal S}_{\nu}$ in the space of the
parameters ${\bf U}$, such that the wave vectors
$(k^{1}({\bf U}),\dots,k^{m}({\bf U}))$ are Diophantine
with index $\nu$ if ${\bf U}\in{\cal S}_{\nu}$.
In this case all ${\cal S}_{\nu}$ with index $\nu>m-1$
are everywhere dense, and also have the full measure in the parameter
space ${\bf U}$.

It is easy to see also that the system~\eqref{BAcondNextChapt}
is resolvable in the space of $2\pi$-periodic in all $\theta^{\alpha}$
functions for each ${\bf U}\in{\cal S}_{\nu}$.
Indeed, due to the
absence of the zero Fourier harmonic of the function
$A_{[1][{\bf q}]}(\bm{\theta},X)$ and conditions of decreasing
of Fourier coef\/f\/icients $A_{[1][{\bf q}]}^{n_{1}\dots n_{m}}(X)$
we can put
\begin{gather*}
B_{[{\bf q}](1)}^{n_{1}\dots n_{m}}(X)={1\over
i n_{1}k^{1}({\bf U})+\cdots+
i n_{m}k^{m}({\bf U})}\,
A_{[1][{\bf q}]}^{n_{1}\dots n_{m}}(X)
\end{gather*}
for $(n_{1},\dots,n_{m})\neq(0,\dots,0)$ and restore
the solution
\begin{gather*}
B_{[{\bf q}](1)}(\bm{\theta},X)=
\sum_{n_{1},\dots,n_{m}}^{\prime}
B_{[{\bf q}](1)}^{n_{1}\dots n_{m}}(X)
\exp(i n_{1}\theta^{1}+\cdots+i n_{m}\theta^{m}),\\
 (n_{1},\dots,n_{m})\neq(0,\dots,0)
\end{gather*}
as a~smooth $m$-periodic function of $\bm{\theta}$.

Assuming, therefore, ${\cal S}={\cal S}_{\nu}$ for any
$\nu>m-1$, we meet all the conditions of
Theorem~\ref{Theorem3.1.}.

We can thus formulate the following proposition.

\begin{proposition}
The Dubrovin--Novikov procedure is well justified in
the averaging of the Gardner--Zakharov--Faddeev bracket on the
\mbox{$m$-phase} solutions of KdV for any $m$ and provides a~local
Hamiltonian structure of hydrodynamic type for the corresponding
regular Whitham system.
\end{proposition}

We note here that both local and weakly nonlocal Hamiltonian
structures of the Whitham hierarchy for KdV were investigated in
detail in the papers~\cite{Alekseev, Pavlov2}. 

In the more general case the constructing of the set ${\cal S}$
requires the study of the eigenfunctions and the eigenvalues of the
Hamilton operator ${\hat B}^{ij}_{[0]}({\bf U})$ on the manifold
of \mbox{$m$-phase} solutions depending on the values of the parameters
${\bf U}$.
For \mbox{$m$-phase} solutions, given by the algebraic-geometric
families, in this case are very convenient methods similar to
those used in~\cite{KricheverDAN, krichev2, VorobevDobr, dobr1, dobr2}
to study the spectrum of auxiliary linear operators of
integrable hierarchies.

\subsection*{Acknowledgements}

In conclusion the author expresses his deep gratitude to
Professors S.P.~Novikov, B.A.~Dubrovin, I.M.~Krichever,
S.Yu.~Dobrokhotov, and M.V.~Pavlov for fruitful discussions.
This work was f\/inancially supported by the Russian Federation Government
Grant No.~2010-220-01-077, Grant of the President of Russian Federation
NSh-4995.2012.1, and Grant RFBR No.~11-01-12067-of\/i-m-2011.

\pdfbookmark[1]{References}{ref}
\LastPageEnding

\end{document}